%% file: iscdm.tex
\documentclass[11pt,onecolumn,twoside,notitlepage,fleqn,a4paper,final]
	      {article}[]
\usepackage{graphicx}
\usepackage[dvips]{epsfig}
\usepackage{pstcol,pst-text}
\usepackage{xspace}
\usepackage{bm}
\usepackage{pstricks}
\usepackage{latexsym}
\usepackage{amsfonts}
\usepackage{amsmath}
\usepackage{amssymb}
\usepackage{cite}
\usepackage[mathscr]{eucal}
\newcommand{\bit}{\begin{itemize}}
\newcommand{\eit}{\end{itemize}}
\newcommand{\bc}{\begin{center}}
\newcommand{\ec}{\end{center}}
\newcommand{\bq}{\begin{quote}}
\newcommand{\eq}{\end{quote}}
\newcommand{\bi}{\begin{itemize}}
\newcommand{\ei}{\end{itemize}}
\newcommand{\be}{\begin{equation}}
\newcommand{\ee}{\end{equation}}
\newcommand{\bea}{\begin{eqnarray}}
\newcommand{\eea}{\end{eqnarray}}
\newcommand{\bt}{\begin{tabular}}
\newcommand{\et}{\end{tabular}}
\newcommand{\bmp}{\begin{minipage}}
\newcommand{\emp}{\end{minipage}}
\newcommand{\btab}{\begin{tabbing}}
\newcommand{\etab}{\end{tabbing}}
\newcommand{\spc}{$\,$}
\newcommand{\x}{\:\!}
\newcommand{\z}{\;\!}

\newcommand{\vs}{\varsigma}
\newcommand{\td}{\tilde}
\def\nn{\nonumber}

\def\mr{\mathrm}
\def\cal{\mathcal}
\def\eg{\rm e.g.\ }
\def\ie{\rm i.e.\ }
\def\il{\mr{i}}
\def\fl{\mr{f}}

\def\sd#1{\scalebox{0.77}{#1}}
\def\eq#1{eq.~(\ref{#1})}

\def\eqs#1{eqs.~(\ref{#1})}
\def\eqss#1#2{eqs.~(\ref{#1}) and (\ref{#2})}
\def\eqsss#1#2#3{eqs.~(\ref{#1}), (\ref{#2}) and (\ref{#3})}

\def\eqsspss#1#2#3#4{eqs.~(\ref{#1}), (\ref{#2}) and (\ref{#3}), (\ref{#4})}

\def\fig#1{fig.~\ref{#1}}
\def\Fig#1{Fig.~\ref{#1}}

\def\Figs#1{Figs.~\ref{#1}}
\def\tab#1{tab.~\ref{#1}}

\def\sec#1{sec.~\ref{#1}}

\def\secss#1#2{secs.~\ref{#1} and \ref{#2}}
\def\secsss#1#2#3{secs.~\ref{#1}, \ref{#2} and \ref{#3}}
\def\smufp{\mbox{$\mu^2_{\mr F}$\xspace}}
\def\smur{\mbox{$\mu^2_{\mr R}$\xspace}}
\def\muf{\mbox{$\tilde\mu_{\mr F}$\xspace}}
\def\mufp{\mbox{$\mu_{\mr F}$\xspace}}
\def\mur{\mbox{$\mu_{\mr R}$\xspace}}

\newcommand{\abs}[1]{\lvert#1\rvert}

\newcommand{\coo}[1]{\ensuremath{\left|{\hphantom{A}\atop\sd{\rm{#1}}}\right.}}
\newcommand{\con}[2]
	   {\ensuremath{#1\left|{\hphantom{A}\atop\sd{\rm{#2}}}\right.}}
\newcommand{\myfigcaption}[2]{%
  \refstepcounter{figure}
  \vspace{0.0mm}
  \begin{center}
    \small
    \parbox{#1}{\rm{\bf Figure \thefigure: }#2}
  \end{center}
}

\newcommand{\mytabcaption}[2]{%
  \refstepcounter{table}
  \vspace*{0.0mm}
  \begin{center}
    \small
    \parbox{#1}{\rm{\bf Table \thetable: }#2}
  \end{center}
}

\newcommand{\ariadne}{A\scalebox{0.8}{RIADNE}\xspace}
\newcommand{\herwig}{H\scalebox{0.8}{ERWIG}\xspace}
\newcommand{\herwigpp}{H\scalebox{0.8}{ERWIG++}\xspace}
\newcommand{\pythia}{P\scalebox{0.8}{YTHIA}\xspace}
\newcommand{\sherpa}{S\scalebox{0.8}{HERPA}\xspace}
\newcommand{\apacic}{A\scalebox{0.8}{PACIC++}\xspace}
\newcommand{\delphi}{D\scalebox{0.8}{ELPHI}\xspace}
\begin{document}
%
%
\title{\ \\[17mm]Initial-state showering based on colour dipoles connected
  to incoming parton lines}
\author{\hphantom{MisterX}\\[1mm]
  Jan-Christopher Winter
  \thanks{jwinter@fnal.gov}
  \\[2mm]
  {\small Fermi National Accelerator Laboratory, Batavia, IL, USA}\\[7mm]
  Frank Krauss
  \thanks{frank.krauss@durham.ac.uk}
  \\[2mm]
  {\small Institute for Particle Physics Phenomenology, Durham University, UK}
  \\[10mm]}
\date{\today}
\maketitle
\thispagestyle{empty}
\begin{flushright}
  \vspace*{-130mm}
  {\small FERMILAB-PUB-07-661-T}\\
  {\small DCPT/07/90}\\
  {\small IPPP/07/45}\\[117mm]
\end{flushright}
\renewcommand{\baselinestretch}{1.1}
\begin{abstract}
\noindent
A parton-shower model for hadronic collisions based on the emission
properties of QCD dipoles is proposed.  This proposal therefore
extends the well-known radiation pattern of pure final-state colour
dipoles to QCD initial-state radiation, both of which are treated
perturbatively.  Corresponding dipole splitting functions are derived
and the kinematics of all dipole splittings is discussed.  Application
to hadron production in electron-positron annihilation, to Drell-Yan
lepton-pair and QCD jet production yields encouraging results.
\end{abstract}
\clearpage
\renewcommand{\baselinestretch}{1.1}
\tableofcontents
\clearpage
\renewcommand{\baselinestretch}{1.0}
%
%
%
%
%
%
\input{tex/dsintro.tex}
\input{tex/dscdm.tex}
\input{tex/dsnew.tex}
\input{tex/dsff.tex}
\input{tex/dsii.tex}
\input{tex/dsfi.tex}
\input{tex/dsmod.tex}
\input{tex/dsres.tex}
\input{tex/dscon.tex}
\input{tex/dsackno.tex}
\bibliographystyle{JHEP-2}
{\raggedright\bibliography{bibo/MCcomp,bibo/ps,bibo/exa,bibo/clu}}
%
%
%
%
%
%
\end{document}

%% file: tex/dsintro.tex
\section{Introduction}
%
\label{sec:dsintro}
In the past decades, Monte Carlo event generators including QCD
parton-shower routines, such as \pythia
\cite{Sjostrand:1993yb,Sjostrand:2000wi,Sjostrand:2001yu,Sjostrand:2003wg,Sjostrand:2006za},
\herwig \cite{Corcella:2000bw,Corcella:2002jc}, or \ariadne
\cite{Lonnblad:1992tz} have been very successful in correctly
describing, both qualitatively and quantitatively, a large range of
QCD-related phenomena at different colliders, at different energies,
and with different initial states.  The success of these programs is
based on good approximations in their treatment of logarithmically
enhanced emission of QCD particles in soft and/or collinear regions of
phase space.
\\
In conventional parton showers such as the ones in \pythia
\cite{Sjostrand:1985xi,Bengtsson:1986gz,Bengtsson:1987kr} and \herwig 
\cite{Marchesini:1987cf}, this is achieved by an expansion around the
collinear limit.  This manifests itself in the ordering of subsequent
emissions by virtual masses supplemented with an explicit veto on
increasing emission angles or by an ordering by emission angles,
respectively.\spc\footnote{
  The current parton shower implementation of \sherpa
  \cite{Gleisberg:2003xi}, \apacic \cite{Krauss:2005re}, is very
  similar to the well-established virtuality-ordered \pythia shower.}
Alternatively, perturbative QCD cascades can be formulated in terms of 
splitting colour dipoles rather than partons.  This has been realized
in the shower algorithm in \ariadne \cite{Lonnblad:1992tz}, which is
based on the Colour Dipole Model (CDM)
\cite{Gustafson:1986db,Gustafson:1987rq,Andersson:1989ki}.  Splitting the
dipoles and ordering the emissions in relative transverse momenta of 
subsequent splittings is equivalent to an expansion around the soft limits 
of the radiation process.  In \cite{Gustafson:1986db} it has been argued that 
such a dipole shower quite naturally fulfils the requirements of quantum 
coherence, which, for the parton showers, lead to angular ordering of
subsequent emission, see \eg \cite{Ellis:1991qj}.  It is interesting 
to note that the \ariadne shower yields results, which show a similar
or even better agreement with data from electron--positron
annihilation into hadrons \cite{Weierstall:1995uz,Siebel:2006np,Akrawy:1990yx,Abreu:1996na,Knowles:1995kj}.
However, in the CDM initial-state radiation (ISR), \ie parton emission
off incoming partons, is not treated explicitly but taken into account
by redefining ISR as final-state radiation (FSR) off hadron remnants
\cite{Andersson:1989gp}.  To correctly model ISR in this picture,
non-perturbative corrections have to be applied, cf.\ \sec{sec:CDM}.
Equipped with such non-perturbative components in its modelling of 
initial-state associated radiation, \ariadne also succeeded in
describing a wealth of DIS data in a very reassuring way, see for
instance \cite{Magnussen:1991xy}.  To some extent, the reason for this
excellent performance in describing $e^+e^-$ and DIS data is not
entirely understood.  The cause could be better treatment of
small-$x$\/ effects in the DIS case, which are assumed to be an issue
also for the forthcoming LHC.  Equally well, it could be just the effect 
of a careful tuning of the additional non-perturbative parameters in the
case of DIS.  Another idea is related to a supposedly improved simulation 
of single, potentially non-global, potentially large logarithms stemming 
from soft corners of emission phase space.  This appears as a consequence 
of the fact that the leading $1/N_\mr{C}$ terms of such contributions are
better accounted for if the description of the radiation is based on
the dipole structure in both matrix element and phase space
\cite{Banfi:2006gy}.  This blurred picture of, on the one hand,
delivering overwhelming agreement with data of various measurements
and, on the other hand, lacking clear determination of the reason for
this success provides a fair, but not the only motivation for trying
out an alternative path in modelling ISR arising from colour dipoles.
\\
In view of the upcoming LHC era, Monte Carlo event generators are
undergoing an intensive overhaul, leading essentially to complete
rewrites of the codes
\cite{Gieseke:2003hm,Gieseke:2006ga,Lonnblad:2006hm,Sjostrand:2007gs}
or to the construction of entirely new programs such as \sherpa
\cite{Gleisberg:2003xi} in the modern, object-oriented programming
language {\tt C++}.  Apart from issues related to maintenance, a
number of improvements concerning physics simulation motivated the
construction of new event generators.  First of all, the shower
algorithms themselves, forming an essential part of the event
generators, have been improved:
in \pythia, a $k_\perp$ ordered parton shower has been introduced 
\cite{Sjostrand:2003wg,Sjostrand:2004ef} in order to better account 
for coherence effects. There is also a dramatically extended model of
multiple parton interactions. In \herwig, a new formulation of angular
ordering \cite{Gieseke:2003rz} better embeds Lorentz invariance and
provides an improved treatment of those regions, where the original
\herwig shower over- or undercounted parton emissions. In addition, 
a new parton-shower formulation has been developed based on
Catani--Seymour dipole factorization
\cite{Nagy:2005aa,Nagy:2006kb,Dinsdale:2007mf,Schumann:2007mg},
and steps have been undertaken in the development of yet another
QCD shower formulation, which uses antenna functions \cite{Giele:2007di}.
For all these recent developments, a common denominator has been to
put more emphasis on the notion of a colour-connected partner of the
splitting parton and thus a reduction of the difference between
parton and dipole showers. Especially, the showers based on either
Catani--Seymour \cite{Catani:1996vz,Catani:2002hc} or antenna
subtraction kernels \cite{Kosower:1997zr,GehrmannDeRidder:2005cm}, aim
at an improved matching with exact higher-order QCD matrix elements.
In fact, considering the need for increased precision, this systematic
inclusion of higher orders in the perturbative expansion of QCD has
been a dramatic and recent improvement of the paradigm underlying
building and using multipurpose Monte Carlo event generators.
\\
In the matching approach, the exact next-to-leading order
matrix-element result is consistently combined with the resummation of
the parton shower
\cite{Frixione:2002ik,Frixione:2003ei,Nason:2006hf,Frixione:2007vw}
such that the overall result correctly reproduces the corresponding 
NLO total cross section and the first additional hard QCD emission.  
This has been first implemented for specific processes in {\tt MC@NLO}
\cite{Frixione:2006gn} on the basis of the Frixione--Kunszt--Signer
subtraction \cite{Frixione:1995ms}.  This method depends to some extent 
on the details of the parton shower and also has some residual dependence 
on the process in question.  With the {\tt POWHEG}-approach a 
shower-independent matching solution \cite{Nason:2006hf,Frixione:2007vw} 
extends the original {\tt MC@NLO} proposal and the appearance of
negative weights, which are present in the former method, can be
circumvented.
\\
In alternative approaches, sequences of tree-level multileg matrix
elements with increasing final-state multiplicity are merged with the
parton shower to yield fully inclusive samples correct at leading
logarithmic accuracy by avoiding double-counting and missing phase-space
regions.  A first approach, known as the CKKW merging approach, has
been presented for the case of electron--positron annihilations into
jets in \cite{Catani:2001cc}; later it has been extended to hadronic
collisions \cite{Krauss:2002up} and it has been reformulated to a
merging procedure -- called LCKKW -- in conjunction with a dipole
shower \cite{Lonnblad:2001iq}.  A further method, the MLM method, 
has been developed also aiming at a merging of matrix elements and parton
showers.  It uses a different way in generating the inclusive samples 
based on a geometric interpretation of the full radiation pattern in 
terms of cone jets \cite{Mangano:2001xp,Mangano:2006rw}.  These different 
algorithms have been implemented in different variations on different 
levels of sophistication in conjunction with various matrix-element 
generators or already in full-fledged event generators, see \eg
\cite{Gleisberg:2003xi,Krauss:2004bs,Schalicke:2005nv,Lavesson:2005xu,%
  Mrenna:2003if,Mangano:2002ea,Stelzer:1994ta,Maltoni:2002qb,Alwall:2007st,%
  Kanaki:2000ey,Papadopoulos:2005ky}.
Despite their differences they exhibit an assuring level of agreement
\cite{Alwall:2007fs}.
\\
In all these new approaches, parton emissions from matrix elements at
a given perturbative order have to be balanced with corresponding
emissions from a shower algorithm.  Intuitively one may anticipate
that dipole-like kinematics, leaving all particles of the splitting on
their mass shells, may facilitate simpler procedures for this
balancing.  Furthermore, concerning matching, the CDM seems to be the
more natural partner to the matrix-element part of calculations based
on a subtraction method using antenna factorization \cite{Kosower:1997zr}.  
\\
In this publication, therefore, an extension of the ``perturbative''
dipole shower \cite{Gustafson:1987rq} as implemented in \ariadne
\cite{Lonnblad:1992tz} to truly perturbative initial-state radiation 
is proposed, in contrast to the original ISR Lund CDM.  Hence, the goal 
is to formulate the QCD evolution of a hard process initiated through 
a hadronic collision entirely perturbatively as a sequence of
colour-dipole emissions.  In particular, emissions associated to the
initial state are treated as to directly emerge from colour dipoles
spanned by the external parton lines.  The beam remnants are kept 
completely outside the perturbative evolution, their connection to the 
evolved cascade is left to the hadronization to deal with.  As a direct 
consequence, three types of dipoles and, hence, of associated radiation 
contribute to the full development of the final cascade, namely 
emissions from initial--initial (II), final--initial (FI), and 
final--final (FF) dipoles.  Consequently, the emissions are denoted as 
initial--initial, final--initial- and final-state radiation 
(ISR, FISR, FSR), respectively.  In order to model ISR and FISR in the 
fully perturbative version of the CDM proposed here, a backward
evolution of the initial-state related radiation pattern of the
shower is mandatory and automatically necessitates the inclusion of
parton distribution functions (PDFs).
\\
The outline of the paper is as follows: in \sec{sec:CDM} the basics 
of the dipole-shower model as implemented in \ariadne will briefly 
be introduced.  In addition, the treatment of ISR through final-state 
dipole splittings involving the beam remnants will be discussed.  In 
the next section, \sec{sec:New}, the basic ideas of the newly proposed 
dipole shower are highlighted, especially its different ansatz for
ISR simulation.  This includes the generalization of the kinematical 
framework and of the evolution variables.  In the following three 
sections, the kinematics and single-emission cross sections of all 
dipole splittings in various configurations of initial- and final-state 
partons will be detailed, recapitulating the case of dipoles 
consisting of final-state partons, the case implemented in \ariadne,
in \sec{sec:FF}.  The kinematics and splitting functions of the new
dipole types present in this model are discussed in
\secss{sec:II}{sec:FI}.  The complete shower algorithm will be
presented in \sec{sec:Shower} and its performance for various physics
processes will be highlighted in \sec{sec:Res}.  
%

%% file: tex/dscdm.tex
\section{The Colour Dipole Model}
%
\label{sec:CDM}
\subsection{Physical background of the CDM}
The Lund colour dipole model (CDM) has strong connections to the
semiclassical method of virtual quanta
\cite{Fermi:1924tc,vonWeizsacker:1934sx,Williams:1934ad}, which
equates the electromagnetic energy flux associated with the fields
emitted by fast moving charges with an energy flux of equivalent
photons.  Owing to the large Lorentz boosts of the charged emitter, the
corresponding electric and magnetic fields are orthogonal to each
other and they populate a plane orthogonal to the direction of motion
of the emitter only.  This amounts to a pulse of electromagnetic
energy, given by
\be
  dI(\omega,b)\;\simeq\;
  \hbar\alpha\,d\omega\;\frac{2\pi\,b\,db}{\pi^2}\,,
\ee
where $b$\/ denotes the impact parameter, \ie the distance w.r.t.\ the
emitter; $\omega$\/ is the frequency of the field component.  It is 
bound from above by $\omega<p/mb$, where $p$\/ and $m$\/ are the
momentum and mass of the emitter, respectively.  Equating this energy
pulse $I$\/ with a number of equivalent quanta $n$,
\be
  dI\;=\;\hbar\,\omega\,d\omega\,,
\ee
and replacing the impact parameter with transverse momentum yields
\be\label{Eq:WeiszaeckerWilliams}
  dn\;\simeq\;
  \frac{\alpha}{\pi}\,\frac{dk_\perp^2}{k_\perp^2}\,\frac{d\omega}{\omega}\,.
\ee
\\
A similar result emerges when considering bremsstrahlung off a charged
particle, changing its otherwise straight direction of motion through
a sudden ``kick'', or connected with the pair production of charged
particles.  Then, in the Breit-frame of the former process, or in the
centre-of-mass frame of the latter, a rapidity $y$\/ can be defined
w.r.t.\ the axis of motion of the charged particle(s).  A short
calculation based on a full quantum-mechanical treatment shows that,
neglecting spin effects, the number of bremsstrahlung-photons is well
approximated by
\be\label{Eq:SimpleBrems}
  dn\;=\;\frac{2\alpha}{\pi}\,\frac{d\omega}{\omega}\,dy\,,
\ee
cf.\ \cite{Andersson:1998tv}.  Here, the rapidity must satisfy
\be
  \abs{y}\;<\;\abs{y_0}\,,
\ee
and $y_0$ is the rapidity of the emitter(s).  Rewriting energy through
transverse momentum,
\be
  k_\perp\;=\;\frac{\omega}{\cosh y}\,,
\ee
then leads to
\be\label{Eq:SimpleDipole}
  dn\;=\;\frac{\alpha}{\pi}\,\frac{dk_\perp^2}{k_\perp^2}\,dy\,.
\ee
Because of its equivalence to \eq{Eq:SimpleBrems}, this equation
exhibits the dominance of soft radiation in the semi-classical limit.
In this context it is worth to note that the same limit is used in
eikonal-type factorization of matrix elements employed, \eg, in
antenna subtraction methods \cite{Kosower:1997zr,Campbell:1998nn}
for the calculation of perturbative higher-order corrections to
scattering cross sections in QCD.
\\
The simple formula for the semi-classical limit of photon radiation
off a charged dipole, \eq{Eq:SimpleDipole}, can be refined through a
full quantum-mechanical treatment, including spin effects, see also
later sections.  However, the dominant features of the radiation
pattern are already fixed by the simple formula, which in turn denotes
the starting point for a shower simulation based on individual dipole
emissions. The differential probability for such an emission to occur 
in an interval $dp^2_\perp$ and $dy$\/ is related to 
$d\cal{P}=d\sigma/d\sigma_0$ given by
\be\label{eq:GenXsec}
  d\cal{P}\;\simeq\;
  \frac{\alpha_s}{\pi}\;\frac{dp^2_\perp}{p^2_\perp}dy\,.
\ee
Here $p_\perp$ denotes a transverse momentum, which in the CDM is
constructed out of Lorentz invariant quantities.  Numbering the momenta
of the particles after emission such that the newly emitted particle
is labelled with ``2'', and, denoting the momenta before and after the
emission with $\tilde p_i$ and $p_i$, respectively, the emission can
be symbolized as
\be
  \td p_1+\td p_3\;=\;p_1+p_2+p_3\,.
\ee
The squared invariant masses of sets of momenta are denoted as
\be
  s_{ij\ldots}\;=\;(p_i+p_j+\ldots)^2
  \qquad\mbox{\rm and}\qquad
  \td s_{ij\ldots}\;=\;(\td p_i+\td p_j+\ldots)^2\,.
\ee
A Lorentz invariant transverse momentum can be defined as
\be\label{eq:LundP2t}
  p^2_\perp
  \;=\;\frac{s_{12}\,s_{23}}{s_{123}}
  \;=\;\frac{s_{12}\,s_{23}}{\tilde s_{13}}\,,
\ee
in agreement with \cite{Gustafson:1987rq,Andersson:1998tv} and the
\ariadne implementation. Moreover, a rapidity can then be computed
through
\be\label{eq:LundY}
  y\;=\;\frac{1}{2}\,\ln\frac{s_{12}}{s_{23}}\,.
\ee
The Lorentz invariant choice guarantees a frame-independent
description of the dipole splitting process.  Using $p_\perp$ as the
ordering parameter for subsequent emissions, a Sudakov form factor
encodes the non-emission probability between two scales
$p^2_{\perp,\mr{high}}$ and $p^2_{\perp,\mr{low}}$ in analogy to
conventional parton showers:
\be
  \Delta(p^2_{\perp,\mr{high}},p^2_{\perp,\mr{low}})\;=\;
  \exp\left\{-\int\limits^{p^2_{\perp,\mr{high}}}_{p^2_{\perp,\mr{low}}}
            dp^2_\perp\int\limits^{y_+(p_\perp)}_{y_-(p_\perp)}dy
	    \;\frac{d\cal{P}}{dp^2_\perp dy}\right\}\,.
  \label{eq:GenSud}
\ee
In this form, the leading logarithms are resummed to all orders.
\\
The Sudakov form factor constitutes the basis of the simulation of
parton emission also in the framework of the CDM.  In contrast to
ordinary parton showers, however, here the relevant objects are colour
dipoles, which emerge naturally when considering the large $N_\mr{C}$
limit.  In this limit, colour charges in the fundamental representation
(quarks and antiquarks) have one colour partner, and colour charges in
the adjoint representation (gluons) have two colour partners.  The
dipoles are built from pairs of such colour partners, and the emission 
of a gluon off a dipole effectively amounts to splitting the dipole
into two.  
\\
This self-similar process of dipole splitting, which is described in a
probabilistic fashion, is easily encoded as a Markovian process in
form of a computer program.  Adding in the leading logarithmic
behaviour and colour coherence as a dominant feature of QCD emissions
results in a strict ordering of subsequent emissions such that the
actual $p_\perp$ of a dipole splitting sets the maximal $p_\perp$ for
the splittings of the two resulting dipoles.
\subsection{Initial-state radiation in the original CDM}
\label{sec:AriISR}
In \ariadne, the only complete CDM implementation so far,
initial-state radiation off incoming partons is not explicitly taken
into account. Instead, ISR is {\it redefined}\/ as FSR, where dipoles
are spanned between potential final-state partons and the outgoing
hadron remnants
\cite{Andersson:1989gp,Andersson:1992zu,Lonnblad:1994wk,Lonnblad:1995ex}.
Considering DIS of leptons on hadrons, it can be argued that, as the
hadron is in a bound state, all radiation originates from the colour dipole
between the struck point-like quark and the hadron remnant -- being an
extended object composed of individual valence quarks and sea partons.
Thus, an extended ``antenna'' is formed and from the electro-magnetic
(semi-classical) analogy it follows that radiation of wavelengths
smaller than the extension of the antenna is suppressed.  Therefore, 
the original CDM was modified such that only a $p_\perp$-dependent
fraction $a(p_\perp)$ of the remnant enters the splitting process
\cite{Andersson:1989gp}:
\be
  a(p_\perp)\;=\;\left(\frac{\mu}{p_\perp}\right)^\alpha\,,
\ee
where $\mu$\/ parametrizes the inverse size of the remnant and
$\alpha$\/ refers to the dimensionality of the emitter, both being
parameters to be tuned to data.  In $e^+e^-$ annihilation the
(``triangle'') phase-space boundaries are approximated by
$\abs{y}<\ln(M/p_\perp)$, which now are supplemented by the extra
condition $y<\ln(M\mu/p^2_\perp)$.  This obviously limits the range of
accessible $p_\perp$ values in the splitting of the dipole of mass
$M$.  The strategy of sharing the recoil in such cases was inspired by
the Lund string model, where an extra kink on the string (hadron) is
interpreted as an extra gluon. This led to the introduction of recoil
gluons to compensate for the recoil momentum associated with the part
of the hadron, which participates in the emission.\spc\footnote{For
  the \ariadne rewrite in {\tt C++}, the possibility of discarding
  recoil gluons completely is under consideration.}
Moreover, in cases where a sea quark is hit, the picture experiences
further minor modifications.  Taken together, a good fraction of
phenomenological, non-perturbative modelling enters the Lund CDM for
ISR through all these assumptions.
\\
Next, consider Drell--Yan-like processes; there, a quark--antiquark 
pair annihilates to produce a lepton pair.  In conventional parton 
showers, the two incoming quarks would emit secondary partons, 
typically simulated in a backward evolution algorithm
\cite{Sjostrand:1985xi,Marchesini:1987cf}. The recoil of these
emissions is transferred to colour partners and the final-state
leptons.  In contrast, in \ariadne the incoming quarks do not radiate
but rather the two beam remnants, which are the only two coloured
final-state objects before radiation (cf.\ left panel in
\fig{fig:LundvsAdicDY}). Then, the recoil of the first emission is
compensated for by the final-state leptons
\cite{Lonnblad:1995ex,Lavesson:2005xu}, for all further dipole
emissions, additional recoil gluons are added, if the emission
occurred in phase space significantly away from the vector boson
\cite{Lonnblad:1995ex}.  A further obvious refinement is the correction
of the first emission to the corresponding matrix-element expression.
The sharp phase-space cut-off is then replaced by a softer suppression
function, in order to describe the high transverse-momentum spectrum
of the vector boson.
%

%% file: tex/dsnew.tex
\section{New approach to initial-state radiation using colour dipoles}
%
\label{sec:New}
The principles underlying the proposal of this paper for the
construction of a purely perturbative colour-dipole model are:
\bi
\item the maintenance of the probabilistic interpretation of emissions
  as encoded in the Sudakov form factor, which will be obtained from
  exponentiating single-emission differential cross sections;
\item the large $N_\mr{C}$ limit of the radiation pattern, and the
  restriction to account for the leading terms only, \ie the leading
  dipoles, of this expansion;
\item the generalization of the kinematics and the evolution variables
  used in the original CDM and in \ariadne to the case of ISR and
  FISR;
\item the factorization of the emission phase space and matrix
  elements around the soft limit (the radiation pattern has to be
  factorized in terms of $2\to3$ splittings, to be derived for II and
  FI dipoles);
\item the utilization of crossing symmetries for the determination of
  dipole splitting functions;
\item the construction of on-shell kinematics for each splitting on
  an emission-by-emission basis, allowing stop and restart the
  cascading after any individual emission;
\item the backward evolution description of radiation related to
  incoming partons, and, consequently, the emergence of PDFs in the
  shower algorithm in a way similar to conventional parton showers.
\ei
A number of issues are not at all covered here, which are, however,
straightforward to include in some future work, namely
\bi
\item the comparison of different forms of splitting cross sections;
\item the approach's extension to account for finite, non-zero quark
  masses (here, all partons are treated as massless);
\item an extension to Supersymmetry;
\item the QED radiation off the dipoles.
\ei
\begin{figure}[t!]
  \vspace{0mm}\bc
  \includegraphics[width=61mm,angle=0.0]{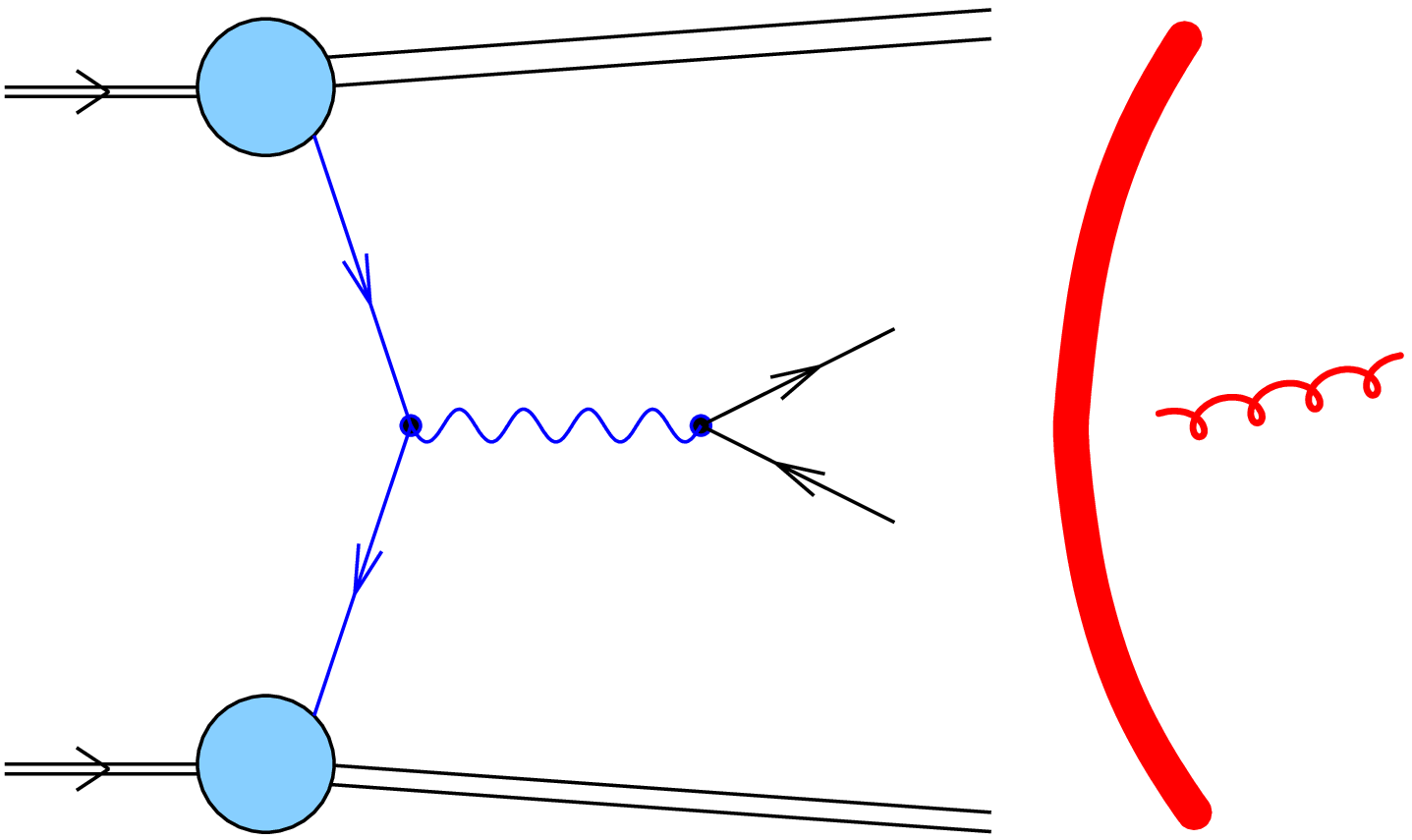}
  \hspace*{27mm}
  \includegraphics[width=44mm,angle=0.0]{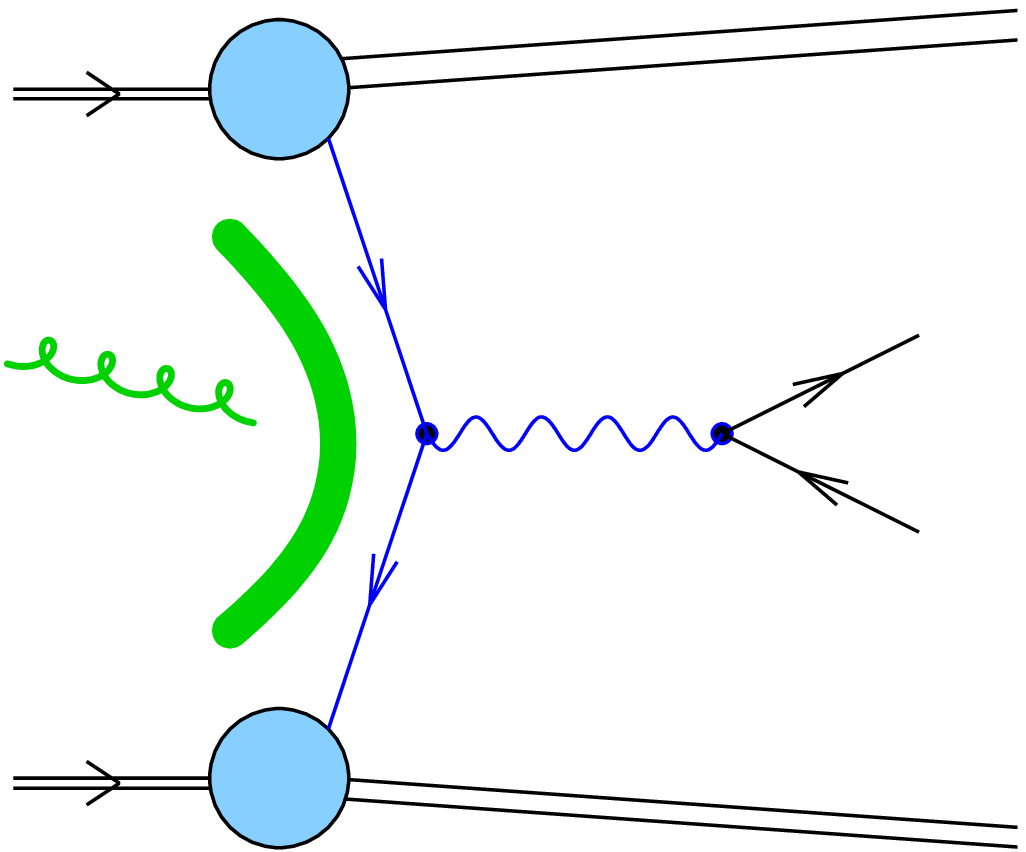}
  \ec\vspace{0mm}
  \myfigcaption{140mm}{The Lund CDM approach to initial-state
    radiation in Drell--Yan processes (left panel) vs.\ the direct,
    perturbative approach as suggested by the new dipole-shower model
    (right panel). The treatment in modelling a first gluon emission
    is illustrated.}
  \label{fig:LundvsAdicDY}
  \vspace{0mm}
\end{figure}
To exemplify the impact of the principle outlined above, consider
Drell--Yan processes; in contrast to the Lund approach, see
\sec{sec:AriISR}, in the new dipole picture the primary dipole $\bar
q_\il q'_\il$ is directly formed by the two incoming quarks, and the
emission will be calculated from the competition between gluon (see
\fig{fig:LundvsAdicDY}), quark and antiquark bremsstrahlung.
The real-emission matrix-element information will directly enter,
through the corresponding dipole splitting functions, as in the FF
counterpart of emitting a first gluon off the $q\bar q'$\/ dipole.
The boson's transverse momentum will be naturally generated, because
the new initial-state momenta will be oriented along the beam
direction.  In case of an actual gluon emission, a system of two
colour-connected successor FI dipoles emerges, namely a $\bar q_\il
g_\fl$ and a $g_\fl q'_\il$ dipole.  A further gluon radiated into the
final state will then create a first FF dipole.  If a quark is
produced first, an FI dipole $q_\fl g_\il$ and a successor II dipole
$g_\il q'_\il$ will be generated with again a dual r\^ole played by
the gluon, here $g_\il$.  Hence, the QCD evolution of the
leading-order Drell--Yan pair production process eventually will
involve all possible dipole types.
\subsection{Kinematic framework}
\label{sec:GenEvoKins}
The occurrence of new dipoles and corresponding splittings is an
immediate consequence of the suggested new CDM approach. A list
summarizing the principal dipoles of QCD is shown in
\tab{tab:DipTypes}.
\begin{table}[t!]
  \vspace{0mm}\bc
  \bt{|c||c||c|}\hline    
  \multicolumn{3}{|c|}{}\\[-2.5mm]
  \multicolumn{3}{|c|}{QCD dipoles,\quad$\td k\,\td\ell$}\\[1.5mm]\hline
  \phantom{FF dipoles} & \phantom{FF dipoles} & \phantom{FF dipoles}\\[-2.5mm]
  II dipoles,\quad$\bar\imath'\,i$ & FI dipoles,\quad$f\,i$ &
  FF dipoles,\quad$f\,\bar f'$\\[1.5mm]\hline
  &&\\[-2.5mm]
  $\bar q'_\il q_\il$&$q_\fl q'_\il$&$q_\fl\bar q'_\fl$\\[1mm]
  $g_\il q_\il$      &$q_\fl g_\il$ &$q_\fl g_\fl$\\[1mm]
  $g_\il g_\il$      &$g_\fl g_\il$ &$g_\fl g_\fl$\\[2.5mm]\hline
  \et
  \ec\vspace{0mm}
  \mytabcaption{140mm}{All dipole types appearing in QCD (the
    supplemental indices $\il$ or $\fl$ label whether the parton is in
    the initial or final state, respectively; if clear from the
    context, the index $\fl$ will be left out). The primes indicate
    that different quark flavours may constitute the dipole.}
  \label{tab:DipTypes}
  \vspace{0mm}
\end{table}
Dipoles are labelled by $\td k\td\ell$, thus, in the most general way
a splitting triggered by the emission of a (new) gluon $g$\/ is
expressed as
\be
  \td k\,\td\ell\;\to\;k\,g\,\ell\,.
\ee
The notation is chosen such that the flavour and colour flow of all
particles is outgoing.  Three types of gluon emission emerge, related
by crossing symmetry; any such splitting will leave the number of
initial-state partons constant:
\be
  \td k\,\td\ell\;\to\;
  \left\{\begin{array}{r@{\quad:\quad}l}
    \td k\,g_\fl\,\td\ell
    &{\rm gluon\ emission}\,,\\
    q\,g_\il\,\td\ell
    &{\rm quark\ emission,\ provided\ that}\ \ \td k=\bar q_\il\,,\\
    \td k\,g_\il\,\bar q
    &{\rm antiquark\ emission,\ provided\ that}\ \ \td\ell=q_\il\,.
  \end{array}\right.
\ee
Here, the subscripts indicate whether the gluon emerges in the initial
or final state. In the former case, this requires to replace the
initial (anti)quark of the original dipole by the initial gluon and
emit the corresponding antiquark (quark) in the final state.
\\
Having clarified the notations used for the dipoles and their
splittings, the kinematic objects will be introduced. First of all,
the momenta are defined as incoming/outgoing if they are associated
with the physical initial/final state. Those before and after the
emission are denoted by $\td p_{\td m}$ and $p_m$, respectively, such
that, expressed through the momenta alone the dipole splitting process
can be written as
\be
  \td\vs_{\td k}\,\td p_{\td k}+\td\vs_{\td\ell}\,\td p_{\td\ell}
  \quad\longrightarrow\quad
  \vs_k\,p_k+\vs_g\,p_g+\vs_\ell\,p_\ell\,.
\ee
Here and in the following the signature factors $\td\vs_{\td m}=\pm1$
and $\vs_m=\pm1$ for partons in the final ($+$) and initial ($-$)
state. The before- and after-emission total momenta $\td p_0$ and
$p_0$ then read
\bea
  -\vs_0\,\td p_0 &=& \td\vs_{\td k}\,\td p_{\td k}+
  \td\vs_{\td\ell}\,\td p_{\td\ell}\,,
  \label{eq:GenMombf}\\[1mm]
  -\vs_0\,p_0     &=& \vs_k\,p_k+\vs_g\,p_g+\vs_\ell\,p_\ell\,,
  \label{eq:GenMomaf}
\eea
with the requirement that $\td p^2_0=p^2_0$. Furthermore
$\td\vs_0\equiv\vs_0$, and the signature factor $\vs_0$, \ie the
association of the total momenta with the initial or final state is
chosen such that the after-emission configuration refers to a
production ($\vs_0=-1$, FF dipoles), scattering ($\vs_0=-1$, FI
dipoles), or annihilation ($\vs_0=1$, II dipoles) process.
Consequently, the four-vector $\td p_0$ then corresponds to the
four-momentum of the decaying parent dipole having mass $\abs{M}$ such
that
\be
  \td p^2_0\;=\;M^2\equiv-Q^2\;=\;p^2_0\,,
\ee
with $Q^2$ arranged to be positive definite for FI dipole emissions.
Accordingly, Lorentz-invariant energy fractions w.r.t.\ $p_0$ are
defined through\spc\footnote{The notion ``energy fraction'' is clear
  in the centre-of-mass frame of a parent FF dipole, where
  $x_m=2E_m/M$.}
\be\label{eq:XkDef}
  x_m\;=\;\frac{2\,p_m p_0}{p_0^2}\,.
\ee
The squared invariant masses of two- and three-parton systems are
denoted by
\be\label{eq:SmnDef}
  s_{mn}\;=\;(\vs_m\,p_m+\vs_n\,p_n)^2
  \qquad{\rm and}\qquad
  s_{mnr}\;=\;(\vs_m\,p_m+\vs_n\,p_n+\vs_r\,p_r)^2
\ee
where the inclusion of $p_0$ and the expressions related to the
momenta before the emission are understood.  Concerning all gluon
emissions considered here, the identity
\be
  M^2\;=\;s_{kg}+s_{g\ell}+s_{k\ell}\;=\;s_{kg\ell}\;=\;-Q^2\,,
  \label{eq:GenstuSum}
\ee
holds true in general, since all partons are consistently treated as
massless.
\subsection{Towards generalized evolution variables}
\label{sec:GenEvoVars}
Next, the dipole evolution variables have to be generalized such that
all emissions of all dipole types can be treated on equal footing and
embedded in a consistent CDM-like evolution.  The generalized
variables should have the property of leaving the well-established FSR
treatment unchanged and they should satisfy the constraint that all
splitting cross sections, \ie those involving initial-state partons as
well, will follow the approximate form given in \eq{eq:GenXsec}.  This
would just manifest the universal features of QCD radiation in the
soft limit, reproduced by eikonal distributions factorizing off the
squared matrix elements in emissions off initial and final states
alike,
\be
  -\frac{1}{2}\left(\frac{p_k}{p_k p_g}-\frac{p_\ell}{p_g p_\ell}\right)^2
  \;=\;
  \frac{2\,s_{k\ell}}{s_{kg}\,s_{g\ell}}\,.
\ee
Note that the right-hand side of this equation explicitly assumes
massless partons.  Following \eq{eq:LundP2t}, the factor $2/p^2_\perp$
becomes identical to the eikonal factor in the soft limit, and the
collinear limits manifest themselves in the two-parton squared masses
appearing in this $p^2_\perp$ definition.  Then, the generalized
kinematic variables should exhibit the same singular behaviour in the
soft/collinear limits for all dipole types, reflecting the crossing
symmetry.  Therefore, in this paper a generalized transverse momentum
and rapidity are proposed in the form
\be
  p^2_\perp\;=\;\left|\frac{s_{kg}\,s_{g\ell}}{s_{kg\ell}}\right|\,,
  \label{eq:GenP2t}
\ee
and
\be
  y\;=\;\frac{1}{2}\,\ln\left|\frac{s_{g\ell}}{s_{kg}}\right|\,.
  \label{eq:GenY}
\ee
Here, the invariant masses $s_{mn(r)}$ are calculated including the
signature factors $\vs_{m,n,r}$, \ie through \eqs{eq:SmnDef}.  Clearly,
for FF dipole cascading, all invariant masses are positive and hence
the original CDM evolution variables of \eqss{eq:LundP2t}{eq:LundY}
are trivially recovered.  For the other cases, the generalized form
suggested here is a minimal Lorentz-invariant extension, guaranteeing
a frame-independent evolution of the colour dipole.  Moreover, these
shower variables allow a global simultaneous ordering of all
emissions.  Given these generalized definitions, the identities
\be
  \abs{s_{kg}}\;=\;\abs{M}\,p_\perp e^{-y}
  \qquad{\rm and}\qquad
  \abs{s_{g\ell}}\;=\;\abs{M}\,p_\perp e^{+y}
  \label{eqs:GenSfromEvoVars}
\ee
are found, indeed showing the similarity to the original Lund CDM.
\bigskip\\
So, having introduced the kinematic framework, in the following
sections, \secsss{sec:FF}{sec:II}{sec:FI}, the derivation of the
splitting kinematics for each dipole type (FF, II, and FI/IF) will be
always pursued in four steps:
\bi
\item First, the evolution variables $p_\perp$ and $y$\/ are
  identified.
\item Then, the limits of the emission phase space are deduced, which
  for the rapidity typically read $y_-\le y\le y_+$.  They guarantee
  that the evolution takes place within the physical region of phase
  space.  These limitations are imposed through constraints on the
  evolution variables and, thus, determine the Sudakov form factor,
  see \eq{eq:GenSud} and \sec{sec:Shower}.
\item Together with the strict limits, approximate ones are stated,
  denoted \eg for the rapidity by $Y_-\le y\le Y_+$, which allow an
  analytical evaluation of approximate Sudakov form factors.
  They are used in a Monte Carlo procedure, which finally corrects for
  the true form of the exact Sudakov form factors by means of a veto
  algorithm, see \eg \cite{Sjostrand:2001yu}.
\item The on-shell three-parton kinematics of the splittings
  characterized by the central variables $p_\perp$ and $y$\/ are
  constructed from the original two-parton configurations.  Remaining
  degrees of freedom are fixed with a few additional assumptions, \ie
  through four-momentum conservation and a splitting-specific recoil
  strategy.  The Lorentz invariant definition of the evolution
  variables guarantees the frame-independence of the actual
  construction.
\ei
\subsection{Dipole splitting cross sections and functions for QCD radiation}
\label{sec:Dsfuncs}
In order to construct a parton shower as a Markovian process, the
emission of any additional parton has to be factorized from the
radiation of partons produced so far, such that the full radiation
pattern can be built as a sequence of individual, mostly independent
emissions.  For the actual construction of a dipole shower, the
individual parton emission should be modelled as being coherently
shared between the two partons forming the dipole.  The asymptotic
form of these $2\to3$ dipole splitting cross sections has been
presented in \sec{sec:CDM}, cf.\ \eq{eq:GenXsec}, which constitutes
the limiting case of soft emissions.  In order to extrapolate to
harder regions of emission phase space and to include spin effects,
the differential cross section for a $\td k\td\ell$\/ dipole
developing into a $kg\ell$\/ colour-connected state is more
conveniently written as
\be
  d\cal{P}_{\td k\td\ell\to kg\ell}
  \;\equiv\;\frac{d\sigma_{0\to kg\ell}}{d\sigma_{0\to\td k\td\ell}}\;=\;
  \frac{\alpha_s}{2\pi}\;D_{\td k\td\ell\to kg\ell}(p_\perp,y)\;
  \frac{dp^2_\perp}{p^2_\perp}dy\,.
  \label{eq:SplFuncDef}
\ee
This defines the {\it dipole splitting function} $D_{\td k\td\ell\to kg\ell}$
in analogy to the splitting kernels employed in conventional parton
showers. In both cases, the splitting kernels of dipole or parton
showers incorporate refinements, which go beyond the corresponding
eikonal or collinear approximation, respectively.
Here, the $D_{\td k\td\ell\to kg\ell}$ may be deduced by analyzing
differential cross sections for additional real emission of partons in
comparison to the corresponding Born level processes.  This then
yields the single-dipole phase-space and matrix-element factorization,
which has to work at least in the singular domains of the
real-emission phase space.  Accordingly, first-order real corrections
are fully or partially encoded in the splitting kernels automatically.
\\
For the timelike case, the reasoning outlined above is realized by
starting from the three-body decay rate of an object with mass $M$,
\be
  d\Gamma_{0\to fg\bar f'}\;=\;
  \frac{(2\pi)^4}{2\,M}\;\abs{{\cal M}_{0\to fg\bar f'}}^2\;
  d\Phi_{0\to fg\bar f'}(p_0;\,p_f,p_g,p_{\bar f'})\,,
\ee
where, in the massless limit, the dipole phase space and
matrix element factorize according to
\be
  d\Phi_{0\to fg\bar f'}(p_0;\,p_f,p_g,p_{\bar f'})\;=\;
  d\Phi_{0\to f\bar f'}(\td p_0=p_0;\,\td p_f,\td p_{\bar f'})\;
  \frac{ds_{fg}\,ds_{g\bar f'}}{16\pi^2\,M^2}\;\frac{d\varphi}{2\pi}
\ee
and
\be
  \abs{{\cal M}_{0\to fg\bar f'}}^2\;\simeq\;8\pi\alpha_s\,C\;
  \hat D_{f\bar f'\to fg\bar f'}\;\abs{{\cal M}_{0\to f\bar f'}}^2\,,
  \label{eq:FFdipoleME}
\ee
respectively.  Here, taking $N_{\mr C}=3$ for the number of colours,
the colour factor labelled $C$\/ has been introduced explicitly.  It
typically takes one of the following values,
$C=C_F=\frac{N^2_{\mr C}-1}{2N_{\mr C}}=\frac{4}{3}$ for gluons
emitted off quarks, $C=C_A=N_{\mr C}=3$ for gluons emitted off gluons,
or $C=T_R=\frac{1}{2}$ for gluon splittings into quarks.  $\hat D$\/
denotes the dipole matrix element, which by definition correctly
reproduces the singular terms of the parton emission process with
potential differences in finite terms.  Therefore,
$d\Gamma_{0\to fg\bar f'}$ can be expressed as
\be
  d\Gamma_{0\to fg\bar f'}\;\simeq\;d\Gamma_{0\to f\bar f'}\;
  \frac{C\alpha_s}{2\pi}\;\hat D_{f\bar f'\to fg\bar f'}(p_\perp,y,\varphi)\;
  dp^2_\perp dy\,\frac{d\varphi}{2\pi}
\ee
with $p_\perp$ and $y$\/ taken from \eqss{eq:GenP2t}{eq:GenY}.  In
most cases, the dependence of $\hat D$\/ on the azimuthal angle
$\varphi$\/ is neglected, thus, integrated out, such that the
connection of the dipole splitting functions to the dipole matrix
elements in the FF case reads
\be
  D_{f\bar f'\to fg\bar f'}(p_\perp,y)\;=\;
  \xi\,C\,p^2_\perp\;\hat D_{f\bar f'\to fg\bar f'}(p_\perp,y)\,.
  \label{eq:SplFuncFF}
\ee
Note that here a gluon-sharing factor $\xi$\/ has been introduced
because each gluon is contained by two dipoles.
\\
For the class of II dipoles, \ie those consisting of colour-connected
incoming partons, the extraction of dipole splitting cross sections
has to be accomplished on the level of hadronic cross sections to
correctly account for PDF effects and possible phase-space
(suppression) factors.  In this case, a $2\to2$ scattering process
rather than an $1\to3$ decay has to be considered.  The differential 
cross section (using massless partons and having already integrated
out the $\varphi$\/ dependence) reads
\be
  d\sigma_{\bar\imath'i(gi)\to0g(0q)}\;=\;
  f_{\bar\imath'(g)}(x_\pm,\mufp)\,f_i(x_\mp,\mufp)\,\frac{1}{S}\,
  \frac{\abs{{\cal M}_{\bar\imath'i(gi)\to0g(0q)}}^2}{16\pi\hat s^2}\,
  d\hat s\,d\hat t\,dy_{\mr{cm}}\,,
\ee
where the usual $2\to2$ process Mandelstam variables, 
$\hat s=s_{\bar\imath'i(g_\il i)}$\/ and $\hat t=s_{\bar\imath'g(qg_\il)}$
have been employed.  Note that the quark emission case is signified by
the parentheses.  The $f_k(x_\pm,\mufp)$ are the PDFs.  At leading
order they can be interpreted as the probability of resolving a parton
$k$\/ inside the nucleon with light-cone momentum fraction $x$\/ taken
w.r.t.\ the nucleon's momentum; \mufp\ names the factorization scale
(defined in energy units), at which, pictorially speaking, the partonic
substructure is probed.  $S$\/ and $y_{\mr{cm}}$ denote the
centre-of-mass energy and rapidity of the collider system,
respectively.  The $2\to1$ hadronic differential Born cross section for
creating a particle of mass $M$\/ through the matrix element
$\cal{M}_{\bar\imath'i(\bar qi)\to0}$ characterizes the
$\bar\imath'i$\/ dipole's situation before the emission.  Then,
similarly to the FF case, the $2\to 2$ matrix element squared can be
cast into a factorized form, reading
\be
  \abs{{\cal M}_{\bar\imath'i(gi)\to0g(0q)}}^2\;\simeq\;8\pi\alpha_s\,C\;
  \hat D_{\bar\imath'i(\bar q_\il i)\to\bar\imath'gi(qg_\il i)}\;
  \abs{{\cal M}_{\bar\imath'i(\bar qi)\to0}}^2\,.
  \label{eq:IIdipoleME}
\ee
This hence allows to write the $2\to2$ differential scattering cross
section in terms of the $2\to1$ Born term
$d\sigma_{\bar\imath'i(\bar qi)\to0}\z$ where the tilde variables
refer to the Born configuration:
\be\begin{split}
  d\sigma_{\bar\imath'i(gi)\to0g(0q)}\;\simeq\;\vphantom{.}&
  d\sigma_{\bar\imath'i(\bar qi)\to0}\;
  \left(\frac{dy_{\mr{cm}}}{d\td y_{\mr{cm}}}\right)\;
  \frac{f_{\bar\imath'(g)}(x_\pm,\mufp)\,f_i(x_\mp,\mufp)}
       {f_{\bar\imath'(\bar q)}(\td x_\pm,\muf)\,f_i(\td x_\mp,\muf)}\;
       \frac{M^4}{\hat s^2(p_\perp,y)}\\[2mm]&\!\!\times\;
  \frac{C\alpha_s}{2\pi}\;
  \hat D_{\bar\imath'i(\bar q_\il i)\to\bar\imath'gi(qg_\il i)}(p_\perp,y)\;
  dp^2_\perp dy\,.
\end{split}\ee
In contrast to the FF case, there is some additional freedom in
arranging the actual recoils, since in principle the total energy of
the splitting parton system will increase with each emission --
additional momentum can be taken off the incoming nucleons.  However,
fixing the new centre-of-mass rapidity $y_{\mr{cm}}$ removes this
ambiguity.  The choice in this paper is to ensure constant rapidity
derivatives, thus, set up a recoil handling, which eventually shifts
the original $\td y_{\mr{cm}}$ through some function $\hat y$\/ that
exclusively depends on the variables associated to the emission,
\be
  y_{\mr{cm}}\;=\;\td y_{\mr{cm}}+\hat y(M^2,\hat s,\hat t)\,.
\ee
Provided that these requirements can be satisfied, the II dipole
splitting functions finally read
\be
  D_{\bar\imath'i(\bar q_\il i)\to\bar\imath'gi(qg_\il i)}(p_\perp,y)\;=\;
  \frac{f_{\bar\imath'(g)}(x_\pm,\mufp)\,f_i(x_\mp,\mufp)}
       {f_{\bar\imath'(\bar q)}(\td x_\pm,\muf)\,f_i(\td x_\mp,\muf)}\;
       \frac{M^4\;\xi\,C\,p^2_\perp}{\hat s^2(p_\perp,y)}\;
       \hat D_{\bar\imath'i(\bar q_\il i)\to\bar\imath'gi(qg_\il i)}(p_\perp,y)
  \label{eq:SplFuncII}
\ee
and can be used to specify the associated differential splitting cross
sections.  In comparison to the gluon emission processes of FF dipoles,
cf.\ \eq{eq:SplFuncFF}, additional terms arise in each of the II
dipole functions, namely a PDF weight, $\cal{W}_{\mr{PDF}}$, which
contains a ratio of PDFs taken at the respective momentum fractions
and factorization scales before and after the emission, and a
phase-space weight, $\cal{W}_{\mr{PSP}}=M^4/\hat s^2$, which accounts
for the altered incoming flux of the parton-level differential cross
section.
\\
Similarly, the generic structure of final--initial dipole splitting
cross sections $d\cal{P}_{fi(fq_\il)\to fgi(fg_\il\bar q)}$ for gluon
(antiquark) emissions into the final state is fixed through
\be
  D_{fi(fq_\il)\to fgi(fg_\il\bar q)}(p_\perp,y)\;=\;
  \frac{f_{i(g)}(x_\pm,\mufp)}{f_{i(q)}(\td x_\pm,\muf)}\;
  \frac{Q^4\;\xi\,C\,p^2_\perp}{\left[\hat s(p_\perp,y)+Q^2\right]^2}\;
  \hat D_{fi(fq_\il)\to fgi(fg_\il\bar q)}(p_\perp,y)\,,
  \label{eq:SplFuncFI}
\ee
where the Mandelstam variable $\hat s$\/ is related to the two-parton
squared masses via $\hat s=s_{fg(f\bar q)}$.  Note that the right-hand
side of \eq{eq:SplFuncFI} exhibits, as for emissions off II dipoles,
the additional PDF and phase-space weights.  The formula has been
derived along the lines already employed for the II case, again in
the limit of zero quark masses. As before, by comparing the
differential hadronic cross sections before and after the emission,
the emission part can be factored out when relying on dipole
matrix-element factorization,
\be
  \abs{{\cal M}_{0i(0g)\to fg(f\bar q)}}^2\;\simeq\;8\pi\alpha_s\,C\;
  \hat D_{fi(fq_\il)\to fgi(fg_\il\bar q)}\;
  \abs{{\cal M}_{0i(0q)\to f}}^2\,.
  \label{eq:FIdipoleME}
\ee
\smallskip\\
Like invariant squared amplitudes the dipole matrix elements $\hat
D$\/ obey crossing symmetry.  Therefore, as an alternative to the
direct calculation of II/FI dipole matrix elements, cf.\ 
\eqss{eq:IIdipoleME}{eq:FIdipoleME}, these $\hat D$\/ expressions can
be easily derived using crossing relations given that the FF dipole
matrix elements have been worked out, see \eq{eq:FFdipoleME}.  In
cases where there are different particles in the final state, there is
more than one possible crossing and, therefore, more than one
corresponding dipole matrix element.  The only remaining issues are
the determination of the associated gluon-sharing and colour factors,
$\xi$\/ and $C$, respectively.  The latter are assigned according to
the generic, large $N_\mr{C}$, colour structure of the emission.
\\
Concerning dipole matrix-element factorization, there are two possible
approaches to specify the dipole shower presented in this work.  These
approaches are:
\bi
\item Extract the $\hat D$\/ terms from the splitting cross sections
  employed in the Lund colour-dipole model
  \cite{Gustafson:1986db,Gustafson:1987rq,Andersson:1989ki,Lonnblad:1992tz}
  for remnant-free dipole cascading. Results for II/FI dipoles are
  then derived from the corresponding FF dipole ones exploiting the
  crossing symmetry.  In \cite{Gustafson:1987rq,Pettersson:1988zu} the
  Lund differential cross sections have been shown to obey the correct
  QCD behaviour in the soft and/or collinear (Altarelli--Parisi)
  limit.\spc\footnote{Strictly speaking, this matching in the singular
    domains of QCD has been demonstrated omitting the influence of
    colour factors, \ie it has been actually shown for the FF dipole
    matrix elements $\hat D$.}
  This reasoning therefore applies to the new cases as well.
\item Use the tree-level antenna functions presented in
  \cite{GehrmannDeRidder:2005cm}.  Their crossing symmetry has been
  exploited already while considering antenna subtraction with
  hadronic initial states, see \cite{Daleo:2006xa}.  Thus, the
  utilization of antenna functions, instead of the Lund kinematic
  functions, has the clear advantage of constructing a dipole shower
  out of subtraction terms that form the basis of the antenna
  subtraction method \cite{Kosower:1997zr,GehrmannDeRidder:2005cm}; it
  therefore constitutes a very attractive alternative to the first
  approach.
\ei
In the subsequent sections of this publication the first approach is
being followed in order to allow for direct comparison with \ariadne
in the FF case.  All relevant gluon emission types of $2\to3$ dipole
splitting functions will be listed and their parton-radiation
characteristics in the various cases will be discussed.  Finally,
the performance of the full model is tested focussing on comparisons
with experimental data.  It is worth stressing, however, that the
implementation of the second approach is straightforward and will be
subject of a forthcoming study.
%

%% file: tex/dsff.tex
\section{Final-state colour dipoles}
%
\label{sec:FF}
In this section emissions emerging from FF dipoles are discussed.
This is the traditional case already present within the original
version of the CDM, implemented in \ariadne.  The dipole splitting
process can be specified by
\be
  f(\td k)\,\bar f'(\td\ell)\;\to\;f(k)\,g\,\bar f'(\ell)\,.
\ee
\subsection{Final--final dipole single-emission phase space and kinematics}
\label{sec:FFlims}
\begin{figure}[t!]
  \vspace{0mm}\bc
  \includegraphics[width=77mm,angle=0.0]{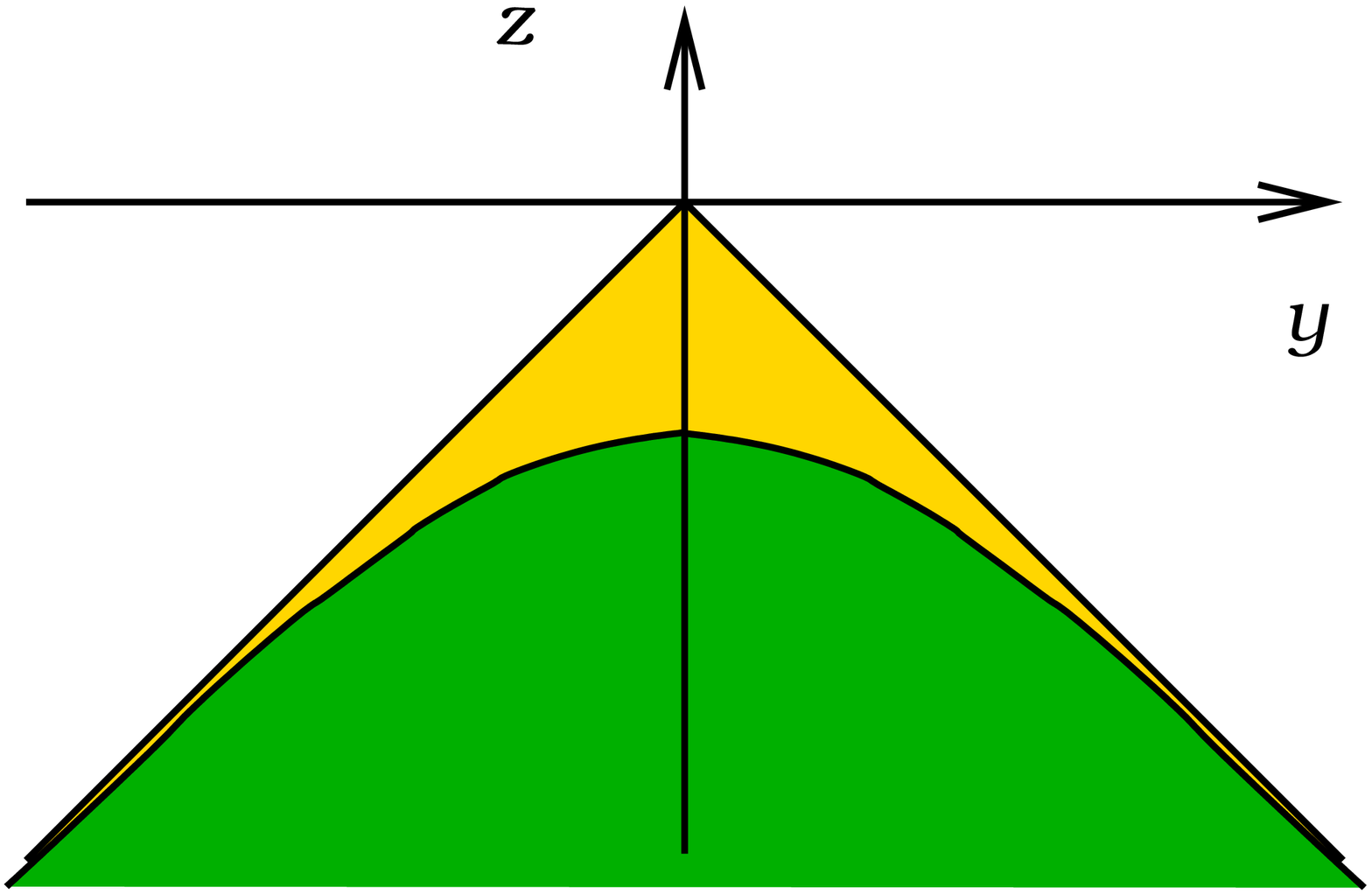}
  \ec\vspace{0mm}
  \myfigcaption{140mm}{The phase space for gluon emission off FF
    dipoles; the dark-coloured region visualizes the available phase
    space. The bright colour is used to show the overestimation as
    given by the approximate limits w.r.t.\ the strict ones. Note that
    $z=\ln\frac{p_\perp}{M}$.}
  \label{fig:ffpsp}
  \vspace{0mm}
\end{figure}

Since the recoil of the emission will be completely shared between the
three new partons, momentum conservation,
\be
  \td p_0=\td p_f+\td p_{\bar f'}\;=\;p_f+p_g+p_{\bar f'}=p_0\,,
\ee
is realized between the momenta present before and after the emission.
Note that apart from $\vs_0=-1$, all other signature factors equal one.
Neglecting parton masses, the relations
\be
  0\;\le\;s_{mn}\;=\;s_{0r}\;=\;M^2(1-x_r)\;\le\;M^2\,,
  \qquad\qquad m\not=n\not=r\in\{f,g,\bar f'\}\,,
  \label{eqs:stu_ff}
\ee
and the identity
\be\label{eq:stuSum_ff}
  M^2\;=\;s_{fg}+s_{g\bar f'}+s_{f\bar f'}\,,
  \qquad{\rm also\ expressed\ by}\qquad
  2\;=\;x_f+x_g+x_{\bar f'}
\ee
hold true.  All energy fractions fall into the range $0\le x_r\le1$,
and, hence, the physics constraints imposed on the kinematic
invariants $s_{mn}$ are satisfied.  Following the steps outlined in
\sec{sec:GenEvoVars}, the $(p^2_\perp,y)$ phase-space parametrization
can be characterized:
\bi
\item The two Lorentz invariant dipole evolution variables are
  \be
    p^2_\perp\;=\;\frac{s_{fg}\,s_{g\bar f'}}{M^2}
             \;=\;M^2(1-x_{\bar f'})(1-x_f)\,,
  \ee
  cf.\ \eqss{eq:LundP2t}{eq:GenP2t}, and the associated rapidity $y$,
  \be
    y\;=\;\frac{1}{2}\,\ln\frac{s_{g\bar f'}}{s_{fg}}
     \;=\;\frac{1}{2}\,\ln\frac{1-x_f}{1-x_{\bar f'}}\,,
  \ee
  cf.\ \eqss{eq:LundY}{eq:GenY}.  Therefore, the invariant masses can
  be re-expressed as,
  \bea
    s_{g\bar f'} &=& Mp_\perp e^{+y}\,,\nn\\[1mm]
    s_{fg}       &=& Mp_\perp e^{-y}\,,\nn\\[1mm]
    s_{f\bar f'} &=& M^2-2Mp_\perp\cosh y\,,
  \eea
  cf.\ \eqs{eqs:GenSfromEvoVars}.  As expected, the dominant
  phase-space regions are characterized by $p_\perp\to0$, which points
  at $p_\perp$'s utilization as the ordering variable.
\item The kinematic phase-space boundaries given through the relations
  in \eqs{eqs:stu_ff} determine the (maximal) integration limits
  $p^2_{\perp,\mr{high}}$ and $y_\pm$ stated in \eq{eq:GenSud}.
  The determination of the precise boundaries is determined by the
  constraint
  \be
    s_{fg}+s_{g\bar f'}\;=\;M^2-s_{f\bar f'}\;\le\;M^2\,,
    \label{eq:strict_ff}
  \ee
  leading to the following symmetric rapidity limits
  \be
    \abs{y}\;\le\;{\mr{arcosh}}\frac{M}{2p_\perp}\;=\;
	          \ln\left(\frac{M}{2p_\perp}+
                  \sqrt{\frac{M^2}{4p^2_\perp}-1\z}\,\right)\,.
    \label{eq:ypm_ff}
  \ee
  The largest possible value for $p^2_\perp$ can also be read off
  these bounds,
  \be
    p^2_{\perp,\mr{max}}\;=\;\frac{M^2}{4}\,.
  \ee
\item Simple rapidity bounds overestimating the more exact interval
  are obtained, for example, from $s_{fg},s_{g\bar f'}\le M^2$; this
  yields
  \be
    Y_-=-\ln\frac{M}{p_\perp}\;\le\;y\;\le\;\ln\frac{M}{p_\perp}=Y_+\,,
  \ee
  which is nothing but the $(y,z=\ln\frac{p_\perp}{M})$ ``triangle''
  commonly used to illustrate a dipole emission phase space. The
  effect of the sharper bounds now becomes apparent: they sizeably 
  reduce the ``triangle'' area particularly in the central rapidity
  region, see \fig{fig:ffpsp}.
\item Splitting kinematics:\qquad here the ideal frame to set up the
  new momenta is the centre-of-mass system of the parent FF dipole.
  Light-cone momenta\spc\footnote{
    In this work, light-cone momenta are defined as follows:
    $q=(q_+,q_-,\vec q_\perp)$ where $q_{\pm}=E_q\pm q_\parallel$;
    on-shell conditions can be intrinsically satisfied, if $q=(m_\perp
    e^z,m_\perp e^{-z},\vec q_\perp)$ is chosen, using
    $m^2_\perp=q^2+q^2_\perp$ and $z=\ln(q_+/q_-)/2$ such that
    $E_q=m_\perp\cosh z$\/ and $q_\parallel=m_\perp\sinh z$.}
  w.r.t.\ the axis of the two original partons can conveniently be
  used. They yield
  \bea
    \td p_0\;=\;\left(M,\,M,\,\vec0\right)
    &\;\,\to\;\,&
    p_0\:\;=\;\left(M,\,M,\,\vec0\right)\,,
    \nn\\[1mm]
    \td p_f\;=\;\left(M,\,0,\,\vec0\right)
    &\;\,\to\;\,&
    p_f\:\;=\;\left(f_\perp\,e^{y_f},\,f_\perp\,e^{-y_f},\,
                    \vec f_\perp\right)\,,
    \nn\\[1mm]
    \td p_{\bar f'}\;=\;\left(0,\,M,\,\vec0\right)
    &\;\,\to\;\,&
    p_{\bar f'}\;=\;\left(f'_\perp\,e^{y'_f},\,f'_\perp\,e^{-y'_f},\,
                          \vec f'_\perp\right)\,,
    \nn\\[1mm]
    &&
    p_g\:\;=\;p_0-p_f-p_{\bar f'}\,.
    \label{eqs:cmskin_ff}
  \eea
  The $\vec f^{(')}_\perp$ and $y^{(')}_f$ are specified by the
  particular recoil strategies that are used for the different types
  of FF dipoles. The choices taken here closely follow the approach
  presented within the Lund CDM, see \eg \cite{Lonnblad:1992tz}.
  Thus, for gluon emissions off $q\bar q$\/ dipoles, the Kleiss trick
  \cite{Kleiss:1986re} has been implemented to treat the recoils: the
  (anti)quark will retain its direction after the emission with a
  probability $x^2_{q(\bar q)}/(x^2_q+x^2_{\bar q})$. For $qg$\/
  dipoles, the recoil of the emitted gluon will be compensated by the
  quark only. Specifying the kinematics of these cases (assuming, for
  example to preserve the direction of the $\bar f'$) leads to
  \be
    f_\perp\;=\;\frac{x_f M}{2}\sin\vartheta\,,\qquad
    \vec f_\perp\;=\;(f_\perp\cos\varphi,f_\perp\sin\varphi)
  \ee
  and
  \be
    y_f\;=\;\frac{1}{2}\ln\frac{1+\cos\vartheta}{1-\cos\vartheta}
       \;=\;\ln\left(\cot\frac{\vartheta}{2}\right)\,,
  \ee
  where the polar angle $\vartheta$\/ is given through
  $\cos\vartheta=(x^2_g-x^2_f-x^2_{\bar f'})/(2\,x_fx_{\bar f'})$ and
  the azimuthal angle $\varphi$\/ is taken to be uniformly distributed
  between $0$ and $2\pi$. Moreover,
  \be
    \vec f'_\perp\equiv\vec 0\,,\qquad
    y'_f\equiv\infty\,,\qquad\mbox{preserving\ the\ product}\qquad
    f'_\perp e^{y'_f}\;=\;x_{\bar f'} M\,.
  \ee
  For the distribution of recoils arising from $gg$\/ dipole
  splittings, the simple specification will be corrected by rotating
  around the $\hat{\mr{y}}$ axis in a way that the $\sum p^2_T$ of the
  parent gluons will be minimized, however, small perturbations
  introduced by an additional rotation around the $\hat{\mr{x}}$ axis
  are allowed.
\ei
\subsection{Splitting functions for final-state QCD radiation}
\label{sec:FFdips}
In this section the refinements introduced by the Lund CDM
\cite{Gustafson:1986db,Gustafson:1987rq,Andersson:1989ki,Lonnblad:1992tz}
to the simple eikonal splitting cross sections are briefly reviewed,
in particular those for gluons arising from a $q_\fl\bar q_\fl$
dipole.  Following the reasoning of \sec{sec:Dsfuncs}, the dipole
splitting function for the $2\to3$ splitting $q\bar q'\to qg\bar q'$\/
is worked out from the comparison of the real-emission process $V\to
q\bar q'g$\/ to the Born contribution for the vector boson decay
$V\to q\bar q'$.\spc\footnote{
  The inclusion of various correlations depends on exactly
  which processes are selected.  For example, also the correlation of
  the leptons, producing the vector boson $V$, with the quarks could
  be accounted for by taking the processes $\ell\bar\ell'\to q\bar
  q'(g)$ instead \cite{Kleiss:1986re,Seymour:1994we}.}
For massless partons, the respective squared matrix elements averaged
(summed) over colour and spin initial (final) states are
\bea
  \overline{\abs{\cal{M}_{\gamma^\ast\to q\bar q}}^2} &=&
  8\;N_{\mr C}\;e^2e^2_q\;\td p_q\td p_{\bar q}\;=\;
  8\;N_{\mr C}\;e^2e^2_q\;\frac{M^2}{2}\,,\nn\\[1mm]
  \overline{\abs{\cal{M}_{\gamma^\ast\to q\bar qg}}^2} &=&
  8\;\frac{N^2_\mr{C}-1}{2}\;e^2e^2_q\;4\pi\alpha_s\;
  \frac{x^2_q+x^2_{\bar q}}{(1-x_q)(1-x_{\bar q})}\,.
\eea
The Lorentz invariant energy fractions of the emission are defined in
\eq{eq:XkDef}, and, for simplicity, $V=\gamma^\ast$ has been chosen,
see \eg \cite{Field:1989uq}. Therefore,
\be
  \frac{d\Gamma_{\gamma^\ast\to qg\bar q}}{dx_q dx_{\bar q}}\;=\;
  d\Gamma_{\gamma^\ast\to q\bar q}\;
  \frac{C_F\alpha_s}{2\pi}\;
  \frac{x^2_q+x^2_{\bar q}}{(1-x_q)(1-x_{\bar q})}\,,
  \label{eq:xsec_xterms_ff}
\ee
where $C_F=(N^2_\mr{C}-1)/(2N_\mr{C})$ is the colour factor of this
emission.  Obviously, in this case the matrix-element factorization is
{\it exact}\/ allowing to read off the corresponding differential
dipole splitting cross section according to \eq{eq:SplFuncDef},
\be
  \frac{d\cal{P}_{q\bar q\to qg\bar q}}{dp^2_\perp dy}\;=\;
  \frac{C_F\alpha_s}{2\pi}\;
  \frac{(1-\frac{p_\perp}{M}\,e^{+y})^2+(1-\frac{p_\perp}{M}\,e^{-y})^2}
       {p^2_\perp}
  \;=\;
  \frac{C_F\alpha_s}{2\pi}\;
  \frac{x^2_q(p_\perp,y)+x^2_{\bar q}(p_\perp,y)}{p^2_\perp}\,.
  \label{eq:qqdip}
\ee
The obvious overestimation of the exact result,
\be
  \frac{d\cal{P}^{\mr{approx}}_{q\bar q\to qg\bar q}}{dp^2_\perp dy}\;=\;
  \frac{C_F\alpha_s}{2\pi}\;
  \frac{2}{p^2_\perp}\,,
\ee
in fact corresponds to a soft-gluon approximation neglecting quark
spins, and allows for a direct implementation in a veto algorithm.
Moreover, the dipole matrix element identified for the exact splitting
can be cast into the following form:
\be
  \hat D_{q\bar q\to qg\bar q}\;=\;
  \frac{x^2_q+x^2_{\bar q}}{p^2_\perp}\;=\;
  \frac{1}{s_{qg\bar q}}\left(
  \frac{s_{qg}}{s_{g\bar q}}+\frac{s_{g\bar q}}{s_{qg}}+
  \frac{2s_{q\bar q}s_{qg\bar q}}{s_{qg}s_{g\bar q}}\right)\,,
\ee
where the rightmost expression exactly reproduces the three-parton
tree-level antenna function $A^0_3(1_q,3_g,2_{\bar q})$ as stated in
\cite{GehrmannDeRidder:2005cm}. This case constitutes the easiest
example for the compatibility of the two matrix-element
factorization approaches.
\\
Similar reasoning can be applied to yield the splitting functions for
gluon emission off quark--gluon and gluon--gluon FF dipoles
\cite{Pettersson:1988zu}.  It should be stressed, however, that in
these cases the dipole matrix-element factorization is correctly
achieved only in the singular limits of the emission.  Taken together,
the dipole splitting functions for gluon emission off final-state
dipoles in the Lund CDM
\cite{Gustafson:1986db,Gustafson:1987rq,Andersson:1989ki} read
\be\begin{split}
  D_{f\bar f'\to fg\bar f'}(p_\perp,y)&\;=\;
  \xi_{\left\{{F\atop A}\right\}}
  C_{\left\{{F\atop A}\right\}}\,
  \left[x_f^{n_f}(p_\perp,y)+x_{\bar f'}^{n_{\bar f'}}(p_\perp,y)\right]
  \\[2mm]&\;\le\;
  2\;\xi_{\left\{{F\atop A}\right\}}
  C_{\left\{{F\atop A}\right\}}\;\equiv\;
  D^{\mr{approx}}_{f\bar f'\to fg\bar f'}(p_\perp,y)\,.
  \label{eq:Dfunc_fffg}
\end{split}\ee
They are all implemented in \ariadne and will be used in the model
presented here as well.  The invariant energy fractions are given by
\be
  x_{f,\bar f'}\;=\;1-\frac{p_\perp}{M}\,e^{\pm y}\,.
\ee
Here and in the following, the parton-dependent exponents are defined
as $n_{q,g}=2,3$ and the curly-brackets notation is understood as
$\left\{\ldots\quad{\rm for\ quark\ dipoles}\atop
\ldots\quad{\rm else}\hphantom{\rm quark\ dipoles}\right\}$.  Note
that $\xi_F=1$; in the splitting functions for dipoles consisting
of at least one gluon the factor of $\xi_A=\frac{1}{2}$ enters, since
gluons are shared among two dipoles.
The $D^{\mr{approx}}_{f\bar f'\to fg\bar f'}$ not only give upper
bounds to the exact splitting functions, they also imply eikonal
approximations to the splitting cross sections of \eq{eq:SplFuncDef}.
\\
A subtle issue in the formulation of a dipole shower is the assignment
of colour factors.  Obviously, for quark--quark and purely gluonic
dipoles there are no problems, and, unambiguously, $C=C_F$ and
$C=C_A$, respectively.  For dipoles consisting of a(n) (anti)quark and
a gluon, it is known that the colour factor cannot be pinpointed as
straightforwardly as in the other cases, since \eg for collinear
radiation, the considered gluon emission can be traced back to either
the (parent) quark or the (parent) gluon, such that in this limit the
emission is therefore governed by $C_F$ or $C_A$, respectively.
Literally taken, these different colour-factor regimes have to be
taken into account. This will lead to modifications of the
corresponding dipole splitting functions and, possibly, to a
decomposition (partitioning) of them into sub-contributions
(subantenn\ae) addressing these different regimes unambiguously.
However, in the large $N_\mr{C}$ limit, underlying the construction of
shower codes, this issue triggers subleading effects only, since
$2\,C_F,\,C_A\to N_\mr{C}$ keeping in mind that the gluon-sharing
factor is $\xi_A=\frac{1}{2}$.
%

%% file: tex/dsii.tex
\section{Initial-state colour dipoles}
%
\label{sec:II}
The first case, which goes beyond the original CDM, is radiation off
an initial-state dipole $\bar\imath'i$\/ of mass $M$.  Two generic
splittings based on gluon emission are available, namely
\be
  \bar\imath'(\td k)\,i(\td\ell)\;\to\;\bar\imath'(k)\,g\,i(\ell)
  \qquad{\rm and}\qquad
  \bar q_\il(\td k)\,i(\td\ell)\;\to\;q(k)\,g_\il\,i(\ell)\,.
\ee
\subsection{Single-emission kinematics}
\label{sec:IIlims}
Restating \eqss{eq:GenMombf}{eq:GenMomaf} for the II dipole scenario,
\ie setting $\vs_0=\vs_{g/q}=1$ and all other signature factors equal
to $-1$, yields
\be
  \td p_{\bar\imath'/\bar q_\il}+\td p_i\;=\;\td p_0
  \qquad{\rm and}\qquad
  p_{\bar\imath'/g_\il}+p_i\;=\;p_0+p_{g/q}\,,
  \qquad{\rm with}\qquad
  p_0^2\;=\;\td p_0^2\;=\;M^2\,.
  \label{eq:p_ii}
\ee
As already noted, the kinematics of the emission process here
corresponds to that of a $2\to2$ scattering process rather than to
that of a $1\to3$ decay.  The recoil of the emitted
parton $p_{g/q}$ cannot be absorbed by $p_{\bar\imath'/g_\il}$ and
$p_i$, since they are fixed to the beam axis.  Thus, in contrast to
the previously presented case, $\td p_0\not=p_0$.\spc\footnote{Unless
  the converse is enforced by a recoil handling in the
  $\bar\imath'i$\/ dipole's rest frame.}
For the scattering process, Mandelstam variables are defined as
\bea
  \hat s &=& (p_0+p_{g/q})^2\;=\;(p_{\bar\imath'/g_\il}+p_i)^2
  \;=\;M^2(1+x_{g/q})\hphantom{.}\;\ge\;M^2\;\equiv\;\hat s_0\,,\nn\\[1mm]
  \hat t &=& (p_0-p_i)^2\;=\;(p_{\bar\imath'/g_\il}-p_{g/q})^2
  \;=\;M^2(1-x_i)\hphantom{x..}\;\le\;0\,,\nn\\[1mm]
  \hat u &=& (p_0-p_{\bar\imath'/g_\il})^2\;=\;(p_i-p_{g/q})^2
  \;=\;M^2(1-x_{\bar\imath'/g_\il})\;\le\;0\,,
  \label{eqs:stu_ii}
\eea
where, again for massless partons, the bounds on $\hat s$, $\hat t$\/
and $\hat u$\/ together with their parametrizations in terms of energy
fractions are simple, and, furthermore,
\be
  \hat s+\hat t+\hat u\;=\;M^2
  \qquad{\rm as\ well\ as}\qquad
  x_{\bar\imath'/g_\il}+x_i\;=\;2+x_{g/q}\,.
  \label{eq:stuSum_ii}
\ee
As already indicated, the emission of a parton here requires an
increase in $\hat s$\/ w.r.t.\ $\hat s_0$, related to an increase of
the ``Bj{\o}rken-$x$''.  This is in contrast to the FF case, where the
system's centre-of-mass energy remains constant.  To deal with this
issue, a generic parametrization is introduced, which relates the
maximal partonic centre-of-mass squared energy to the squared mass of
the parent dipole,
\be
  \hat s_{\mr{max}}\;=\;a\,M^2\;\ge\;\hat s\,,
  \qquad{\rm such\ that}\qquad
  1\;\le\;\hat s/M^2\;\le\;a\;\le\;S/M^2\,,
  \label{eq:shatmax_ii}
\ee
where $\sqrt{S}$\/ is the centre-of-mass energy of the colliding
hadrons.  The limits on the invariants, detailed in
\eqss{eqs:stu_ii}{eq:shatmax_ii}, clearly differ from the ones of the
FF scenario, cf.\ \sec{sec:FFlims}.  This implies that the II dipole
splittings arise in phase-space regions being distinct from the FF
case and thus with a different kinematics.  Consequently, the energy
fractions populate new ranges compared to the FF splittings, viz.\
\bea
  0\;\le & x_{g/q}                     & \le\;a-1\,,\nn\\[1mm]
  1\;\le & x_{\bar\imath'/g_\il},\,x_i & \le\;1+x_{g/q}\,,\nn\\[1mm]
  2\;\le & x_{\bar\imath'/g_\il}+x_i   & \le\;1+a\,.
  \label{eqs:xbounds_ii}
\eea
The phase-space parametrization is better worked out separately for
both relevant II dipole splitting channels.
\subsubsection{Gluon emission phase space of initial--initial dipoles}
\label{sec:gf_ii}
\begin{figure}[t!]
  \vspace{0mm}\bc
  \includegraphics[width=61mm,angle=0.0]{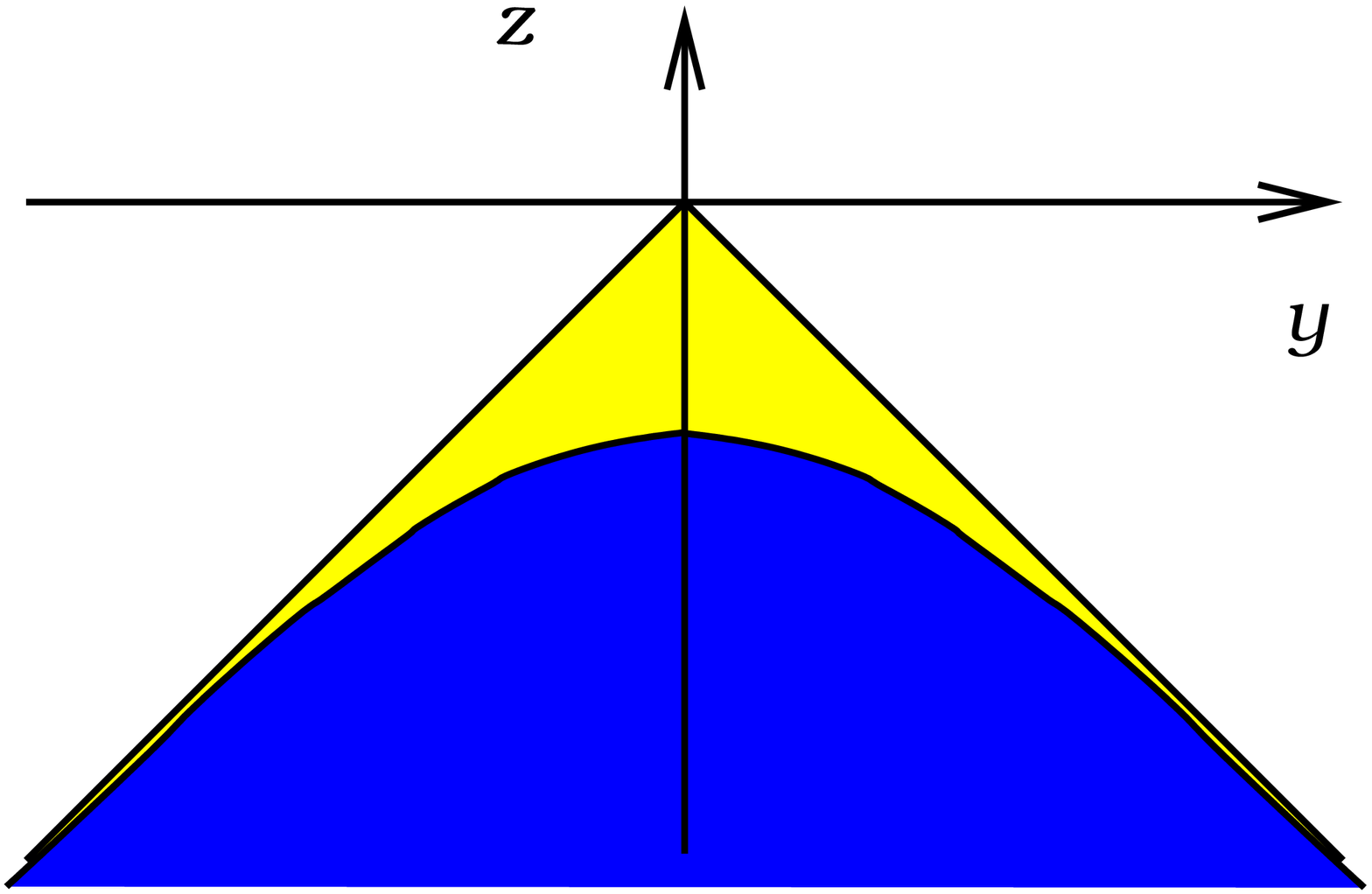}
  \hspace*{24mm}
  \includegraphics[width=65mm,angle=0.0]{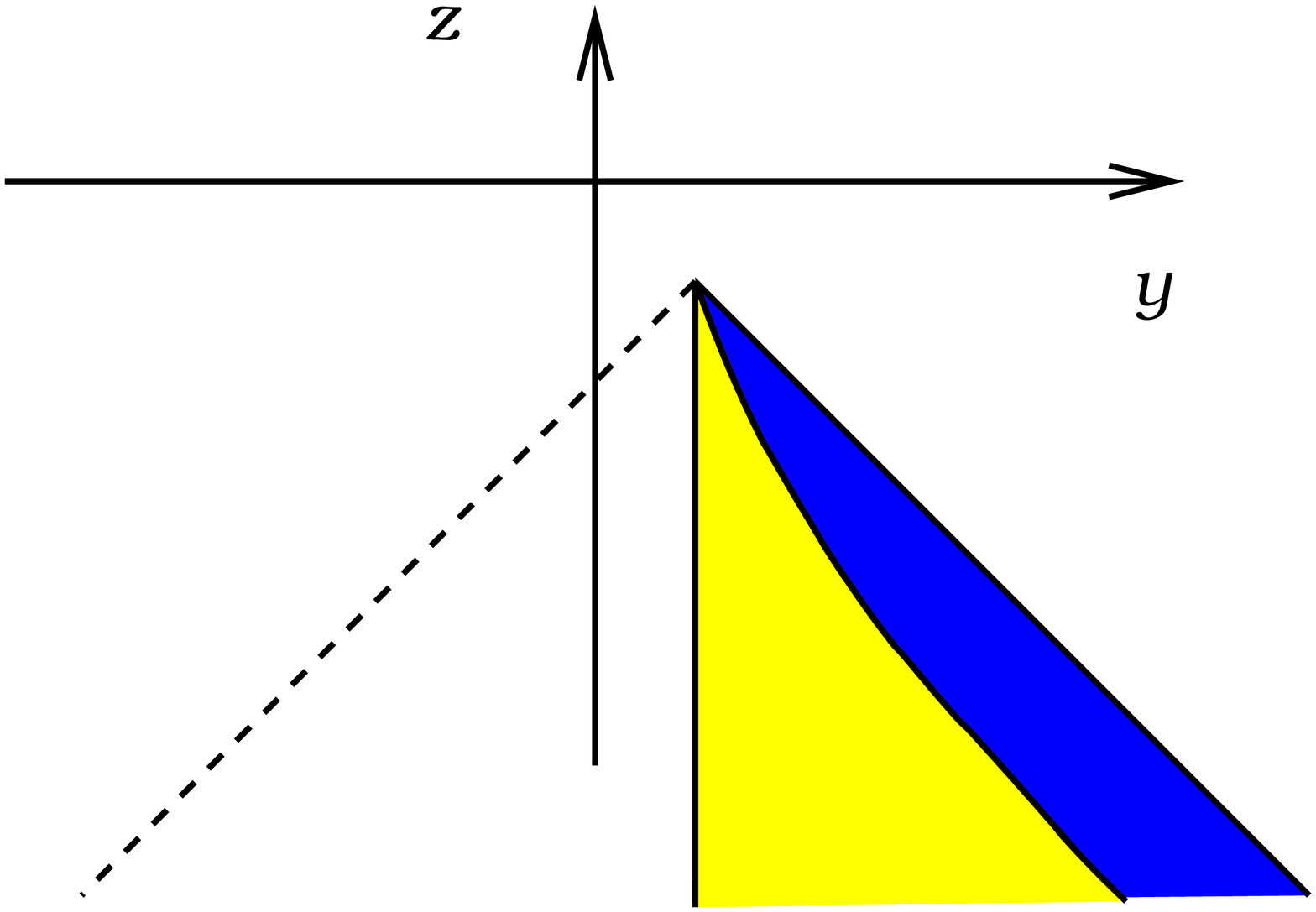}
  \ec\vspace{0mm}
  \myfigcaption{140mm}{The accessible phase space for final-state
    (left panel) and initial-state (right panel) gluon emission off II
    dipoles. Bright colours indicate the phase-space fractions, which
    overestimate the respective allowed phase-space regions, which are
    shown in dark colours. The definitions of $z$\/ are,
    $z=\ln\frac{p_\perp}{(a-1)M}$ and $z=\ln\frac{p_\perp}{aM}$ for
    $g_\fl$ and $g_\il$ emissions, respectively. Notice that the
    visualization of the $g_\il$ emission phase space is for $a=2$.}
  \label{fig:iipsp}
  \vspace{0mm}
\end{figure}

First, the case of final-state gluon ($g_\fl$) emission, \ie
$\bar\imath'i\to\bar\imath'gi$, is discussed:
\bi
\item The evolution variables are taken as suggested by
  \eqss{eq:GenP2t}{eq:GenY}.  The Lorentz invariant $p^2_\perp$ thus
  reads
  \be
    p^2_\perp\;=\;
    \left|\frac{s_{\bar\imath'g}\,s_{gi}}{s_{\bar\imath'gi}}\right|\;=\;
    \frac{\hat t\,\hat u}{M^2}\;=\;
    M^2(1-x_i)(1-x_{\bar\imath'})\,,
  \ee
  and the Lorentz invariant $y$\/ is given by
  \be
    y\;=\;\frac{1}{2}\,\ln\left|\frac{s_{gi}}{s_{\bar\imath'g}}\right|
    \;=\;\frac{1}{2}\,\ln\frac{\hat u}{\hat t}
    \;=\;\frac{1}{2}\,\ln\frac{1-x_{\bar\imath'}}{1-x_i}\,,
  \ee
  such that the kinematic invariants can be re-written as
  \bea
    \hat s &=& s_{0g}\,=\,s_{\bar\imath'i}
    \;=\;M^2+2M\,p_\perp\cosh y\;\ge\;M^2\,,\nn\\[1mm]
    \hat t &=& s_{0i}\,=\,s_{\bar\imath'g}
    \;=\;-M\,p_\perp e^{-y}\;\le\;0\,,\nn\\[1mm]
    \hat u &=& s_{0\bar\imath'}\,=\,s_{gi}
    \;=\;-M\,p_\perp e^{+y}\;\le\;0\,.
    \label{eqs:stu_iifg}
  \eea
\item The bounds on the Mandelstam variables -- or equally well -- on
  the invariant energy fractions translate, of course, into bounds on
  the evolution variables.  As for FF dipoles emitting gluons, the
  more restrictive requirement is obtained from
  \be
    (a-1)\,M^2\;=\;\hat s_{\mr{max}}-M^2
    \;\ge\;\hat s-M^2\;=\;-\hat u-\hat t\;=\;2M\,p_\perp\cosh y\,.
    \label{eq:sharplim_iifg}
  \ee
  Hence, the allowed phase space, which is depicted in the left part
  of \fig{fig:iipsp}, is described quantitatively through
  \be
    \abs{y}\;\le\;\mr{arcosh}\,\frac{(a-1)M}{2\,p_\perp}\,,
    \label{eq:ypm_iifg}
  \ee
  and\spc\footnote{The maximal rapidity range is determined by the
    overall cut-off on $p_\perp$: $\hat s_{\mr{max}}-M^2=
    2M\,p_{\perp,\mr{cut}}\cosh\abs{y}_\mr{max}$.}
  \be
    p^2_{\perp,\mr{max}}\;=\;\frac{(\hat s_{\mr{max}}-M^2)^2}{4\,M^2}
    \;=\;\frac{(a-1)^2M^2}{4}\,.
    \label{eq:pperpmax_iifg}
  \ee
\item Weaker constraints are obtained from
  \be
    \hat s_{\mr{max}}-M^2\;\ge\;\hat s-M^2\;\ge\;-\hat u,-\hat t
  \ee
  and, as in the FF case, they result in symmetric rapidity limits,
  \be
    \abs{y}\;\le\;\ln\frac{\hat s_{\mr{max}}-M^2}{M\,p_\perp}\;=\;
    \ln\frac{(a-1)M}{p_\perp}\,.
  \ee
  These estimates again can be visualized by a ``triangle'' in the
  $(y,z=\ln\frac{p_\perp}{(a-1)M})$ plane.
\item The splitting kinematics will be detailed in \sec{sec:IIKins}
  together with that of the quark-emission process.
\ei
Compared to the FF case, a new issue emerges: the maximal partonic
centre-of-mass scale $\hat s_{\mr{max}}$ is not fixed and can be
chosen.  The actual choice then regulates the maximal size of the
allowed emission phase space.  This will be discussed together with
the shower algorithm in \sec{sec:Shower}.
\subsubsection{Quark emission phase space of initial--initial dipoles}
\label{sec:gi_ii}
Along the lines of the previous section, the phase-space
parametrization and its consequences are now discussed for gluon
emission into the initial state ($g_\il$), \ie (massless) quark
emission into the final state: $\bar q_\il i\to qg_\il i$.
The details of the kinematics as outlined in \sec{sec:GenEvoVars} are
as follows.
\bi
\item The Lorentz invariant shower variables expressed through the
  Mandelstam variables read
  \be
    p^2_\perp\;=\;\left|\frac{s_{qg_\il}\,s_{g_\il i}}{s_{qg_\il i}}\right|
    \;=\;-\frac{\hat t\,\hat s}{M^2}
    \;=\;M^2(x_i-1)(1+x_q)\,,
  \ee
  \be
    y\;=\;\frac{1}{2}\,\ln\left|\frac{s_{g_\il i}}{s_{qg_\il}}\right|
    \;=\;\frac{1}{2}\,\ln\frac{\hphantom{x}\hat s\hphantom{x}}{-\hat t}
    \;=\;\frac{1}{2}\,\ln\frac{1+x_q}{x_i-1}\,.
  \ee
  This allows to rewrite the kinematic invariants as
  \bea
    \hat s &=& s_{0q}\,=\,s_{g_\il i}
    \;=\;+M\,p_\perp e^{+y}\;\ge\;M^2\,,\nn\\[1mm]
    \hat t &=& s_{0i}\,=\,s_{qg_\il}
    \;=\;-M\,p_\perp e^{-y}\;\le\;0\,,\nn\\[1mm]
    \hat u &=& s_{0g_\il}\,=\,s_{qi}
    \;=\;M^2-2M\,p_\perp\sinh y\;\le\;0\,,
    \label{eqs:stu_iiig}
  \eea
  implying, compared to the case of $g_\fl$ emission, a different
  shape of the valid $(p_\perp,y)$ phase space covered by this type of
  emission.
\item The accessible rapidity range is (cf.\ right panel of \fig{fig:iipsp})
  \be
    \ln\left(\frac{M}{2p_\perp}+
    \sqrt{\frac{M^2}{4p^2_\perp}+1\z}\,\right)\;=\;
    \mr{arsinh}\,\frac{M}{2\,p_\perp}\;\le\;
    y\;\le\;
    \ln\frac{a\,M}{p_\perp}\,,
    \label{eq:ypm_iiig}
  \ee
  where the left and right bounds result from $\hat u\le0$ and 
  $\hat s\le\hat s_{\mr{max}}$, cf.\ \eq{eq:shatmax_ii}, respectively.
  This can be visualized in the $(y,z=\ln\frac{p_\perp}{a\,M})$ plane
  as a ``strip'' that emerges in the point
  $(y_{\mr{min}},z_{\mr{max}}=-y_{\mr{min}})$ and is confined between
  $z=-y-\ln a$\/ and $z=-y$. The equations
  $y_{\mr{min}}=\mr{arsinh}\,\frac{M}{2\,p_{\perp,\mr{max}}}$ and
  $\hat s_{\mr{max}}=M\,p_{\perp,\mr{max}}\,e^{y_{\mr{min}}}$ yield
  \be
    y_{\mr{min}}
    \;=\;\frac{1}{2}\ln\frac{\hat s_{\mr{max}}}{\hat s_{\mr{max}}-M^2}
    \;=\;\frac{1}{2}\ln\frac{a}{a-1}\,,
  \ee
  and
  \be
  p^2_{\perp,\mr{max}}
    \;=\;(\hat s_{\mr{max}}-M^2)\;\frac{\hat s_{\mr{max}}}{M^2}
    \;=\;a(a-1)\,M^2\,.
    \label{eq:pperpmax_iiig}
  \ee
\item The allowed phase-space region is safely covered by a
  ``half-triangle'' described through $y_{\mr{min}}\le y\le-z$.
  Accordingly, $\Delta y=-z-y_{\mr{min}}=\ln(p^2_{\perp,\mr{max}}/p^2_\perp)/2$.
\item The splitting kinematics is presented in the next subsection.
\ei
Finally, notice that, as for $g_\fl$ emissions, the single-emission
phase-space maximally available is determined by the actual value
given to $\hat s_{\mr{max}}$.
\subsubsection{Construction of the splitting kinematics}
\label{sec:IIKins}
In the model proposed here the initial--initial dipole kinematics is
directly constructed in the lab-frame.  Particularly, to handle the
recoils for the case of $\bar q'_\il q_\il$ dipoles, the strategy
according to Kleiss \cite{Kleiss:1986re,Seymour:1994we} has been
implemented.
\paragraph{Lab-frame kinematics:}
the fixed orientation of incoming partons implies that the emitted
parton's recoil will directly be transferred to the entire
final-state system, \ie to all QCD and non-QCD final-state particles
that are present before the emission takes place.  As an example,
consider the first emission in a Drell--Yan process, where the
corresponding recoil is compensated for by the lepton pair.  This
recoil transfer results in $\td p_0\not=p_0$, and, therefore, a
Lorentz transformation ${\cal T}$\/ defined through $p_0=\cal{T}\td
p_0$ is necessary and will be applied on all particles
(whose vectors are summed up in $\td p_0$).  For the construction of
the momenta, a light-cone decomposition w.r.t.\ the beam axis is well
suited, such that, for massless partons, the situation before and
after the emission is summarized as
\bea
  \td p_i\;=\;\left(\td x_+\,\sqrt{S},\,0,\,\vec0\right)
  &\;\,\to\;\,&
  p_{i\hphantom{'/g_\il}}\;=\;\left(x_+\,\sqrt{S},\,0,\,\vec0\right)\,,
  \nn\\[1mm]
  \td p_{\bar\imath'/\bar q_\il}\;=\;\left(0,\,\td x_-\,\sqrt{S},\,\vec0\right)
  &\;\,\to\;\,&
  p_{\bar\imath'/g_\il}\;=\;\left(0,\,x_-\,\sqrt{S},\,\vec0\right)\,,
  \nn\\[1mm]
  \td p_0\;=\;\left(M\,e^{\td y_0},\,M\,e^{-\td y_0},\,\vec0\right)
  &\;\,\to\;\,&
  p_{0\hphantom{/q}}\;=\;\left(M_\perp\,e^{y_0},\,M_\perp\,e^{-y_0},\,
                               -\vec\ell_\perp\right)\,,
  \nn\\[1mm]
  &&
  p_{g/q}\;=\;\left(\ell_\perp\,e^{y_e},\,\ell_\perp\,e^{-y_e},\,
                               \vec\ell_\perp\right)\,.
  \label{eqs:labkin_ii}
\eea
Furthermore, $\hat s_0=\td x_+\td x_-S=M^2$ and
$\td y_0=\td y_{\mr{cm}}=\ln(\td x_+/\td x_-)/2$ with $\td
y_{\mr{cm}}$ denoting the centre-of-mass rapidity of the parton
system.  The $\td x_\pm$, here functions of $M$, $S$\/ and $\td
y_{\mr{cm}}$, parametrize the momentum fractions of the partons w.r.t\
their respective hadron.  Employing $M^2_\perp=M^2+\ell^2_\perp$,
after the emission they read
\be
  x_\pm\;=\;\frac{\ell_\perp\,e^{\pm y_e}+M_\perp\,e^{\pm y_0}}{\sqrt S}
       \;\ge\;\td x_\pm\,.
\ee
Clearly, emissions leading to $x_\pm>1$ must be rejected.  The vector
$\vec\ell_\perp=(\ell_\perp\cos\varphi,\ell_\perp\sin\varphi)$ and the
quantity $y_e$ denote the transverse momentum and the rapidity of the
emitted parton w.r.t.\ the beam axis, respectively.  In terms of the
Mandelstam variables, cf.\ \eqs{eqs:stu_ii}, they are:
\be
  \ell^2_\perp\;=\;\frac{\hat t\,\hat u}{\hat s}
  \qquad{\rm and}\qquad
  e^{y_e}\;=\;\frac{e^{y_0}}{M_\perp\ell_\perp}(-\hat t-\ell^2_\perp)\,.
\ee
The azimuthal angle $\varphi$\/ can in first approximation be assumed
to be uniformly distributed, and $\hat s$, $\hat t$, $\hat u$\/ are
determined by the evolution parameters $p^2_\perp$ and $y$\/ through
\eqs{eqs:stu_iifg} and \eqs{eqs:stu_iiig} for $g_\fl$ and $g_\il$
emissions, respectively.  The squared lab-frame transverse
momenta are exemplified below as functions of $p_\perp$ and $y$.  For
gluon emission into the final state,
\be
  \ell^2_\perp\;=\;\frac{M^2\,p^2_\perp}{\hat s}
              \;=\;\frac{p^2_\perp}{2\,p_\perp M^{-1}\cosh y+1}\,,
  \label{eq:labkperp_ii}
\ee
whereas for quark emission,
\be
  \ell^2_\perp\;=\;\frac{M^2\,p^2_\perp}{\hat s}-
                   \frac{(M^2+\abs{\hat t}\z)\,\abs{\hat t}}{\hat s}
	      \;=\;(Me^{-y})^2\,(2\,p_\perp M^{-1}\sinh y-1)\,.
\ee
When comparing both equations for the same ratio $p^2_\perp/\hat s$,
it becomes apparent that the emissions of quarks yield smaller
lab-frame transverse momenta than those of gluons.
\\
To fix the last degree of freedom, an additional assumption is
necessary, which is to preserve the rapidity of the system of outgoing
particles, $y_0=\td y_0=\td y_{\mr{cm}}$.\spc\footnote{
  If $y_{\mr{cm}}=\td y_{\mr{cm}}$ was naively exploited, the ratio of
  momentum fractions would remain constant, $x_+/x_-=\td x_+/\td x_-$,
  which constitutes a rather strange behaviour, since, for instance,
  very asymmetric starting configurations would persist to the end of
  the shower evolution.}
Having the complete emission at hand, $\hat s=x_+x_-S$\/ and
$y_{\mr{cm}}=\ln(x_+/x_-)/2=\ln(\hat u/\hat t\x)/2+y_e\x$.\spc\footnote{
  Particularly, for gluon emissions into the final state,
  $y_{\mr{cm}}-y_e=y$. This simply expresses that rapidity differences
  are invariant under boosts along the beam axis.}
In more detail,
\be
  y_{\mr{cm}}
  \;=\;\td y_{\mr{cm}}+\frac{1}{2}\,\ln\frac{\hat u}{\hat t}+
  \ln\left(\frac{-\hat t\,(\hat s+\hat u)}
	  {\sqrt{M^2\,\hat s\,\hat t\,\hat u+(\hat t\,\hat u)^2}}\right)
  \;=\;\td y_{\mr{cm}}+\frac{1}{2}\,\ln\frac{M^2-\hat t}{\hat s+\hat t}\,,
  \label{eq:ycmprime_ii}
\ee
which exposes the impact of the $y_0=\td y_{\mr{cm}}$ choice and shows
that the system undergoes a rapidity shift during splitting.  In
addition, the new momentum fractions $x_\pm$ can be written down,
\be
  x_\pm\;=\;e^{\pm y_0}
  \sqrt{\frac{\hat s}{S}\left(\frac{M^2-\hat t}{\hat s+\hat t}\right)^{\pm1}}\,.
\ee
Finally the momenta, $\td p^{(j)}_0$, of all final-state particles,
numbered by $j$, have to be transformed in order to account for the
non-trivial change of $\td p_0\to p_0$.  Here, the Lorentz
transformation $\cal{T}$\/ is specified as follows: the particles are
boosted into the original dipole's centre-of-mass frame, afterwards
the boost that forms $p_0$ out of $(M,\vec0\z)$ is applied on them
likewise.  Altogether $p^{(j)}_0=\cal{B}(-\vec
p_0/p^0_0)\cal{B}(\vec{\td p}_0/\td p^0_0)\,\td p^{(j)}_0=\cal{T}\,\td
p^{(j)}_0$ is computed.  This finalizes the construction of the
on-shell kinematics of an individual emission.
\paragraph{Improved description of lepton--hadron correlations (Kleiss trick):}
when analyzing \eqs{eqs:labkin_ii} again, it is noticed that, apart
from the azimuthal angle $\varphi$, which eventually fixes the vector
$\vec\ell_\perp$, all unknown variables are determined by
Lorentz invariants plus the additional assumption $y_0=y_{\mr{cm}}$.\spc
\footnote{Therefore, when neglecting the angle $\varphi$, it makes no
  difference whether the kinematics is arranged in the parent dipole's
  rest frame or in the lab-frame.}
In a first approximation, the choice is to uniformly distribute in
azimuth w.r.t.\ the lab-frame, but more sophisticated schemes can be
introduced correcting this simple ansatz. One such scheme can be
derived from the work presented in \cite{Kleiss:1986re} where it has
been shown how to exactly factorize the first order tree-level
corrections to the electroweak production of quarks. The corresponding
Monte Carlo algorithm in fact is employed within the Lund CDM to
arrange the splitting kinematics of $q_\fl\bar q'_\fl$ dipoles. In
\cite{Seymour:1994we} this factorization was proven for scattering and
annihilation processes involving initial states and corresponding
algorithms were developed. Accordingly, for the $\bar q'_\il q_\il$
dipoles of this model, the suggestion of \cite{Seymour:1994we} has
been employed to improve the splitting kinematics: the new momenta
are constructed in the original dipole's rest frame in a distinct way,
then they are transformed to the lab-frame such that the
$0$-particle's rapidity is preserved. The essence is that the
primitive $\varphi$\/ choice is substituted by a prescription, which
\eg in Drell--Yan processes correctly accounts for correlations
between the radiated parton and the leptons. As before, the particles
associated to the parent II dipole have to be transformed, however
they now undergo a more complicated series of transformations out of
the (before-emission) lab-frame, \ie more accurately
\be
  p^{(j)}_0\;=\;
  \cal{B}_\parallel\left(\left.\beta_3=\frac{p_{0,+}-e^{2y_0}p_{0,-}}
      {p_{0,+}+e^{2y_0}p_{0,-}}
      \right|_{\hat{\mr{z}}\mbox{-}\mr{align.f.}}\right)
  \cal{R}_{\hat{\mr{z}}}\;
  \cal{B}_{\mr{align}}\left(\left.\frac{\vec p_{\bar\imath'/g_\il}+\vec p_i}
      {p^0_{\bar\imath'/g_\il}+p^0_i}\right|_{\mr{dip.rf.}}\right)
  \cal{B}_\parallel\left(\frac{\vec{\td p}_0}{\td p^0_0}\right)
  \,\td p^{(j)}_0\,,
\ee
where starting from the right, one applies to a momentum: the
longitudinal boost into the dipole's rest frame, the alignment boost
followed by the rotation that brings the newly incoming partons onto
the light-cone axis maintaining the initial $\pm$ assignments, and the
final longitudinal boost to satisfy that $y_0$ stays the same as it
was before the emission, \ie $y_0=\td y_0$.
\subsection{Initial--initial dipole splitting functions}
\label{sec:IIdips}
The first QCD-type emission in vector boson production (real-gluon
bremsstrahlung or QCD Compton scattering) can be described as a
coherent emission of a gluon or a(n) (anti)quark off the primary $\bar
q_\il q'_\il$ dipole, cf.\ \fig{fig:LundvsAdicDY}.
\begin{figure}[t!]
  \vspace{0mm}\bc
  \includegraphics[width=25mm,angle=0.0]{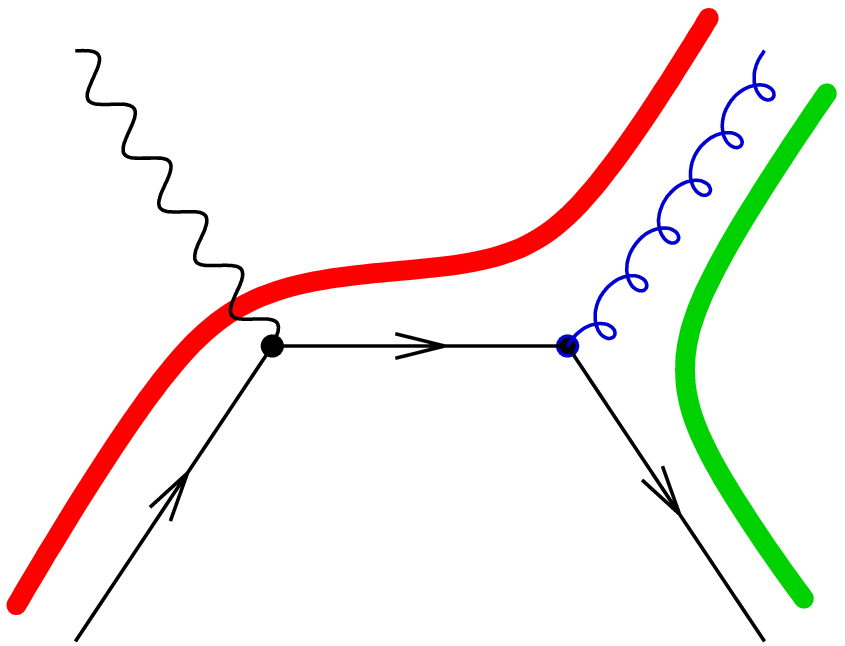}
  \hspace*{7mm}
  \includegraphics[width=25mm,angle=0.0]{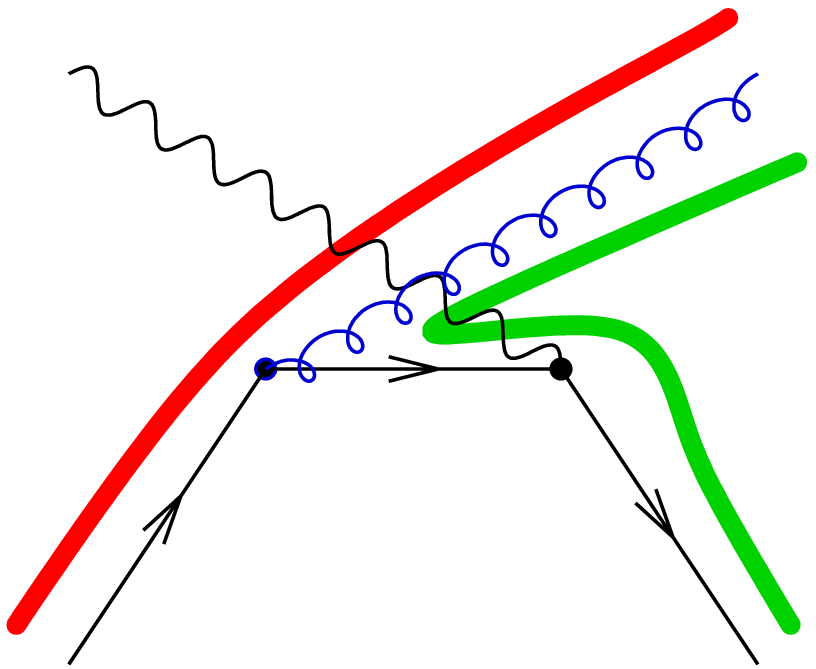}
  \hspace*{17mm}
  \includegraphics[width=25mm,angle=0.0]{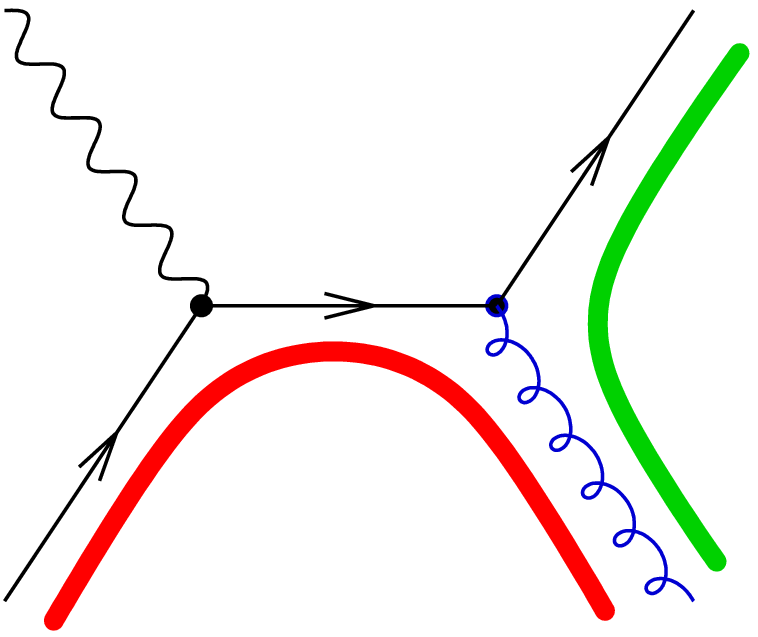}
  \hspace*{7mm}
  \includegraphics[width=25mm,angle=0.0]{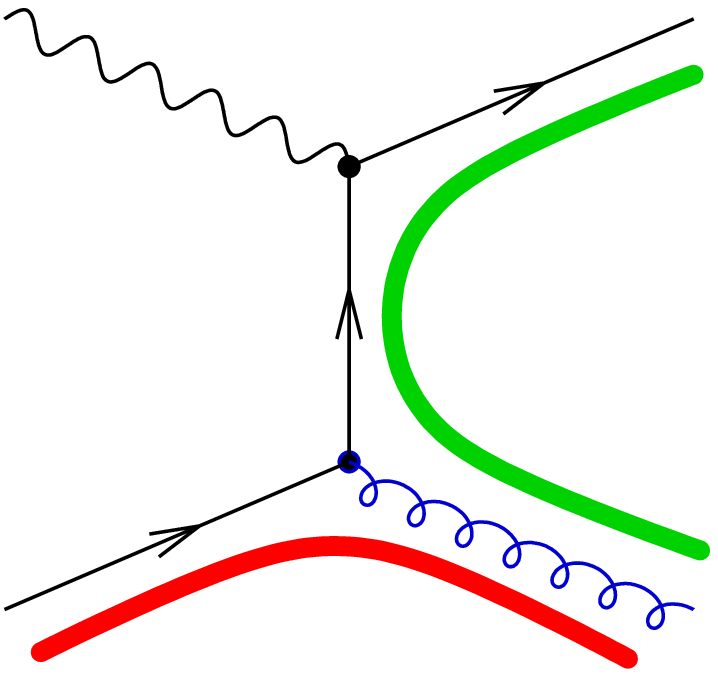}
  \ec\vspace{0mm}
  \myfigcaption{140mm}{Relevant Feynman diagrams contributing to
    vector boson production in association with a gluon (left panel,
    showing the $\hat t$\/ and $\hat u$\/ channel graphs) or with a
    quark (right panel, visualizing the $\hat t$\/ and $\hat s$\/
    channel graphs). The modified colour flows due to the emission are
    illustrated by the thick lines.}
  \label{fig:GandQemitDiags}
  \vspace{0mm}
\end{figure}
For gluon bremsstrahlung $\bar qq'\to Vg$\/ ($g_\fl$ emission) and
QCD Compton scattering $gq'\to Vq$ ($g_\il$ emission), the amplitudes
can be worked out from the Feynman diagrams depicted in
\fig{fig:GandQemitDiags}.  The partonic squared matrix elements, with
colour and spin indices averaged (summed) over initial (final) states,
can then be expressed in terms of the Born amplitude squared as
\be
  \overline{\abs{{\cal M}^{g_\fl}_{\bar qq'\to Vg}}^2}\;=\;
  \frac{N^2_\mr{C}-1}{2N_\mr{C}}\;4\pi\alpha_s\;
  \frac{\overline{\abs{{\cal M}_{\bar qq'\to V}}^2}}{M^2/2}\;
  \frac{M^4+\hat s^2-2\hat u\hat t}{\hat u\hat t}\,,
\ee
and
\be
  \overline{\abs{{\cal M}^{g_\il}_{gq'\to Vq}}^2}\;=\;
  \frac{1}{2}\;4\pi\alpha_s\;
  \frac{\overline{\abs{{\cal M}_{\bar qq'\to V}}^2}}{M^2/2}\;
  \frac{M^4+\hat u^2-2\hat s\hat t}{-\hat s\hat t}\,,
\ee
respectively, whereas the Mandelstam variables have been defined in
\eqs{eqs:stu_ii}.  Again, in this particular case the factorization is
exact and the equations above fix the dipole matrix elements 
$\hat D_{\bar q_\il q'_\il\to\bar q_\il gq'_\il(qg_\il q'_\il)}$,
which are in fact related to $D_{q\bar q'\to qg\bar q'}$ by crossing
symmetry.  The colour factor for the $g_\fl$ emission is
$C_F$. However, if the radiated gluon is assigned to the initial
state, it actually is incoming and splits into a $q\bar q$\/ pair with
one of the quarks entering the hard process.  So, the colour averaging
changes relative to the $g_\fl$ case by $N_\mr{C}/(N^2_\mr{C}-1)$;
therefore, the colour factor amounts to $T_R=\frac{1}{2}$.
\\
For all initial--initial dipoles containing gluons, the dipole matrix
elements may be obtained either directly in a similar way, or they are
inferred from their final--final counterparts of the Lund CDM using
crossing relations, $\hat D^{\mr{II}}={\rm cross}\,\hat D^{\mr{FF}}$,
cf.\ \secss{sec:Dsfuncs}{sec:FFdips}.
The recoil strategies presented in \sec{sec:IIKins} lead to trivial
rapidity Jacobians, $\frac{dy_{\mr{cm}}}{d\td y_{\mr{cm}}}=1$.  The
initial--initial dipole splitting functions of \eq{eq:SplFuncII}
are then fully specified: for gluons $g_\fl$ emitted into the final
state,
\be\begin{split}
  D_{\bar\imath'i\to\bar\imath'gi}(p_\perp,y)&\;=\;
  \frac{f_{\bar\imath'}(x_\pm,\mufp)\,f_i(x_\mp,\mufp)}
       {f_{\bar\imath'}(\td x_\pm,\muf)\,f_i(\td x_\mp,\muf)}\;
  \xi_{\left\{{F\atop A}\right\}}
  C_{\left\{{F\atop A}\right\}}\;
  \frac{x_{\bar\imath'}^{n_{\bar\imath'}}(p_\perp,y)+x_i^{n_i}(p_\perp,y)}
       {\left[x_{\bar\imath'}(p_\perp,y)+x_i(p_\perp,y)-1\right]^2}
  \\[2mm]&\;\le\;
  {\cal N}_{\mr{PDF}}\;
  \xi_{\left\{{F\atop A}\right\}}
  C_{\left\{{F\atop A}\right\}}\;
  \left\{{2\atop a+1}\right\}\;\equiv\;
  D^{\mr{approx}}_{\bar\imath'i\to\bar\imath'gi}(p_\perp,y)\,,
  \label{eq:Dfunc_iifg}
\end{split}\ee
where the energy fractions are given as
\be
  x_{\bar\imath',i}(p_\perp,y)\;=\;1+\frac{p_\perp}{M}\,e^{\pm y}\,,
  \label{eq:xdef_iifg}
\ee
and, for gluons $g_\il$ radiated into the initial state,
\be\begin{split}
  D_{\bar q_\il i\to qg_\il i}(p_\perp,y)&\;=\;
  \frac{f_g(x_\pm,\mufp)\,f_i(x_\mp,\mufp)}
       {f_{\bar q}(\td x_\pm,\muf)\,f_i(\td x_\mp,\muf)}\;
  T_R\;
  \frac{x_q^2(p_\perp,y)+x_i^{n_i}(p_\perp,y)}
       {\left[1+x_q(p_\perp,y)\right]^2}
  \\[2mm]&\;\le\;
  {\cal N}_{\mr{PDF}}\;
  T_R\;
  \left\{{2\atop a+1}\right\}\;\equiv\;
  D^{\mr{approx}}_{\bar q_\il i\to qg_\il i}(p_\perp,y)\,,
  \label{eq:Dfunc_iiig}
\end{split}\ee
where the energy fractions are characterized by
\be
  x_{q,i}(p_\perp,y)\;=\;\mp1+\frac{p_\perp}{M}\,e^{\pm y}\,.
  \label{eq:xdef_iiig}
\ee
In both cases the overestimations $D^{\mr{approx}}_{\ldots}$ finally
determine eikonal approximations to the improved splitting cross
sections.  The ${\cal N}_{\mr{PDF}}$ factors denote estimates for the
respective upper bounds of the PDF ratios. The tilde variables refer
to the before-emission state.
\\
All splitting functions discussed here are finite, \ie the (soft and
collinear) singularities of the various differential splitting cross
sections defined through \eq{eq:SplFuncDef} are entirely contained in
the $1/p^2_\perp$ term of \eq{eq:SplFuncDef}.  This nicely confirms
that each eikonal cross section encodes the full singularity structure
of the exact result.  For $g_\fl$ emissions, the invariant transverse
momentum will tend to zero in either of the collinear limits that the
gluon can have with the parent partons, \ie the $\hat t$\/ or $\hat
u$\/ variables turn independently to zero, or in the soft limit where
$x_g\to0$ and therefore $\hat t$\/ and $\hat u$\/ collectively
approach the limit at zero.  For $g_\il$ emissions, the divergence
pattern is not as rich as for $g_\fl$ emissions off II dipoles, since
$\hat s$\/ is bounded to stay well above zero owing to the mass of the
parent dipole.  So, it only is critical if the emitted quark becomes
soft or collinear with the incoming splitting gluon $g_\il$.  No other
radiating dipole contributes to this singularity, therefore
$\xi\equiv1$, consequently being omitted in the corresponding
formul\ae\ above.
\\
The colour-factor assignment is unproblematic for quark dipoles $\bar
q_\il q'_\il$ (see above) and also for gluon dipoles $g_\il g_\il$,
where $C=C_A$ and $\xi=\xi_A=0.5$.  For II dipoles with a single gluon
leg, the ambiguities beyond the large $N_\mr{C}$ limit appear in the
same way as in the FF case.  The following choices are currently made:
final-state gluons are emitted adopting the Lund CDM choice of $C=C_A$
(and $\xi=\xi_A=0.5$); for initial-state ones, $C=T_R$ (and $\xi=1$)
is selected adopting the result from the calculation for $\bar q_\il
q'_\il$ dipoles.  Since $\hat s\ge M^2>0$, the selection $C=T_R$ at
least ensures the correct behaviour in the singular limit $\hat t\to0$
of $g_\il$ emissions.
%

%% file: tex/dsfi.tex
\section{Dipoles from final--initial colour flows}
%
\label{sec:FI}
The branching of an FI dipole, $fi$, caused by a gluon may occur again
in two ways by either radiating it to the final state, or to the
initial state, releasing an antiquark instead:
\be
  f(\td k)\,i(\td\ell)\to f(k)\,g\,i(\ell)
  \qquad{\rm and}\qquad
  f(\td k)\,q_\il(\td\ell)\to f(k)\,g_\il\,\bar q(\ell)\,.
\ee
\subsection{Single-emission kinematics}
\label{sec:FIlims}
Factorization implies that in deep inelastic scattering the evolution
of the QCD particles proceeds completely independently of the
evolution of the leptonic part. Therefore, not only the squared
momentum transfer $q^2=-Q^2$ from the lepton to the parton, probed by
the scattering of the virtual photon, is a constant, but also
$q^\mu=k^\mu_e-k'^\mu_e$ remains unaltered, when emitting QCD
secondaries.  This is used as the paradigm for the construction of the
FI dipole kinematics in this model.  Hence, similar to FF and in
contrast to II dipoles, here the subsystem kinematically fully
decouples from the rest of the cascade. Thus, in the FI case the
partons directly participating in the splitting are affected only.
Therefore,
\be
  \td p_0\;\equiv\;p_0
\ee
and
\be
  \td p_0+\td p_{i/q_\il}\;=\;\td p_f\,,
  \qquad
  p_0+p_{i/g_\il}\;=\;p_f+p_{g/\bar q}\,,
  \qquad{\rm with}\qquad
  \td p_0^2\;=\;M^2\;\equiv\;-Q^2<0\,,
  \label{eq:p_fi}
\ee
such that $Q$\/ may be interpreted as the ``mass'' of the parent
dipole.  Taking \eqss{eq:GenMombf}{eq:GenMomaf}, the signature factors
are $\td\vs_f=\vs_f=\vs_{g/\bar q}=1$, all other ones equal $-1$. The
underlying $2\to2$ process implies to define the kinematic invariants
for radiating FI dipoles as
\bea
  \hat s &=& (p_0+p_{i/g_\il})^2\;=\;(p_f+p_{g/\bar q})^2
  \;=\;-Q^2(1+x_{i/g_\il})\;\ge\;0\;\equiv\;\hat s_0\,,\nn\\[1mm]
  \hat t &=& (p_0-p_f)^2\;=\;(p_{g/\bar q}-p_{i/g_\il})^2
  \;=\;-Q^2(1-x_f)\hphantom{x.}\;\le\;0\,,\nn\\[1mm]
  \hat u &=& (p_0-p_{g/\bar q})^2\;=\;(p_f-p_{i/g_\il})^2
  \;=\;-Q^2(1-x_{g/\bar q})\z\;\le\;0\,,
  \label{eqs:stu_fi}
\eea
where the identification of the energy fractions and the bounds are
again given for massless partons. The Mandelstam variables then
satisfy
\be
  \hat s+\hat t+\hat u+Q^2\;=\;0
  \qquad{\rm such\ that}\qquad
  2+x_{i/g_\il}\;=\;x_f+x_{g/\bar q}\,.
  \label{eq:stuSum_fi}
\ee
In analogy to the case of II dipoles, the maximal $\hat s$\/ is
parametrized in terms of $Q^2$ as
\be
  \hat s_{\mr{max}}\;=\;a\,Q^2
  \qquad{\rm implying\ that}\qquad
  0\;\le\;\hat s\;\le\;\hat s_{\mr{max}}\;\le\;\mathscr{S}\,.
  \label{eq:shatmax_fi}
\ee
Here, the quantity $\mathscr{S}=(p_0+P)^2$ plays the r\^ole, which the
squared collider energy $S$\/ does for II dipoles, namely representing
the maximal upper bound. The use of $p_0=\td p_0$ and the rigorous
definition of the Bj{\o}rken-$x$ variable,
\be
  x_B\;=\;\frac{Q^2}{2\,\td p_0 P}\,,
\ee
where $P$\/ labels the momentum of the incoming hadron, lead to
\be
  \mathscr{S}\;=\;Q^2\left(\frac{1}{x_B}-1\right)\,.
\ee
This signifies that the Bj{\o}rken-$x$ determines the maximal range
for the parameter $a$, namely $0\le a\le1/x_B-1$. Since parton masses
are neglected, $\hat s_0=(\td p_0+\td p_{i/q_\il})^2=0$ and the
Bj{\o}rken-$x$ is the momentum fraction $\td x$\/ of the original
incoming parton, $\td p_{i/q_\il}=x_B P$. Employing $p_{i/g_\il}=xP$,
it is found that $x_B\le x=-x_{i/g_\il}x_B\le(a+1)\,x_B\le1$ and
the limits on $x_{i/g_\il}$ are clear:
\bea
  -1-a\;\le & x_{i/g_\il} & \le\;-1\,,\nn\\[1mm]
  1+x_{i/g_\il}\;\le & x_f,\,x_{g/\bar q} & \le\;1\,,\nn\\[1mm]
  1-a\;\le & x_f+x_{g/\bar q} & \le\;1\,.
  \label{eqs:xbounds_fi}
\eea
The various other bounds then follow from
\eqss{eqs:stu_fi}{eq:stuSum_fi}.
\subsubsection{Gluon emission phase space of final--initial dipoles}
\label{sec:gf_fi}
\begin{figure}[t!]
  \vspace{0mm}\bc
  \includegraphics[width=67mm,angle=0.0]{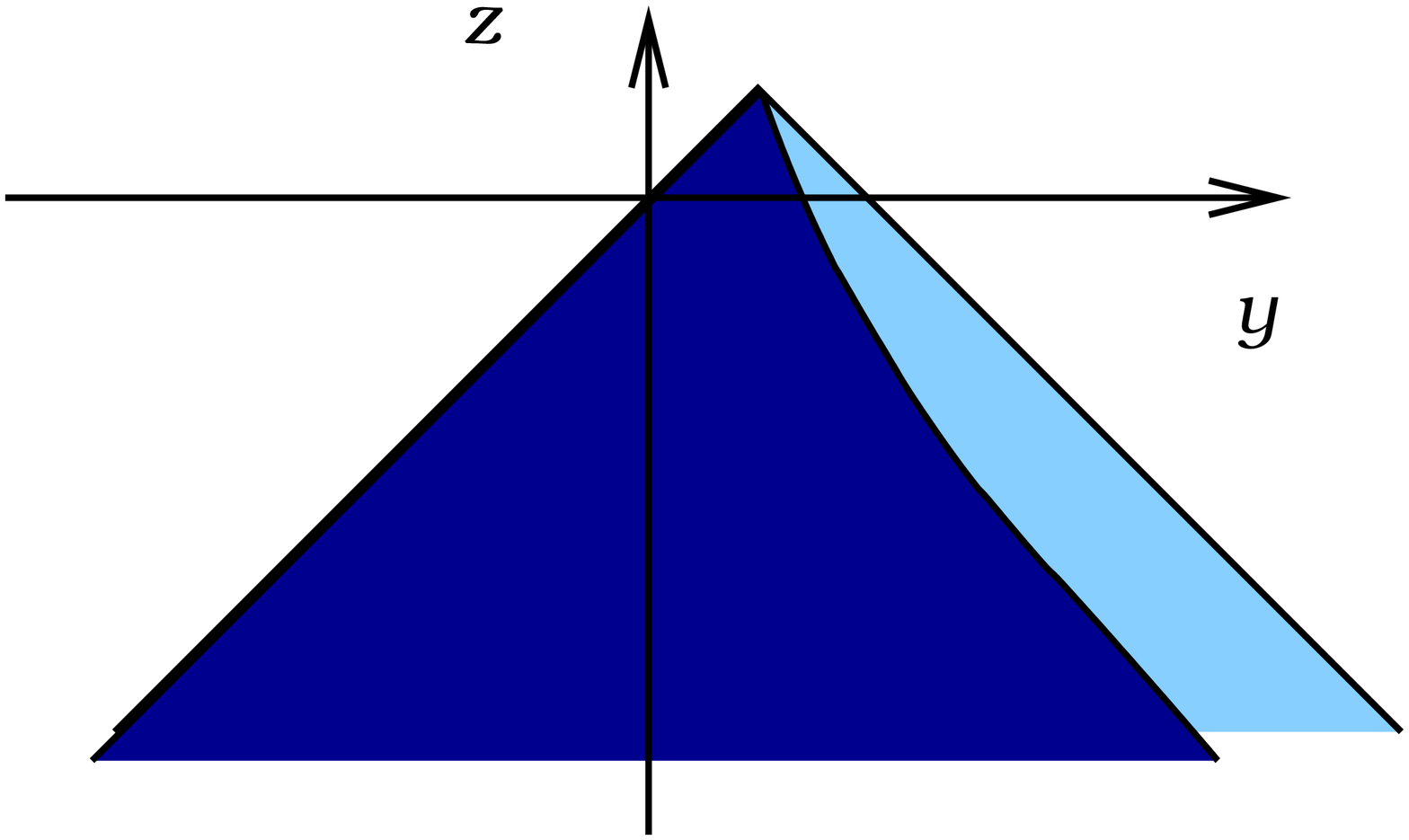}
  \hspace*{24mm}
  \includegraphics[width=61mm,angle=0.0]{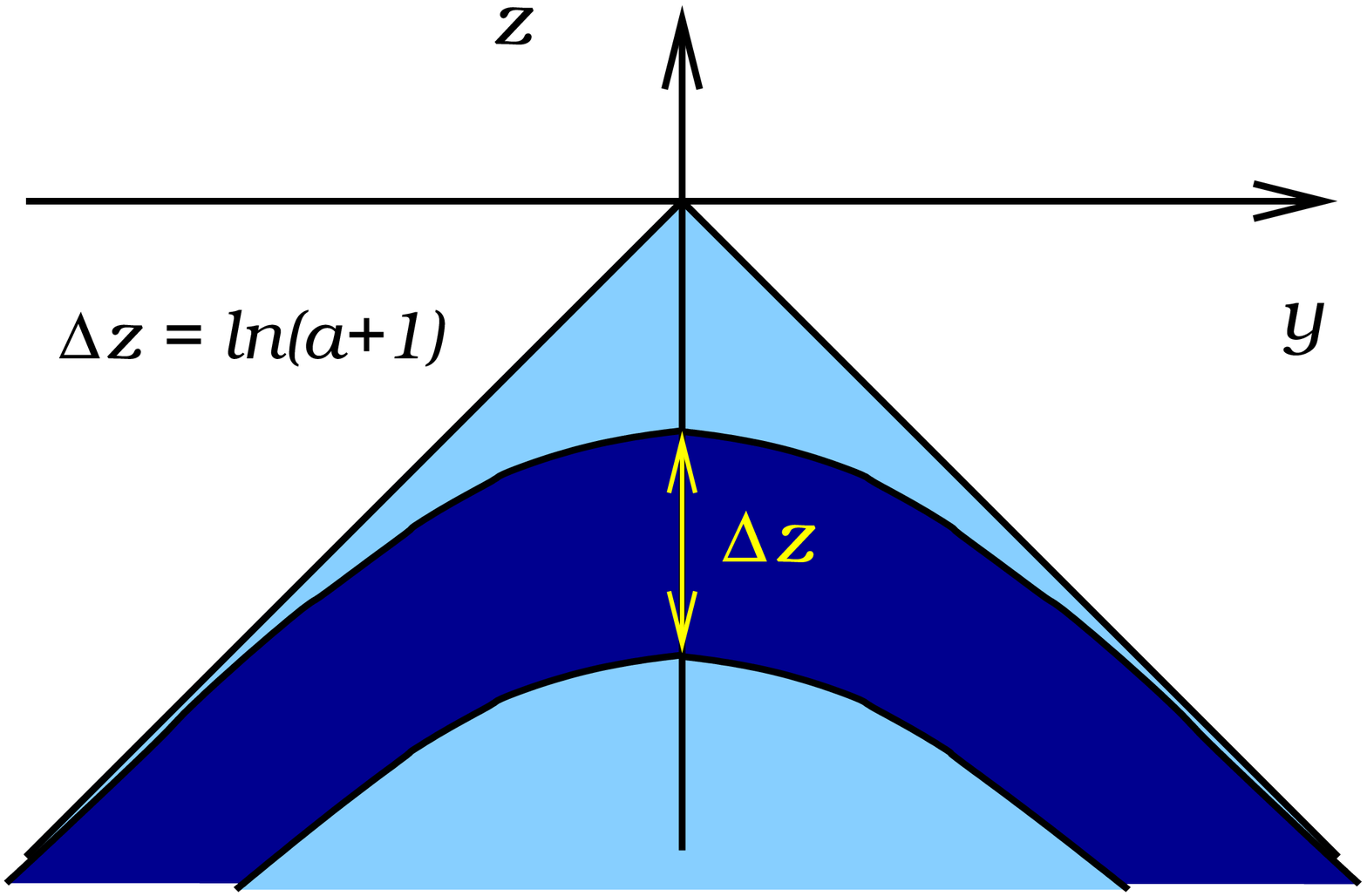}
  \ec\vspace{0mm}
  \myfigcaption{140mm}{Phase-space visualization of the final-state
    (left panel) and initial-state (right panel) gluon emissions off
    FI dipoles. Again, dark colours indicate the available phase space
    for the emissions, whereas the fractions of phase space stemming
    from the overestimations are shown in bright colours. The
    definitions are, $z=\ln\frac{p_\perp}{aQ}$ and
    $z=\ln\frac{p_\perp}{(a+1)Q}$ for $g_\fl$ and $g_\il$ emissions,
    respectively. Note that, for the visualization here, $a=1$ has
    been assumed.}
  \label{fig:fipsp}
  \vspace{0mm}
\end{figure}

First, FI dipole gluon emissions emerging into the final state, $fi\to
fgi$, are discussed according to the steps outlined in
\sec{sec:GenEvoVars}:
\bi
\item The evolution variables are identified as before by
  specifying \eqss{eq:GenP2t}{eq:GenY} for the case at hand.
  They read
  \be
    p^2_\perp\;=\;\left|\frac{s_{fg}\,s_{gi}}{s_{fgi}}\right|
             \;=\;\frac{\hat s\,\hat t}{-Q^2}
	     \;=\;Q^2(\abs{x_i}-1)(1-x_f)\,,
  \ee
  and
  \be
    y\;=\;\frac{1}{2}\,\ln\left|\frac{s_{gi}}{s_{fg}}\right|
     \;=\;\frac{1}{2}\,\ln\frac{-\hat t}{\hphantom{x}\hat s\hphantom{x}}
     \;=\;\frac{1}{2}\,\ln\frac{1-x_f}{\abs{x_i}-1}\,,
  \ee
  whereas, using \eqs{eqs:stu_fi}, the Mandelstam variables can be
  rewritten as
  \bea
    \hat s &=& s_{0i}\,=\,s_{fg}\;=\;+Q\,p_\perp e^{-y}\;\ge\;0
    \,,\nn\\[1mm]
    \hat t &=& s_{0f}\,=\,s_{gi}\;=\;-Q\,p_\perp e^{+y}\;\le\;0
    \,,\nn\\[1mm]
    \hat u &=& s_{0g}\,=\,s_{fi}\;=\;-Q^2+2Q\,p_\perp\sinh y\;\le\;0\,.
    \label{eqs:stu_fifg}
  \eea
  Obviously, the rightmost relations for $\hat s$\/ and $\hat t$\/ are
  trivially fulfilled.
\item The largest phase space available is found from $\hat s\le\hat
  s_{\mr{max}}$, cf.\ \eq{eq:shatmax_fi}, and $\hat u\le0$, hence
  \be
    -\ln\frac{a\,Q}{p_\perp}
    \;\le\;y\;\le\;
    \mr{arsinh}\,\frac{Q}{2\,p_\perp}\,.
    \label{eq:ypm_fifg}
  \ee
  In the $(y,z=\ln\frac{p_\perp}{a\,Q})$ plane, see left part of
  \fig{fig:fipsp}, these bounds manifest
  themselves in a deformed ``triangle'', whose right side is curved to
  the inside diverging for $y\to0$ while approaching $z=-y-\ln a$\/
  for $y\to\infty$. The left side of the ``triangle'' is described by
  $z\le y$\/ and  the intersection is at
  $(y=z_{\mr{max}},z_{\mr{max}}=\ln\sqrt{1+1/a\z}\z)$, suggesting that
  \be
    p^2_{\perp,\mr{max}}
    \;=\;a\,(a+1)\,Q^2
    \;=\;(\hat s_{\mr{max}}+Q^2)\;\frac{\hat s_{\mr{max}}}{Q^2}\,.
    \label{eq:s_fifg}
  \ee
  Similar to II dipole splittings, the maximum size of the emission
  phase space is dictated by the choice of $\hat s_{\mr{max}}$, see
  \eq{eq:shatmax_fi}.  This can easily be understood, since the
  emission implies a new initial state with a larger momentum fraction
  taken off the corresponding hadron.
\item The exact rapidity interval is overestimated through the
  ``triangle'' bounds, which read $z\le y\le-z+2\,z_{\mr{max}}$,
  resulting in $\Delta y=\ln(p^2_{\perp,\mr{max}}/p^2_\perp)$.
\item Again, the construction of the momenta is separately
  detailed, see \sec{sec:FIKins}.
\ei
\subsubsection{Antiquark emission phase space of final--initial dipoles}
\label{sec:gi_fi}
In this subsection, the phase-space parametrization for gluon emission
into the initial state, \ie antiquark emission into the final state,
is discussed for massless (anti)quarks, $fq_\il\to fg_\il\bar q\,$:
\bi
\item According to \eqsss{eq:GenP2t}{eq:GenY}{eqs:stu_fi} the
  evolution variables read
  \be
    p^2_\perp
    \;=\;\left|\frac{s_{fg_\il}\,s_{g_\il\bar q}}{s_{fg_\il\bar q}}\right|
    \;=\;\frac{\hat u\,\hat t}{Q^2}\;=\;Q^2(1-x_{\bar q})(1-x_f)\,,
  \ee
  and
  \be
    y\;=\;\frac{1}{2}\,\ln\left|\frac{s_{g_\il\bar q}}{s_{fg_\il}}\right|
    \;=\;\frac{1}{2}\,\ln\frac{\hat t}{\hat u}
    \;=\;\frac{1}{2}\,\ln\frac{1-x_f}{1-x_{\bar q}}\,.
  \ee
  The Mandelstam variables are then cast into the form
  \bea
    \hat s &=& s_{0g_\il}\,=\,s_{f\bar q}\;=\;2Q\,p_\perp\cosh y-Q^2\;\ge\;0
    \,,\nn\\[1mm]
    \hat t &=& s_{0f}\,=\,s_{g_\il\bar q}\;=\;-Q\,p_\perp e^{+y}\;\le\;0
    \,,\nn\\[1mm]
    \hat u &=& s_{0\bar q}\,=\,s_{fg_\il}\;=\;-Q\,p_\perp e^{-y}\;\le\;0\,,
    \label{eqs:stu_fiig}
  \eea
  where the inequalities for $\hat t$\/ and $\hat u$\/ are satisfied
  by construction.
\item The requirement $0\le\hat s\le\hat s_{\mr{max}}=a\,Q^2$ in
  conjunction with \eqs{eqs:stu_fiig} leads to
  \be
    \mr{arcosh}\,\frac{Q}{2\,p_\perp}
    \;\le\;\abs{y}\;\le\;
    \mr{arcosh}\,\frac{\hat s_{\mr{max}}+Q^2}{2\,Q\,p_\perp}\;=\;
    \mr{arcosh}\,\frac{(a+1)\,Q}{2\,p_\perp}\,,
    \label{eq:ypm_fiig}
  \ee
  where the inner and outer bounds follow from the lower and upper
  limits of the accessible $\hat s$\/ interval, respectively.
  Concerning the former the central rapidity region becomes
  unaccessible for emissions of $p_\perp<Q/2$ (cf.\ right part of
  \fig{fig:fipsp}).
  In the latter case the similarities to $g_\fl$ emissions off II
  dipoles, cf.\ \sec{sec:gf_ii}, are fairly obvious.
  If the available squared energy $\hat s_{\mr{max}}$ is fully used to
  generate the transverse momentum then $\cosh y\equiv1$ and the
  maximal $p_\perp$ is achieved, given by
  \be
    p^2_{\perp,\mr{max}}
    \;=\;\frac{(\hat s_{\mr{max}}+Q^2)^2}{4\,Q^2}
    \;=\;\frac{1}{4}(a+1)^2\,Q^2\,.
  \ee
  As in all cases involving initial-state partons, the adjustment of
  the size of the phase space is triggered by the choice of $\hat
  s_{\mr{max}}$, which will be discussed in \sec{sec:Shower}.
\item Loose constraints stem from $\hat s_{\mr{max}}+Q^2\ge -\hat
  t,-\hat u$ and yield an increased phase space w.r.t.\ the precise
  one discussed above:
  \be
    \abs{y}\;\le\;\ln\frac{\hat s_{\mr{max}}+Q^2}{Q\,p_\perp}\;=\;
                  \ln\frac{(a+1)\,Q}{p_\perp}\,.
    \label{eq:loose_fiig}
  \ee
  This invokes the usual ``triangle'' interpretation in the
  $(y,z=\ln\frac{p_\perp}{(a+1)\,Q})$ plane.
\item The splitting kinematics will be discussed in the next
  subsection.
\ei
\subsubsection{Construction of the emission momenta}
\label{sec:FIKins}
The basic construction principles mentioned throughout
\secss{sec:New}{sec:IIKins} are, of course, taken over when explicitly
establishing the FI splitting kinematics. The kinematic decoupling,
encoded as $p_0\equiv\td p_0$, alleviates the task, since Lorentz
transformations will only affect the local splitting. Thus, including
the fact that the squared dipole momentum $\td p^2_0=(\td p_f-\td
p_{i/q_\il})^2=-Q^2$, the original dipole's Breit-frame constitutes a
suitable frame to set up the three new four-momenta. Using light-cone
notation again, in this frame the momenta read
\bea
  \td p_0\;=\;\left(-Q,\,Q,\,\vec0\right)
  &\;\,\to\;\,&
  p_{0\hphantom{/\bar q}}\;=\;\left(-Q,\,Q,\,\vec0\right)\,,
  \nn\\[1mm]
  \td p_{i/q_\il}\,\;=\;\left(Q,\,0,\,\vec0\right)
  &\;\,\to\;\,&
  p_{i/g_\il}\!\;=\;\left(-x_{i/g_\il}\,Q,\,0,\,\vec0\right)\,,
  \nn\\[1mm]
  \td p_{f\hphantom{/g_\il}}\;=\;\left(0,\,Q,\,\vec0\right)
  &\;\,\to\;\,&
  p_{f\hphantom{/g_\il}}\!\!\;=\;\left(
    \frac{(1-x_f)(-x_{i/g_\il}-1)}{\abs{x_{i/g_\il}}}\,Q,\,
    \frac{x_f-x_{i/g_\il}-1}{\abs{x_{i/g_\il}}}\,Q,\,
    \vec b_\perp\right)\,,
  \nn\\[1mm]
  &&
  p_{g/\bar q}\x\;=\;p_0+p_{i/g_\il}-p_f\,.
  \label{eqs:breitkin_fi}
\eea
The Breit-frame transverse momentum is given through $\vec
b_\perp=(b_\perp\cos\varphi,b_\perp\sin\varphi)$, where
\be
  b_\perp\;=\;\frac{\sqrt{\hat s\,\hat t\,\hat u\,}}{\hat s+Q^2}
         \;=\;\sqrt{(1-x_f)(-x_{i/g_\il}-1)(x_f-x_{i/g_\il}-1)\,}\;\,
	      \frac{Q}{\hphantom{\,}\abs{x_{i/g_\il}}\hphantom{\,}}\,.
\ee
Note that, for $g_\fl$ emissions, it becomes zero for $\hat u\to0$ (in
this limit the rapidity value associated to this emission coincides
with the $y_+$ bound, cf.\ \sec{sec:gf_fi}).
This just happens independently of the actual value for the
evolution variable $p_\perp$, therefore, ordering the emissions in
$p_\perp$ does not impose any ordering in $b_\perp$. Finally, the new
Breit-frame momenta are transformed into the lab-frame.\spc
\footnote{This is done by inverting the transformations that (1) align
  the lab-frame momenta $\td p_{i/q_\il}$ and $\td p_f$ and (2) rotate
  them afterwards onto the $\hat{\mr{z}}$ axis.}
\\
Using \eqs{eqs:breitkin_fi} the recoil strategy can directly be read
off: before and after the splitting the initial-state parton is fixed
to the $+$ direction of the light-cone decomposition, therefore to the
beam axis\spc\footnote{Note that the choice $p_{i/g_\il}=-x_{{i/g_\il}}\,\td
  p_{i/q_\il}=\con{(-x_{{i/g_\il}}x_B\sqrt{S},0,\vec0)}{lab\mbox{-}f.}$
  is Lorentz invariant.},
leaving the recoil to be completely compensated for by the final-state
particle. Of course, more sophisticated recoil strategies following
the ones of the Lund model and/or the Kleiss idea are possible, but
not yet implemented. Especially the prescription for quark scattering
processes given in \cite{Seymour:1994we} seems very attractive, since
it includes the correlations between leptons and partons associated to
the lowest-order DIS process and the first emission.
\subsection{Final--initial dipole splitting functions}
\label{sec:FIdips}
For $qq'_\il$ dipoles emitting gluons, the two respective
matrix-element factorizations of \eq{eq:FIdipoleME} can directly be
specified utilizing the results\spc\footnote{Colour and spin
  final-state summed plus initial-state averaged squared matrix
  elements are \eg given in \cite{Field:1989uq}.} for the two typical
real-correction processes to leading order DIS, namely the QCD Compton
and the boson--gluon fusion processes. These are compared to the sole
scattering of a quark caused by a space-like vector boson; hence, the
dipole's gluon emission will again be treated coherently and,
moreover, exact factorization is achieved as in all other quark--quark
dipole cases. This yields the corresponding dipole matrix elements,
both of which in fact reflecting the crossing symmetry of the $\hat
D_{q\bar q'\to qg\bar q'}$ term:
\be
  \hat D_{qq'_\il\to qgq'_\il}
  \;=\;
  \frac{1}{Q^2}\,\frac{\hat s^2+\hat t^2-2\hat uQ^2}{-\hat s\hat t}
  \;=\;
  -\,\frac{x^2_{q'_\il}+x^2_q}{Q^2(1+x_{q'_\il})(1-x_q)}
  \;=\;
  \frac{x^2_{q'_\il}+x^2_q}{p^2_\perp}\,,
\ee
for $g_\fl$ radiation off the $qq'_\il$ dipole where $C=C_F$, and
\be
  \hat D_{qq'_\il\to qg_\il\bar q'}
  \;=\;
  \frac{1}{Q^2}\,\frac{\hat u^2+\hat t^2-2\hat sQ^2}{\hat u\hat t}
  \;=\;
  \frac{x^2_{\bar q'}+x^2_q}{Q^2(1-x_{\bar q'})(1-x_q)}
  \;=\;
  \frac{x^2_{\bar q'}+x^2_q}{p^2_\perp}\,,
\ee
for $g_\il$ radiation off the $qq'_\il$ dipole with $C=T_R$. For the
definitions of the Mandelstam variables etc., see \eqs{eqs:stu_fi} and
the previous section.
\\
The dipole matrix elements of the FI dipoles containing gluon(s) are
calculated either following the above procedure, or, alternatively,
exploiting the crossing symmetry of the respective FF dipole matrix
element taken from the Lund CDM. This completely determines the
final--initial dipole splitting functions as introduced in
\eq{eq:SplFuncFI}: for $g_\fl$ emissions,
\be\begin{split}
  D_{fi\to fgi}(p_\perp,y)&\;=\;
  \frac{f_i(-x_ix_B,\mufp)}{f_i(x_B,\muf)}\;
  \xi_{\left\{{F\atop A}\right\}}
  C_{\left\{{F\atop A}\right\}}\;
  \frac{\abs{x_f(p_\perp,y)}^{n_f}+\abs{x_i(p_\perp,y)}^{n_i}}
       {x_i^2(p_\perp,y)}
  \\[2mm]&\;\le\;
  {\cal N}_{\mr{PDF}}\;
  \xi_{\left\{{F\atop A}\right\}}
  C_{\left\{{F\atop A}\right\}}\;
  \left\{{2\atop 2(a+1)}\right\}\;\equiv\;
  D^{\mr{approx}}_{fi\to fgi}(p_\perp,y)\,,
  \label{eq:Dfunc_fifg}
\end{split}\ee
with the energy fractions as functions of $p_\perp$ and $y$\/ reading
\be
  x_{f,i}(p_\perp,y)\;=\;\pm1-\frac{p_\perp}{Q}\,e^{\pm y}\,,
  \label{eq:xdef_fifg}
\ee
and, for $g_\il$ emissions (for antiquarks emitted into the final state),
\be\begin{split}
  D_{fq_\il\to fg_\il\bar q}(p_\perp,y)&\;=\;
  \frac{f_{g_\il}(-x_{g_\il}x_B,\mufp)}{f_{q_\il}(x_B,\muf)}\;
  T_R\;
  \frac{\abs{x_f(p_\perp,y)}^{n_f}+x_{\bar q}^2(p_\perp,y)}
       {x_{g_\il}^2(p_\perp,y)}
  \\[2mm]&\;\le\;
  {\cal N}_{\mr{PDF}}\;
  T_R\;
  \left\{{1\atop\max\{2,\,a+1\}}\right\}\;\equiv\;
  D^{\mr{approx}}_{fq_\il\to fg_\il\bar q}(p_\perp,y)\,,
  \label{eq:Dfunc_fiig}
\end{split}\ee
where the energy fractions are then given in terms of the evolution
variables by
\be
  x_{f,\bar q}(p_\perp,y)\;=\;1-\frac{p_\perp}{Q}\,e^{\pm y}\,,
  \qquad{\rm such\ that}\qquad
  x_{g_\il}(p_\perp,y)\;=\;-\frac{2\,p_\perp}{Q}\,\cosh y\,.
  \label{eqs:xdef_fiig}
\ee
The modulus ensures that the terms in the rightmost numerator of the
exact splittings are positive definite. Additionally, the eikonal
approximations are displayed, which again overestimate the true form
of the splitting functions.
\\
As in the previous cases, the dipole splitting functions are finite,
such that the divergences are fully encapsulated in the $1/p^2_\perp$
term. For gluons emitted into the final state, collinear/soft limits
($\hat t\to0$ or/and $\hat s\to0$) appear as before, where the
(collinear) singularities for gluons are again only fully accounted
through the inclusion of the contributions of the neighbouring
dipoles (cf.\ the choice of $\xi$, $\xi=\xi_A=0.5$).
\\
If the gluon is radiated into the initial state, the incoming gluon
may split collinearly and, therefore, in singular domains w.r.t.\ both
the emitted antiquark ($\hat t\to0$) and the ``other'' final-state
parton associated to the emission ($\hat u\to0$), cf.\
\eqs{eqs:stu_fi}. This is in contrast to the situation of II dipoles
where a collinear divergence cannot emerge between the incoming gluon
and the ``other'' parton, since in this case it belongs to the initial
state. Turning to the discussion of the soft infrared limit, the gluon
$g_\il$ itself cannot become soft, since it is coupled to the initial
state. Therefore, $\hat t$\/ and $\hat u$\/ cannot vanish at the same
time, \ie the soft limit is kinematically shielded, which is also
clear from \eqs{eqs:xbounds_fi}. In case a soft antiquark is being
emitted, a singular effect only occurs once it is also collinear
with the splitting gluon such that $\hat t\to0$ (the associated
disappearance of $\hat s$\/ is non-singular).
\\
The colour factors are chosen similarly to the previous cases, with
the same reservations concerning the collinear limits. For FI dipoles,
not only a final-state gluon emerging from a quark--gluon dipole gives
rise to the ambiguities, in this case, also the antiquark emission
into the final state induces them on the same level. This is related
to the fact that this splitting, as already mentioned, is singular
when either the emitted antiquark ($\hat t\to0$) or the ``other''
parton ($\hat u\to0$) in the final state become collinear with the
initial-state gluon, cf.\ \eqs{eqs:stu_fi}. The ambiguity here occurs
when this ``other'' final-state parton is a gluon, apparently
resulting in a collinear splitting governed by $C_A$ rather than
$T_R$.
%

%% file: tex/dsmod.tex
\section{The complete shower algorithm}
%
\label{sec:Shower}
In this section the dipole-shower algorithm is presented, which
models the full QCD radiation picture in terms of initial-state,
final--initial and final-state colour-dipoles on purely perturbative
grounds.  This formulation of the shower aims at resumming effects at
leading logarithmic accuracy while producing exclusive final states of
partons.  These are generated in a Markovian process, iterating
individual emissions.  In analogy to conventional parton showers, a
Sudakov form factor constitutes the central probabilistic quantity
that determines the full development of the cascade.  This will be
discussed first, before the procedure of evaluating the evolution
variables that characterize a single emission briefly will be
explained. Finally, the showering algorithm will be fixed by
specifying its relevant parameters and scale choices.
\subsection{The Sudakov form factor}
\label{sec:SudFactor}
The evolution variables are given by the invariant transverse momentum
$p_\perp$ and the invariant rapidity $y$, defined in
\eqss{eq:GenP2t}{eq:GenY}, respectively. Since the dipole splitting
functions $D(p_\perp,y)$, cf.\ \eq{eq:SplFuncDef}, are finite
throughout, the entire singular structure of each emission cross
section in each case is incorporated as a term $1/p^2_\perp$. The
concept of ``time'' therefore is realized through $p_\perp$, which
thus operates as the (leading) variable ordering the emissions within
the cascade. Consequently, $y$\/ is considered as the associated
variable. Within the generic framework, cf.\ \sec{sec:New}, all
emissions are treated on equal footing, resulting in a competition
between different available channels at each evolution step. The
Sudakov form factor is obtained from integrating the corresponding
differential single-emission cross sections
$\frac{d\cal{P}}{dp^2_\perp dy}$ (which are positive definite) in
suitable boundaries of $p_\perp$ and $y$.  Summing over all allowed
splitting channels $\{\td k\td\ell\to kg\ell\}$ and exponentiating the
negative result finally yields the Sudakov form factor:
\be
  \Delta(p^2_{\perp,\mr{stt}},p^2_\perp)\;=\;
  \exp\left\{-\int\limits^{p^2_{\perp,\mr{stt}}}_{p^2_\perp}
  \frac{d\tilde p^2_\perp}{\tilde p^2_\perp}\;
  \cal I(\tilde p^2_\perp)\right\}\,,
\ee
where
\be
  \cal I(p^2_\perp)\;=\;
  \frac{\alpha_s[\mur(p_\perp)]}{2\pi}\;\sum_{\{\td k\td\ell\to kg\ell\}}\;
  \int\limits^{y_+(p_\perp,\,a)}_{y_-(p_\perp,\,a)}
  dy\;D_{\td k\td\ell\to kg\ell}(p_\perp,y)\,.
\ee
In this form the Sudakov form factor resums the leading logarithms
as encoded in the dipole splittings to all orders, and, hence, can be
interpreted as a no-branching probability. Accordingly, the two
infrared divergent contributions of virtual and unresolvable real
emission cancel each other below the infrared cut-off leaving an
overall finite result.
Thus, $\Delta(p^2_{\perp,\mr{stt}},p^2_\perp)$ quantifies how likely a
state consisting of a number of dipoles will not emit any further
resolvable parton between the start scale $p^2_{\perp,\mr{stt}}$ and a
lower (cut-off) scale $p^2_\perp$. The quantity \mur\ denotes the
renormalization scale in energy units for the evaluation of the
(running) strong coupling. Typically \mur\ is given as a simple
function of the evolution variables to include some higher-order
virtual contributions beyond the leading-logarithmic approximation
\cite{Amati:1980ch,Curci:1980uw}. If reduced to the case of FF dipole
evolution only, the expression for the Sudakov form factor of
course becomes equivalent to \eq{eq:GenSud} of the Lund CDM. Note that
in the more general case, the rapidity limits $y_\pm$ also depend on
the scaling quantity $a$, cf.\ \eqss{eq:shatmax_ii}{eq:shatmax_fi},
\ie on the choice of the maximal available phase space. Additionally,
their actual functional form depends on the particular emission channel.
For notational brevity, this has been omitted, but is clear in view of
\eq{eq:ypm_ff}, \eqsspss{eq:ypm_iifg}{eq:ypm_iiig}{eq:ypm_fifg}{eq:ypm_fiig}.
The presence of the PDF ratios in the I/FI splitting kernels
naturally yields a Sudakov form factor including these ratios. This
resembles the typical backward evolution treatment, where the ratio of 
parton densities ensures that the parton composition of the hadron is 
properly reflected in each evolution step \cite{Sjostrand:1985xi}.
\\
Finally, the actual differential probability (the probability density)
for some branching to occur at $p^2_\perp$ then reads
\be
  \frac{dP}{dp^2_\perp}\;=\;
  \frac{d\Delta(p^2_{\perp,\mr{stt}},p^2_\perp)}{dp^2_\perp}\;=\;
  \frac{\cal I(p^2_\perp)}{p^2_\perp}\,
  \Delta(p^2_{\perp,\mr{stt}},p^2_\perp)\,.
  \label{eq:summedProb}
\ee
Subsequent emissions are ordered in $p_\perp$, \ie their start scale
$p^2_{\perp,\mr{stt}}$ is identical to the $p^2_\perp$ of the last
parton radiation. This generates the Markov chain. Note that this
still leaves the initial starting scale -- dubbed initializing scale
-- for the very first emission, $p^2_{\perp,\mr{ini}}$, to be
selected. The choices made here are detailed in \sec{sec:Scales}.
\subsection{Generation of the emission's Sudakov variables}
\label{sec:SudAlgo}
In the model a valid pair of evolution variables is generated by
exploiting the strict $p_\perp$ ordering, which enables to treat any
dipole and each of its emission channels separately. Therefore, for
each single channel, a trial $(p^2_\perp,y)$ pair is generated
according to its probability density
\be\begin{split}
  \frac{dP_{\td k\td\ell\to kg\ell}}{dp^2_\perp}\;=\;&
  \frac{\alpha_s[\mur(p_\perp)]}{2\pi\,p^2_\perp}\,
  \int\limits^{y_+(p_\perp,\,a)}_{y_-(p_\perp,\,a)}
  dy\,D_{\td k\td\ell\to kg\ell}(p_\perp,y)\\[2mm]&\qquad\;\times\;
  \exp\left\{-\int\limits^{p^2_{\perp,\mr{stt}}}_{p^2_\perp}
  \frac{d\tilde p^2_\perp}{\tilde p^2_\perp}\,
  \frac{\alpha_s[\mur(\td p_\perp)]}{2\pi}
  \int\limits^{y_+(\tilde p_\perp,\,a)}_{y_-(\tilde p_\perp,\,a)}
  dy\,D_{\td k\td\ell\to kg\ell}(\tilde p_\perp,y)\right\}\,.
  \label{eq:singleProb}
\end{split}\ee
A valid $(p^2_\perp,y)$ pair generated according to the distribution
\eq{eq:summedProb} is finally obtained by iterating over all channels
picking the one of largest $p^2_\perp$ from the ensemble of all trial
$p^2_\perp$'s.
\\
The procedure of selecting such a trial $(p^2_\perp,y)$ pair for a
single dipole emission channel follows the standard Monte Carlo
technique (hit-or-miss method) of the veto algorithm
\cite{Sjostrand:2001yu} exploiting that, for any given pair, the
eikonal approximations gathered throughout
\secsss{sec:FFdips}{sec:IIdips}{sec:FIdips} overshoot the respective
dipole splitting cross sections.\spc
\footnote{Both forms are positive definite and describe differential
  cross sections. Therefore, employing them as kernels in Sudakov
  exponentials will always yield Sudakov form factors smaller than
  one, such that these form factors can be interpreted as all-orders
  expressions in leading logarithmic accuracy for emitting no parton
  between two evolution scales.}
Each of which yields a fully integrable and invertible probability
density, for which the $(p^2_\perp,y)$ selection can be solved
analytically using two random numbers.  The respective simpler density
can be easily read off \eq{eq:singleProb} when replacing
$\alpha_s[\mur]$, $y_\pm$ and $D_{\td k\td\ell\to kg\ell}$ by a
sufficiently larger $\alpha^{\mr{max}}_s$, loose rapidity bounds
$Y_\pm$ (overestimating the actual rapidity interval) and approximate
splitting functions $D^{\mr{approx}}_{\td k\td\ell\to kg\ell}$,
respectively.  The correction to the true form of the single-channel
density is then achieved by accepting the trial pair with a
probability equal to the ratio (correction weight) reading
\be
  \cal W\;=\;\frac{\alpha_s(\mur)}{\alpha^{\mr{max}}_s}\;
	     \frac{D_{\td k\td\ell\to kg\ell}}
		  {D^{\mr{approx}}_{\td k\td\ell\to kg\ell}}\;
	     \frac{\Delta y(p_\perp,a)}{\Delta Y(p_\perp,a)}\,.
\ee
However, there are additional kinematical constraints, such as the
demand for valid momentum fractions $x_\pm\le1$ in II dipole
evolution, see \sec{sec:IIKins}; once violated, they translate into
rejection of the trial emission, implying the generation of a new
trial emission for the considered channel, starting over from the
rejected $p_\perp$ value.
\\
Note that the PDF-ratio overestimations, $\cal N_{\mr{PDF}}$, present
in the approximate I/FI splitting functions are taken from a
dynamically self-adapting table in order to improve the generation
efficiency. The last term of above equation exhibits the correction
for the exact rapidity interval, where $\Delta y=y_+-y_-$ and $\Delta
Y=Y_+-Y_-$.
\subsection{Scale choices, starting conditions and iteration principles}
\label{sec:Scales}
Finally, the remaining free scale choices are fixed. This completely
defines the (default) cascade-generating algorithm of this shower
model.
\paragraph{Renormalization scales:}
\be
  \smur\left|{\hphantom{A}\atop\sd{\rm{FF}}}\right.\;=\;\frac{p^2_\perp}{2}
  \qquad{\rm and}\qquad
  \smur\left|{\hphantom{A}\atop\sd{\rm{I/FI}}}\right.\;=\;2\,z\,(1-z)\,p^2_\perp
      \;=\;\frac{p^2_\perp}{1+\cosh(2y)}\;\le\;\frac{p^2_\perp}{2}
\ee
are used for the argument of the running strong coupling, where $z$\/
is defined as a fraction of squared two-particle masses of the
partons partaking in the emission:
\be
  z\;=\;\frac{\abs{s_{kg}}}{\abs{s_{kg}}+\abs{s_{g\ell}}}
   \;=\;\frac{1}{1+e^{2y}}\,.
\ee
An offset of $\cal{O}(1\ \rm{GeV})$ ensures the evolution to proceed
well above the Landau pole $\Lambda_{\mr{QCD}}$.
\paragraph{Factorization scales:}
for a new (trial) emission \mufp\ is calculated according to
\be
  \smufp\;=\;(4\,k^2_\perp)^{d/2}\,\muf^{2-d}\,,
\ee
where $d=1$ and $d=2$ are employed for II and FI dipoles,
respectively. The modified transverse momentum squared, $k^2_\perp$,
is computed from the emission's $p^2_\perp$ and $y$\/ and the mass of
the parent dipole. Based on \eq{eq:labkperp_ii} $k^2_\perp$ is
frame-independently defined as
\be
  k^2_\perp\;=\;\frac{\abs{s_{kg}\,s_{g\ell}}}
                     {\abs{s_{kg\ell}}+\abs{s_{kg}}+\abs{s_{g\ell}}}
           \;=\;\frac{\abs{M}\,p^2_\perp}{\abs{M}+2\,p_\perp\cosh y}\,,
  \label{eq:k2perp}
\ee
and intended to function as a more natural scale for the argument of
the parton densities, since it better compares to the lab-frame squared
transverse momentum. The respective old factorization scales
associated with the state before the emission are encoded in the \muf\
values, whereas, for the very initial case, \muf$_{,\mr{ini}}$ is
adopted from the hard process.
\paragraph{Initializing scales:}
the subsequent cascading off the core process starts at the hardest
scale, $p^2_\perp=p^2_{\perp,\mr{ini}}$, which can not be set
independently of the underlying process. Generally it should
guarantee that the shower strictly evolves in the soft and collinear
phase-space regions only. Here, the following choices are made for
three different scenarios of hard $2\to2$ processes:
\bi
\item Showering off a single $q\bar q$\/ dipole as in $e^+e^-\to q\bar
  q$\/ processes: the start scale is set by the squared mass of the
  parent dipole, $p^2_{\perp,\mr{ini}}\left|{
    \hphantom{A}\atop\sd{\mbox{$q\bar q$-}\rm{prod.}}}\right.=\hat s_0=M^2$.
\item Showering off a single $\bar q_\il q'_\il$ dipole as in
  Drell--Yan processes: here, a $p_{\perp,\mr{ini}}$ estimate is
  gained by inverting \eq{eq:k2perp} for $y=0$, which yields
  \be
    p_{\perp,\mr{ini}}\left|{\hphantom{A}\atop\sd{\rm{DY}}}\right.\;=\;
    k_{\perp,\mr{max}}\left(
    \frac{k_{\perp,\mr{max}}}{M}+\sqrt{\frac{k^2_{\perp,\mr{max}}}{M^2}+1\z}
    \z\right)\,,
    \label{eq:inidy}
  \ee
  and gives $p_{\perp,\mr{ini}}\left|{\hphantom{A}\atop\sd{\rm{DY}}}\right.
  =(1+\sqrt{2})M$, provided that $k^2_{\perp,\mr{max}}=\hat s_0=M^2$.
  For the same reason as in the previous item, this estimate is
  found by restricting the true transverse momentum of final-state
  gluons radiated off $\bar q_\il q'_\il$ dipoles, cf.\
  \eq{eq:labkperp_ii}. For vector boson production, in this model, the
  first emission is matrix-element corrected per construction.
  Therefore, the restricted scale may be discarded, the shower instead
  evolve freely with the initializing scale set as largely as
  kinematically allowed.
\item Showering off a multi-dipole state as in pure QCD jet
  production: recalling the definition
  $p^2_\perp=\abs{s_{kg}\,s_{g\ell}}/\abs{s_{kg\ell}}$, all possible
  combinations for this fraction can be calculated using the strong
  particles provided by the hard process. The combination yielding the
  lowest $p^2_\perp$ should represent a sufficient estimate for the
  initializing scale. Applied to QCD jet production, the minimal
  numerator is given by $\min\{\hat u\hat t,\hat s\hat t,\hat s\hat u\}$ 
  employing the Mandelstam variables of the $2\to2$ QCD core process. 
  For the denominators, a mean squared mass
  determined by $\frac{1}{9}(\hat s+\abs{\hat t}+\abs{\hat u})$ is
  used replacing the vanishing $\abs{s_{kg\ell}}$. Taken together,
  \be
    p^2_{\perp,\mr{ini}}\left|{\hphantom{A}\atop\sd{\rm{Jets}}}\right.\;=\;
    \frac{9\,\min\{\hat u\,\hat t,\,\hat s\,\hat t,\,\hat s\,\hat u\}}
         {\hat s+\abs{\hat t}+\abs{\hat u}}\,.
    \label{eq:jinivfr}
  \ee
\ei
The assignment of large $N_\mr{C}$ colour flows is straightforward and
unique for the first two examples. In the latter case, the most likely
flow among the possible ones for a given set of partons will be
determined by the actual kinematical configuration of the QCD $2\to2$
core process.
\paragraph{Maximal phase space:}
for II and FI dipole regimes, the limits on the evolution variables
vary with the choices of the $\hat s_{\mr{max}}$ parameters
restricting the phase space of a single emission for decreasing $\hat
s_{\mr{max}}$ values, cf.\ \eqss{eq:shatmax_ii}{eq:shatmax_fi}. The
default settings impose no extra restrictions, they hence are
$\hat s_{\mr{max}}\left|{\hphantom{A}\atop\sd{\rm{II}}}\right.=S$
(see \sec{sec:IIlims}) and $\hat
s_{\mr{max}}\left|{\hphantom{A}\atop\sd{\rm{FI}}}\right.=\mathscr S$
(see \sec{sec:FIlims}) and allow full access to the centre-of-mass
energy as given by the collider.
\paragraph{Cascading:}
each chain (colour-singlet), once appeared, is independently evolved,
with the only potential exception of recoil transfer from an II
splitting. This does not spoil the further evolution of the
corrected chain owing to the Lorentz invariance of the shower
formulation. Starting off $p^2_{\perp,\mr{ini}}$ consecutive emissions
are decreasingly ordered in $p^2_\perp$ within a chain. The dipole
splitting that generated the largest $p^2_\perp$ in a certain
evolution step is finalized in its kinematics, its $p^2_\perp$ is used
as the new start scale for the trial emissions of the next round. The
procedure continues until the infrared cut-off has been reached.
\paragraph{Cut-off and hadronization aspects:}
the cut-off is always taken on $p^2_\perp$, hence denoted by
$p^2_{\perp,\mr{cut}}$. With a prescription avoiding the Landau pole
in $\alpha_s$, it can, in principle, be chosen arbitrarily small,
since then the setting of the renormalization scale is safe and the
Sudakov suppression quenches the appearing soft and collinear
divergences.
\\
After the cascading is finished, the interface to the hadronization --
currently described through phenomenological models only -- does not
require any special treatment inside the shower. The conversion of the
shower partons into primary hadrons proceeds similarly to the case of
conventional parton cascades.
%

%% file: tex/dsres.tex
\section{First results}
%
\label{sec:Res}
In this section, the newly developed dipole shower is validated by
comparing its predictions of QCD dynamics to data and other Monte
Carlo calculations. To this end, the following physics processes 
are studied:
\bi
\item the production of vector bosons and their subsequent hadronic
 decays in $e^+e^-$ collisions at LEP1 energies,
\item the inclusive production of Drell--Yan $e^+e^-$ pairs at
  Tevatron and future LHC energies, and,
\item the inclusive QCD production of jets at Tevatron energies.
\ei
The shower model presented here has been implemented into the event
generator \sherpa, and supplemented by an interface to the Lund string
fragmentation routines of \pythia 6.2 \cite{Sjostrand:2001yu}, which
are provided by the \sherpa framework. Cascading starts off the
corresponding hard $2\to2$ processes, which are generated inside
\sherpa utilizing its facilities of evaluating matrix elements. Only
light-quark flavours, \ie massless quarks, are considered. For the
simulation of hadronic collisions, all predictions have been obtained
from the CTEQ6L set of PDFs \cite{Pumplin:2002vw}. In accordance
with the choice in the PDF, the strong coupling constant has been
fixed by $\alpha_s(M_Z)=0.118$ and its running is taken at the
two-loop level. The dipole-shower cut-offs related to final--initial
and initial--initial dipole evolution are both set to $1$ GeV, \ie
$\con{p^2_{\perp,\mr{cut}}}{FI}=
\con{p^2_{\perp,\mr{cut}}}{II}=1\mbox{\ GeV}^2$.
In contrast, $\con{p^2_{\perp,\mr{cut}}}{FF}$ is tuned by hand
together with the Lund string model parameters.
\\
The lower panel in each of the plots presented below visualizes the
(MC$-$reference)/reference ratio, where the ``reference'' (ref) is
given by the data as long as they are available. The bright band
always illustrates the uncertainty of the respective measurement.
\subsection{Hadron production in electron--positron collisions}
\label{sec:Lep}
\begin{figure}[t!]
  \vspace{0mm}
  \hspace*{2mm}
  \includegraphics[width=78mm,angle=0.0]{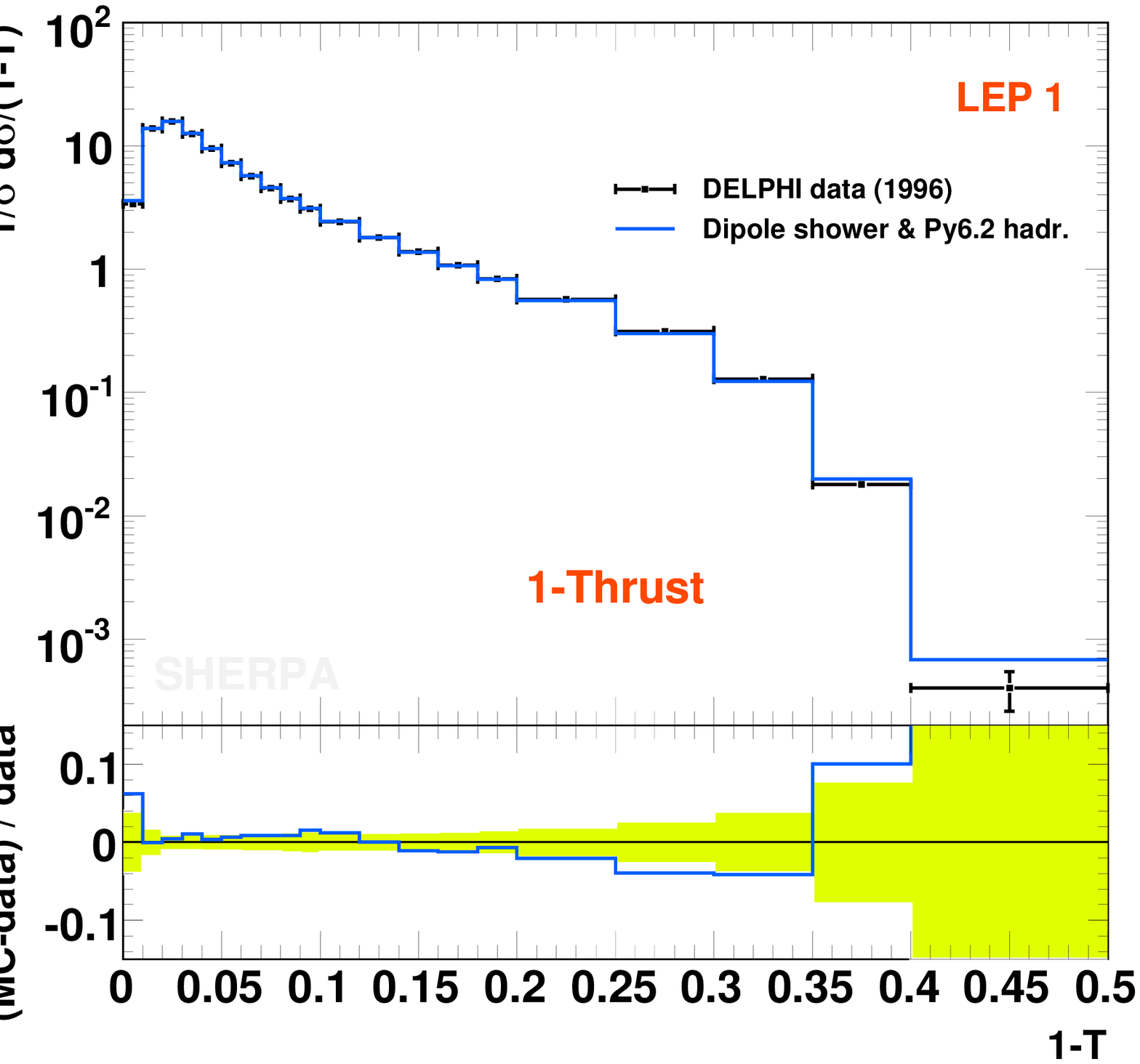}
  \hspace*{4mm}
  \includegraphics[width=78mm,angle=0.0]{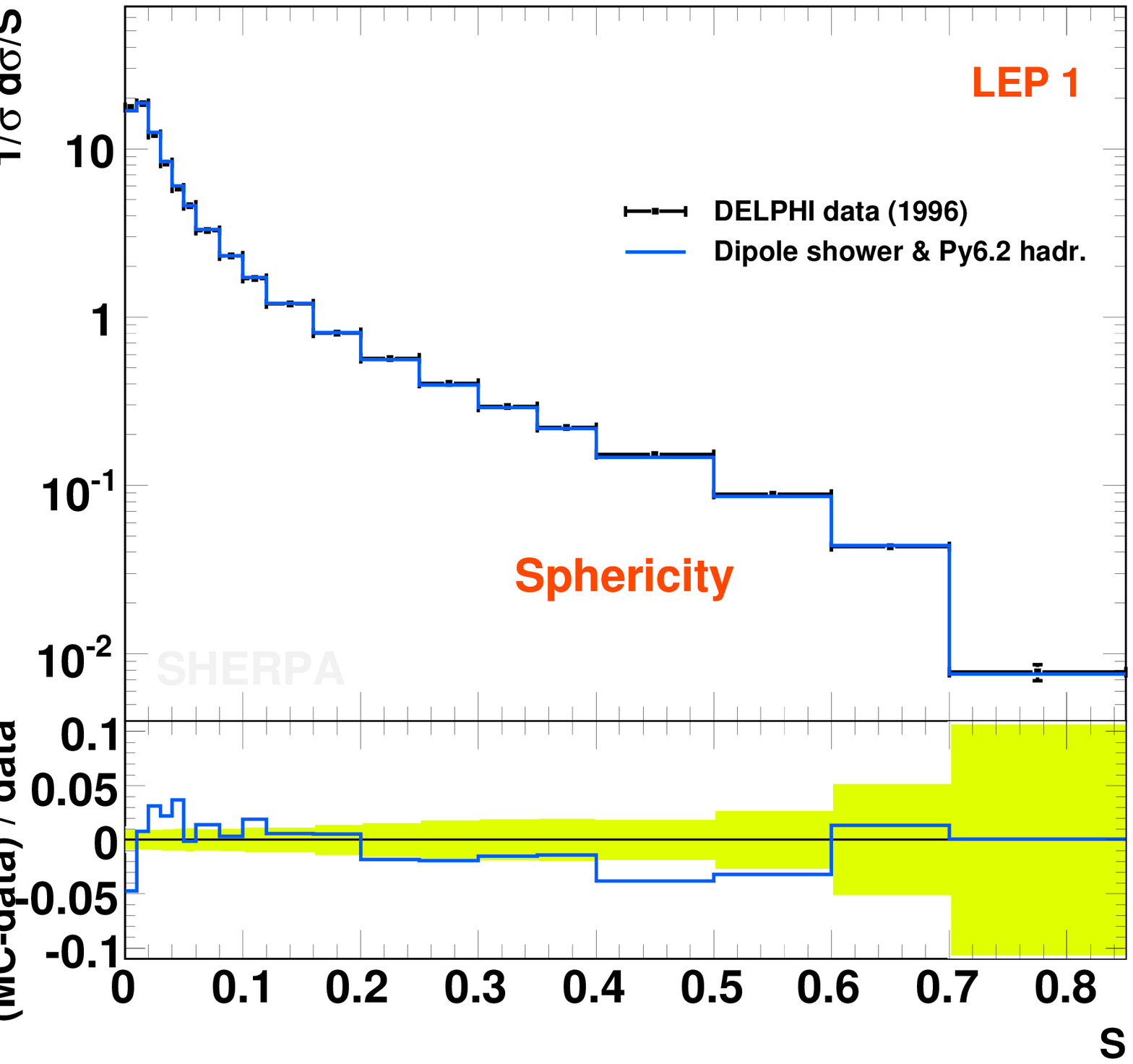}
  \hspace*{2mm}
  \includegraphics[width=78mm,angle=0.0]{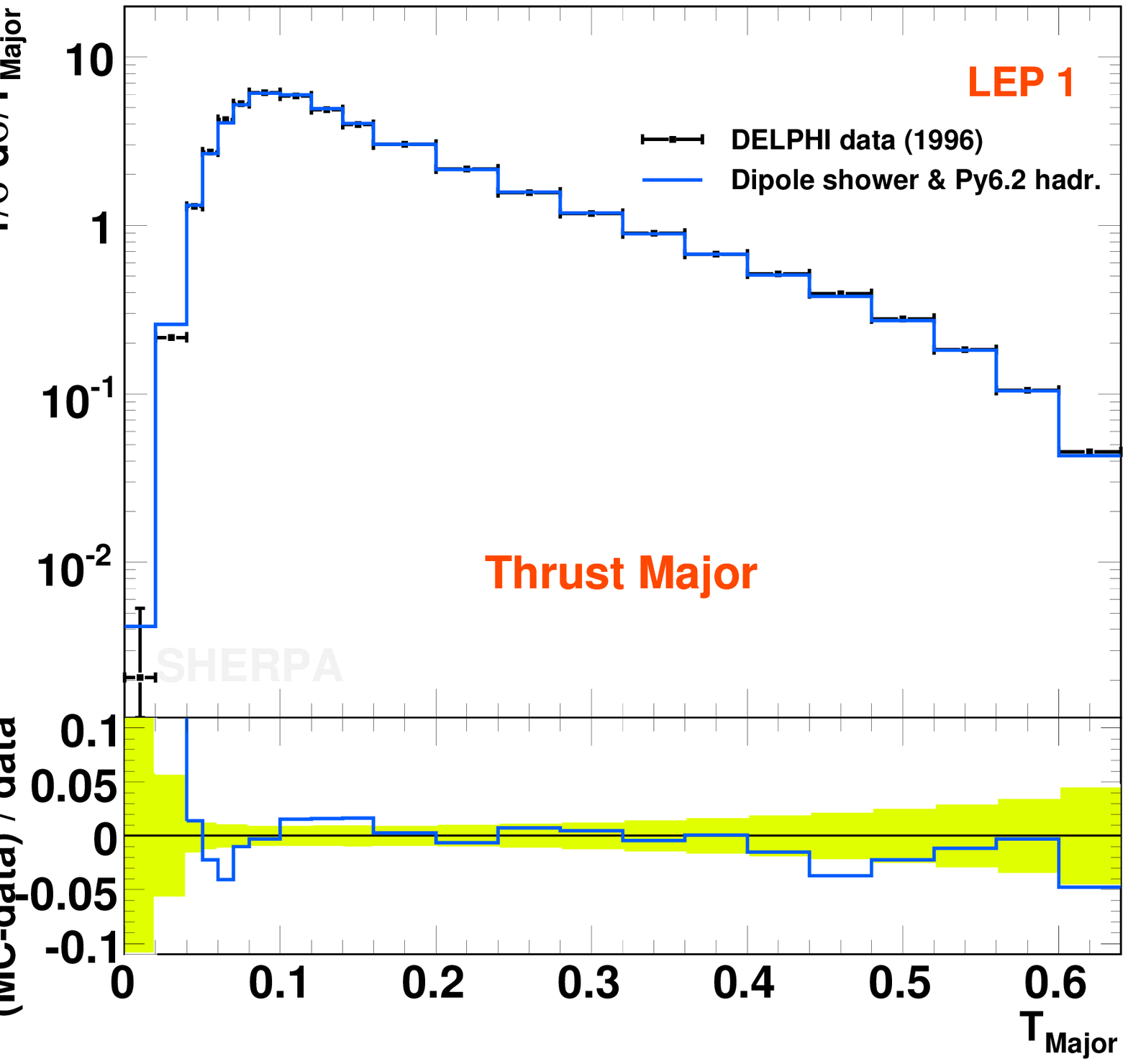}
  \hspace*{4mm}
  \includegraphics[width=78mm,angle=0.0]{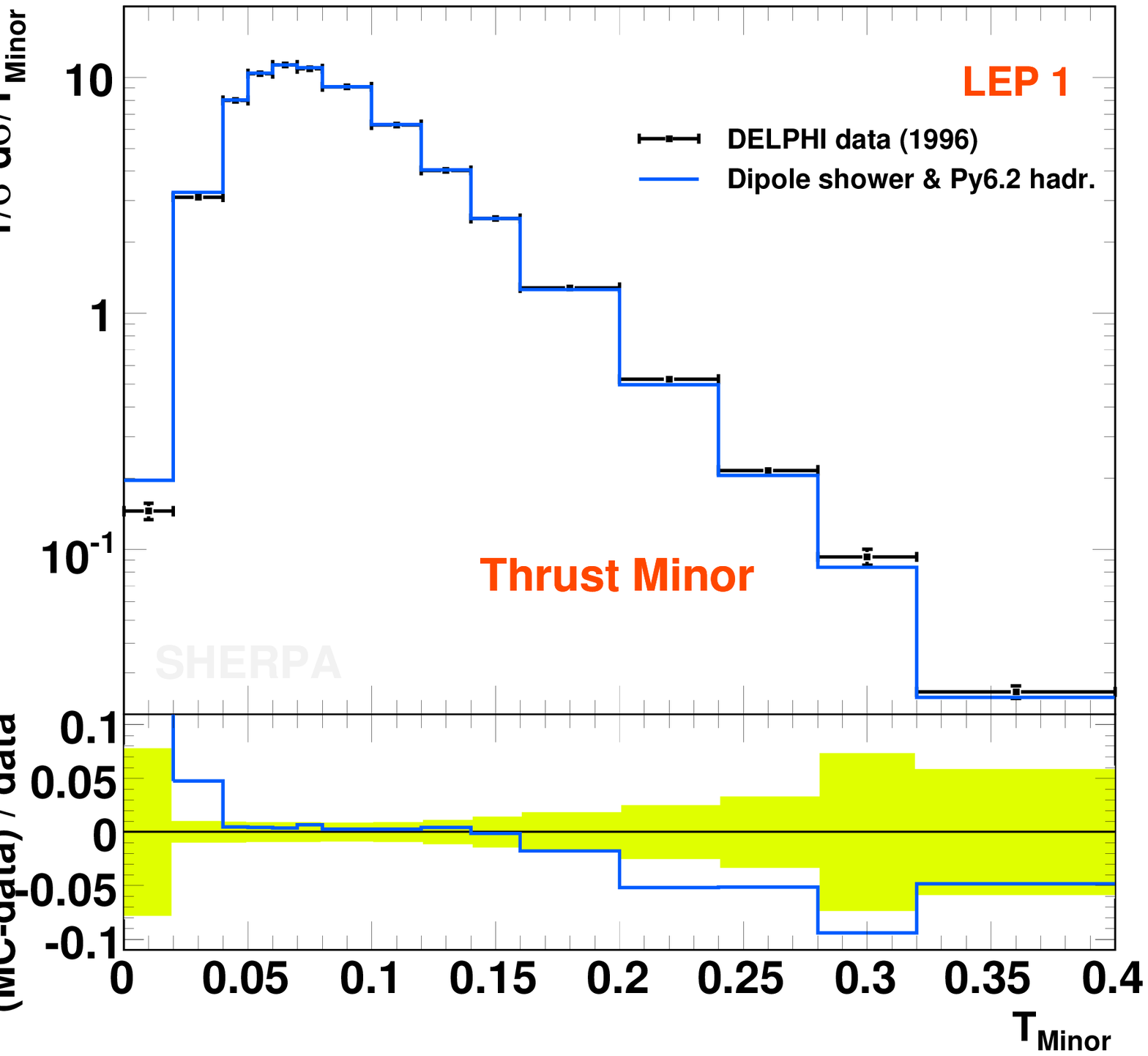}
  \vspace{0mm}
  \myfigcaption{140mm}{The dipole-shower predictions for the event
    shapes 1$-$thrust, sphericity, thrust major and thrust minor vs.\
    \delphi data \cite{Abreu:1996na}. Again, the light-coloured band
    represents the sum of the statistical and systematic errors.}
  \label{fig:LepEvsh1}
  \vspace{0mm}
\end{figure}

The testbed to exclusively validate the performance of the sole
final-state piece of the dipole-shower model\spc\footnote{Note
  that, for pure final-state cascading, the splitting of gluons into
  quark--antiquark pairs already has been included, and the
  implementation is nearly identical to the treatment proposed in the
  Lund CDM.}
is the process $e^+e^-\to Z^0/\gamma^\ast\to\mr{hadrons}$, where the
$q\bar q$\/ pair produced in the hard process will initiate the
cascade. The QCD Monte Carlo predictions can be compared with large
sets of data, which, for example, are available from the LEP1
measurements. The data precisely test the QCD dynamics of hadronic
final states produced at the $Z^0$ pole. 
\begin{figure}[t!]
  \vspace{0mm}
  \hspace*{2mm}
  \includegraphics[width=78mm,angle=0.0]{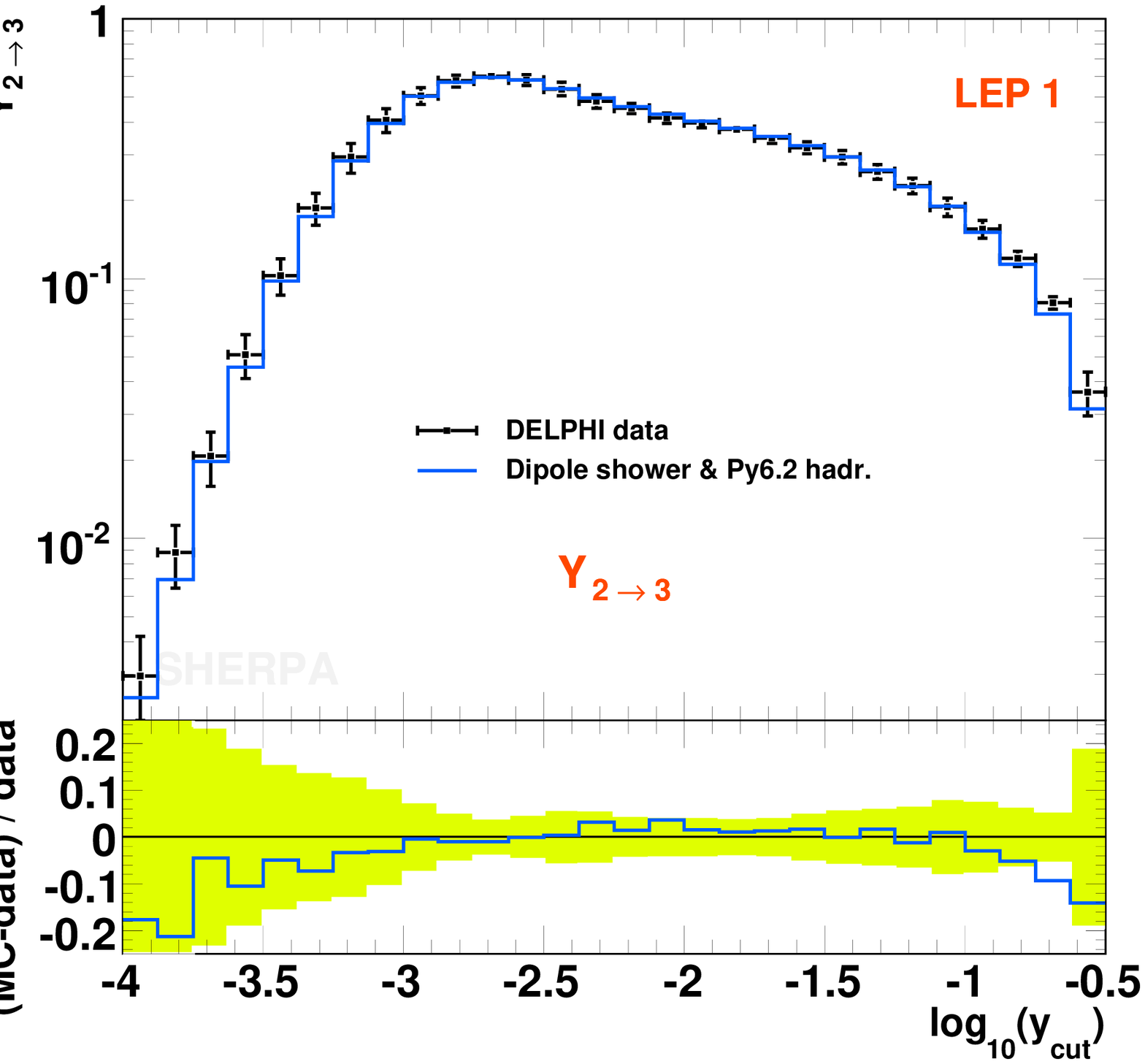}
  \hspace*{4mm}
  \includegraphics[width=78mm,angle=0.0]{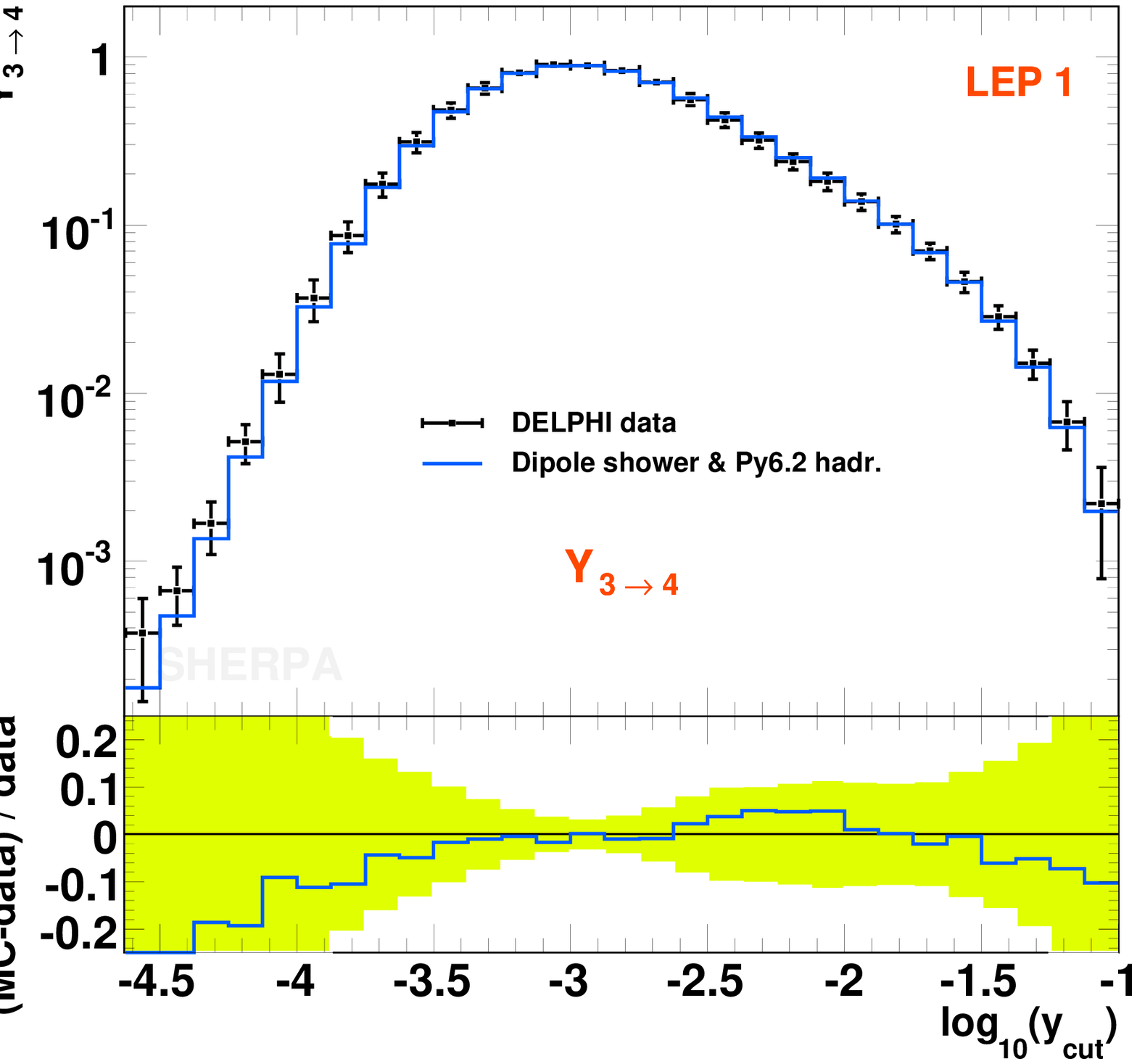}
  \hspace*{2mm}
  \includegraphics[width=78mm,angle=0.0]{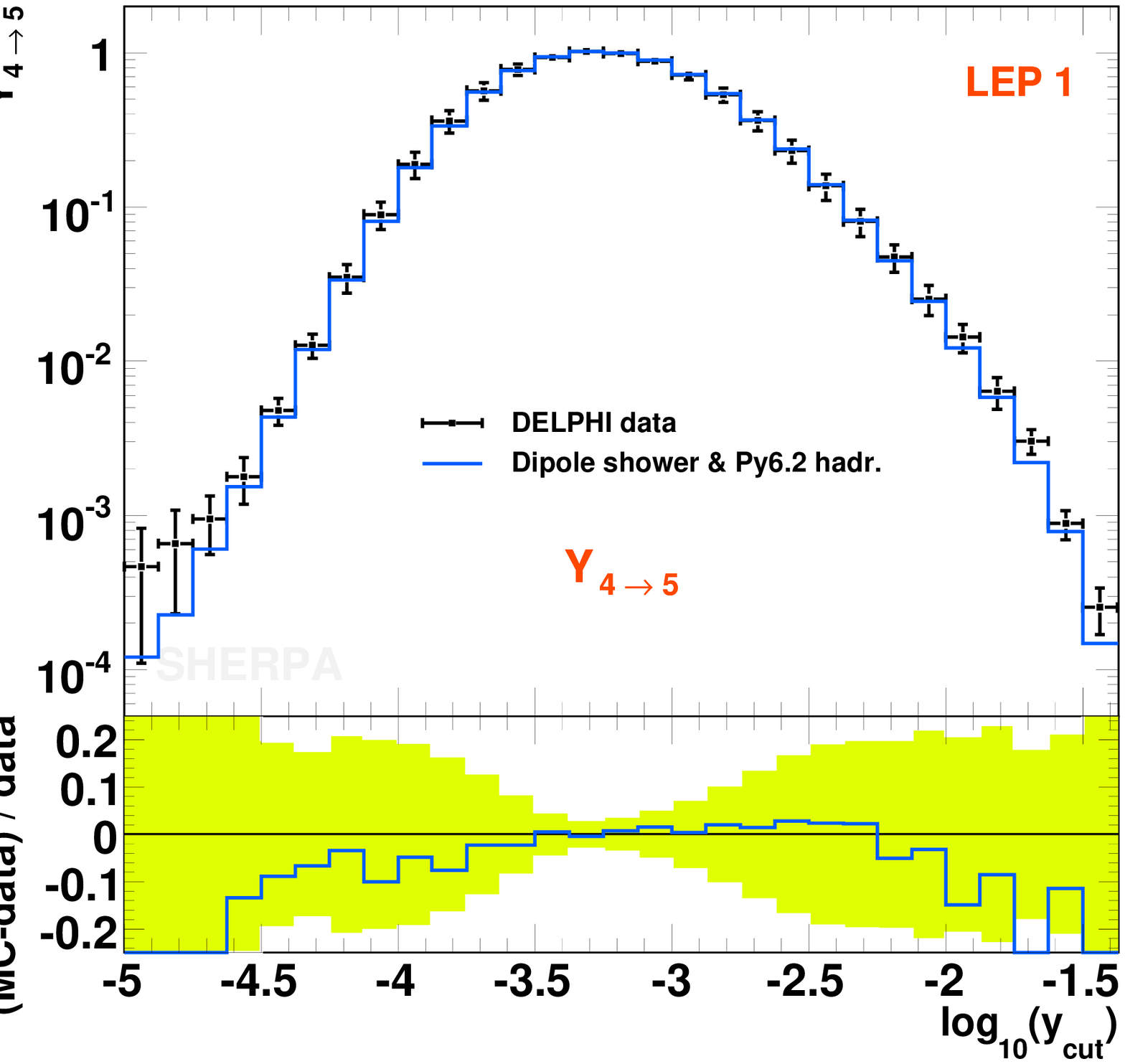}
  \hspace*{4mm}
  \includegraphics[width=78mm,angle=0.0]{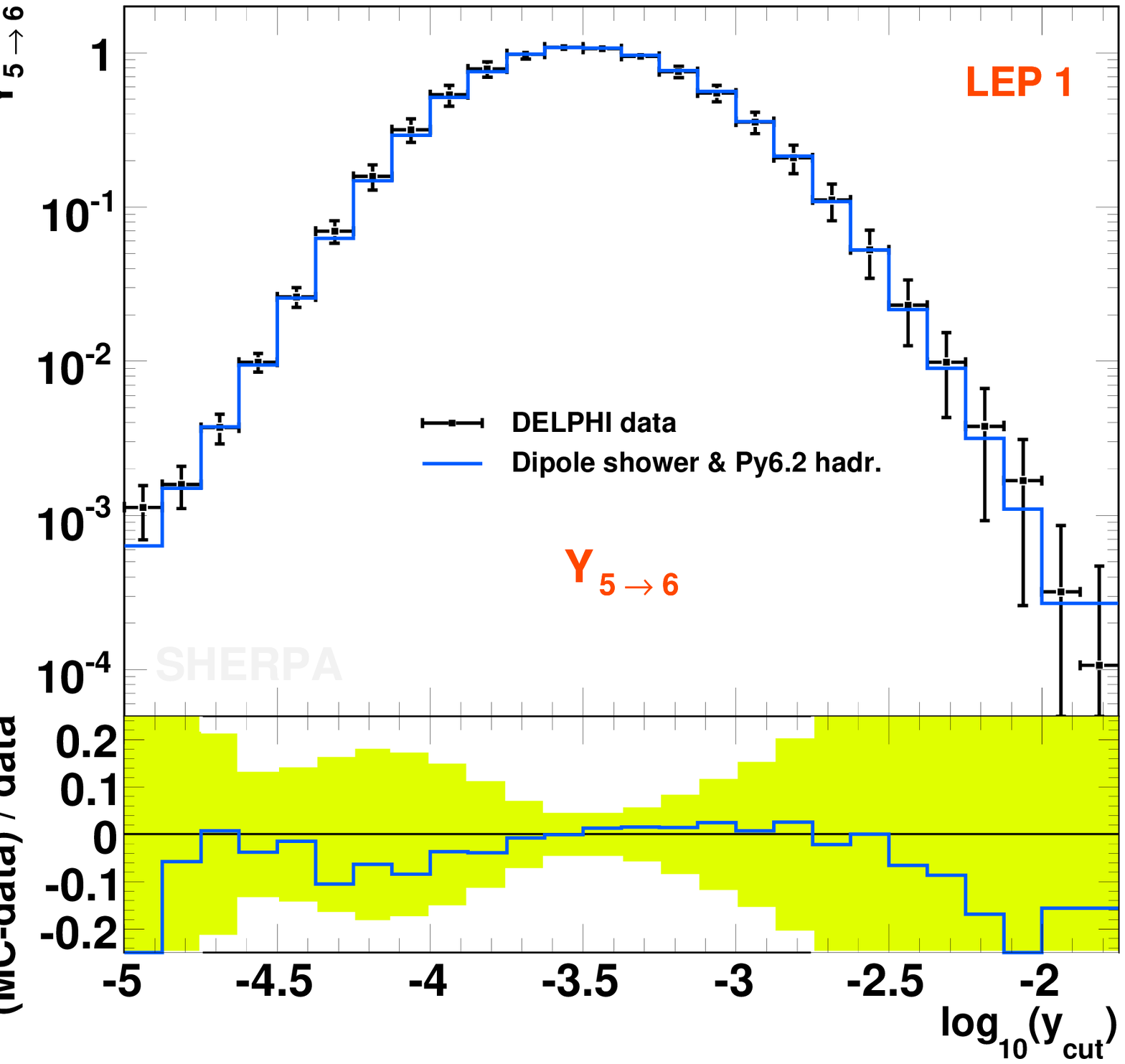}
  \vspace{0mm}
  \myfigcaption{140mm}{Durham differential jet rates as a function of
    the jet-resolution parameter $y_\mr{cut}$; dipole-shower
    prediction vs.\ \delphi data taken from \cite{Hoeth:2003}. Here,
  the light-coloured band represents the sum of the statistical and
  systematic errors.}
  \label{fig:LepDJR}
  \vspace{0mm}
\end{figure}
The parameters of the shower and the hadronization model were tuned by
hand, \ie the value taken for the strong coupling at $M_Z$
was specified, the FF cut-off $\con{p^2_{\perp,\mr{cut}}}{FF}$ of the
dipole shower was adjusted, and, suitable values for the Lund string
model parameters $a$\/ ({\tt PARP(41)}), $b$\/ ({\tt PARP(42)}) and
$\sigma_q$ ({\tt PARP(21)}) were found. The method employed for that is
sufficient to yield first significant results. However, it can not be
compared to the effort of delicate Monte Carlo tuning procedures as
presented in \cite{Hamacher:1995df} and foreseen in
\cite{Buckley:2007hi} in order to automatize the procedure.
The ``naively'' tuned parameters read:
\be\begin{split}
  \alpha_s(M_Z)&\;=\;0.1254\,,\\[1mm]
  \con{p^2_{\perp,\mr{cut}}}{FF}&\;=\;0.54\mbox{\ GeV}^2\,,\\[1mm]
  a&\;=\;0.29\,,\\[1mm]
  b&\;=\;0.76\mbox{\ GeV}^{-2}\,,\\[1mm]
  \sigma_q&\;=\;0.36\mbox{\ GeV}\,.
\end{split}\ee
Since massive quarks are not handled yet, the dipole shower always
started off massless $q\bar q$\/ pairs. At $S=M^2_Z$ a mean parton
multiplicity of $\langle\cal{N}_\mr{parton}\rangle=9.24$ and a mean
charged-particle multiplicity of $\langle\cal{N}_\mr{ch}\rangle=20.47$
are found, where the latter is somewhat below the experimentally
detected value of $\langle\cal{N}_\mr{ch}\rangle=20.92\pm0.24$
\cite{Abreu:1996na}. \Figs{fig:LepDJR} and \ref{fig:LepEvsh1} show a
selection of distributions obtained with the dipole shower and
compared to \delphi data taken at $\sqrt S=91.2$ GeV during the LEP1
run.
In \fig{fig:LepDJR}, Durham differential jet rates, $Y_{n\to n+1}$,
are presented up to $Y_{5\to6}$. They disentangle at which
$y_\mr{cut}=2\min\{E^2_i,E^2_j\}(1-\cos\theta_{ij})/S$\/ values an
$n+1$ jet event is merged into an $n$\/ jet event according to the
Durham jet clustering scheme \cite{Catani:1991hj}. The agreement with
the data taken by the \delphi experiment \cite{Hoeth:2003} is
very good, in particular the description around the peak positions.
All predictions tend to be somewhat below the bin means for low and
high values of the jet-resolution parameter $y_\mr{cut}$.
\\
Event shape variables probe the pattern of QCD radiation for both soft
and hard emissions arising from the primary $q\bar q$ dipole.
Therefore, in \fig{fig:LepEvsh1}, the shape distributions of 1$-$thrust,
$1-T$, thrust major, $T_\mr{major}$, and thrust minor, $T_\mr{minor}$
are displayed together with the sphericity, $\mr{S}$, placed in the
top right corner of the figure. The former are all obtained from a
linear momentum tensor, whereas the latter stems from a quadratic
one, thus, puts more emphasis on high momenta. All dipole-shower
results are compared, once again, to \delphi data \cite{Abreu:1996na}.
The low-value parts, which are sensitive to soft emissions, are all
quite well described, except for larger deviations in thrust major and
minor. Differences in these observables also appear, even 
somewhat larger, for instance for the new shower of \herwigpp 
\cite{Gieseke:2003hm} and the new shower presented in \cite{Schumann:2007mg} 
based on Catani--Seymour dipole factorization. For a very recent 
comparison, please cf.\ \cite{Dissertori:2007xa}, where the value of
the strong coupling constant has been determined at the $Z^0$ pole
using the results from a first NNLO calculation for $e^+e^-\to3$ jets
\cite{GehrmannDeRidder:2007hr}. Although the soft parts of these
distributions are all affected by hadronization corrections and their
careful modelling and tuning, the good behaviour of the dipole shower
in describing soft emissions can be seen as a consequence of
exponentiating the eikonal rather than the collinear limit of QCD
radiation. The predictions for hard emissions agree somewhat worse
with the data. The last two bins of the 1$-$thrust distribution are
overestimated signalling a slight excess of spherical events, whereas
thrust minor is underestimated for high values.
\\
Taken together, the agreement with data is satisfactory. This allows 
to conclude that the final-state piece of the dipole shower
sufficiently is under good control.
%
%
%
\subsection{Inclusive production of Drell--Yan lepton pairs at hadron colliders}
\label{sec:DY}
In the scope of hadronic collisions, the processes $pp(p\bar p)\to
Z^0/\gamma^\ast\to e^+e^-$ constitute the simplest and cleanest
testbed for the further validation of the dipole shower as they form
the initial--initial dipole counterpart of the $q\bar q$ timelike
evolution.
\begin{figure}[t!]
  \vspace{0mm}
  \hspace*{2mm}
  \includegraphics[width=78mm,angle=0.0]{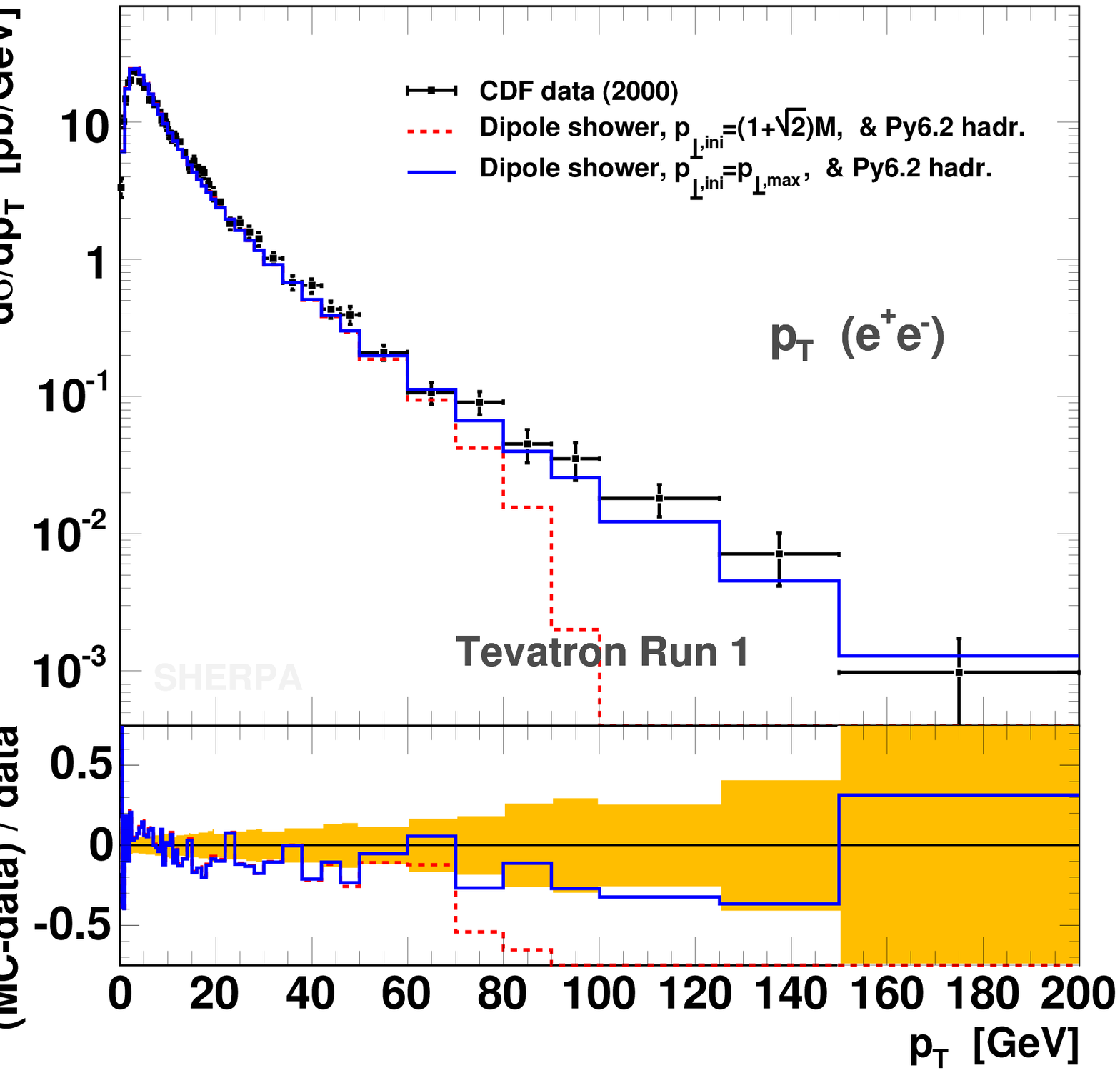}
  \hspace*{4mm}
  \includegraphics[width=78mm,angle=0.0]{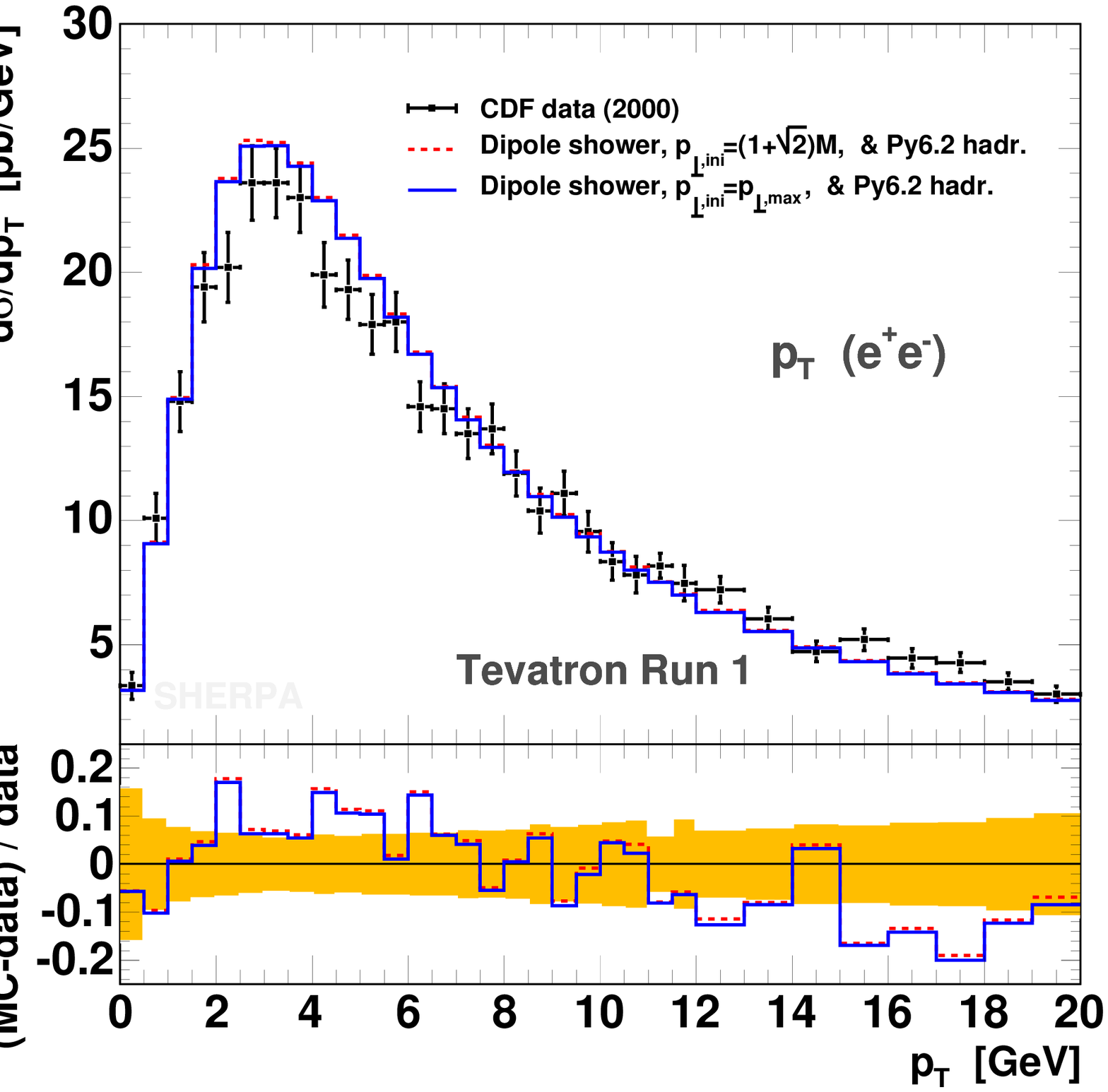}
  \vspace{0mm}
  \myfigcaption{140mm}{Boson transverse-momentum distribution in
    $e^+e^-+X$ as predicted by the dipole shower for two different
    choices concerning the initializing scale. The Monte Carlo
    calculations are compared with CDF data \cite{Affolder:1999jh}
    taken during Run I at the Fermilab Tevatron. The right panel
    depicts the very soft region of the distribution only.}
  \label{fig:Tev1Pt}
  \vspace{0mm}
\end{figure}
\subsubsection*{Tevatron Run I predictions}
The transverse-momentum distribution of the lepton pair is heavily
influenced by additional QCD radiation arising in both soft and hard
phase-space domains. Owing to its clear signal, this spectrum has been
measured with high precision by the Tevatron experiments. It is shown
in \fig{fig:Tev1Pt} for lepton-pair invariant masses in the range
$66\mbox{\ GeV}<M_{ee}<116\mbox{\ GeV}$.
Two hadron-level predictions produced by the dipole shower are
confronted with data from a CDF measurement at $\sqrt S=1.8$ TeV
\cite{Affolder:1999jh} and normalized to the experimental inclusive
cross section. They differ in their choice of the initializing scale,
using, first, $p_{\perp,\mr{ini}}=(1+\sqrt 2)M_{ee}$ and, second,
$p_{\perp,\mr{ini}}=p_{\perp,\mr{max}}$ (cf.\
\eqss{eq:pperpmax_iifg}{eq:pperpmax_iiig}). In the latter case the
shower evolves totally unconstrained, exploiting the fact that the
first emission is corrected for the true matrix element by
construction and may hence appear at a scale exceeding $M^2_{ee}$.
This in turn sets the highest scale for all consecutive emissions.
The whole treatment eventually leads to a good agreement with the
data for large $p_T$. In contrast, the dipole shower with restricted
initializing scale gradually undershoots data above $p_T=60$ GeV before
it dies off rapidly around $p_T=80$ GeV.
\\
The figure's right part contains a close-up of the peak region on a
linear scale, almost identically predicted by both dipole-shower
variants. The turn-on of the distribution is well described. Around
the peak, narrower described by the data, a slight excess is found,
followed by an underestimation of the data for the region above $12$
GeV. The predictions include an additional improvement for very low
$p_T$, namely an intrinsic transverse-momentum smearing, which has
been tuned by hand to these low $p_T$ data according to a Gaussian to
have a mean(width) of $0.3$($0.4$) GeV.\spc\footnote{The assignment of
  an intrinsic transverse momentum to the hard process is a
  non-perturbative correction applied after the shower phase, and,
  therefore, irrelevant for shower kinematics.}
Without this correction the shower $p_T$ spectra would slightly shift
to the left.
\begin{figure}[t!]
  \vspace{0mm}
  \bc\includegraphics[width=127mm,angle=0.0]{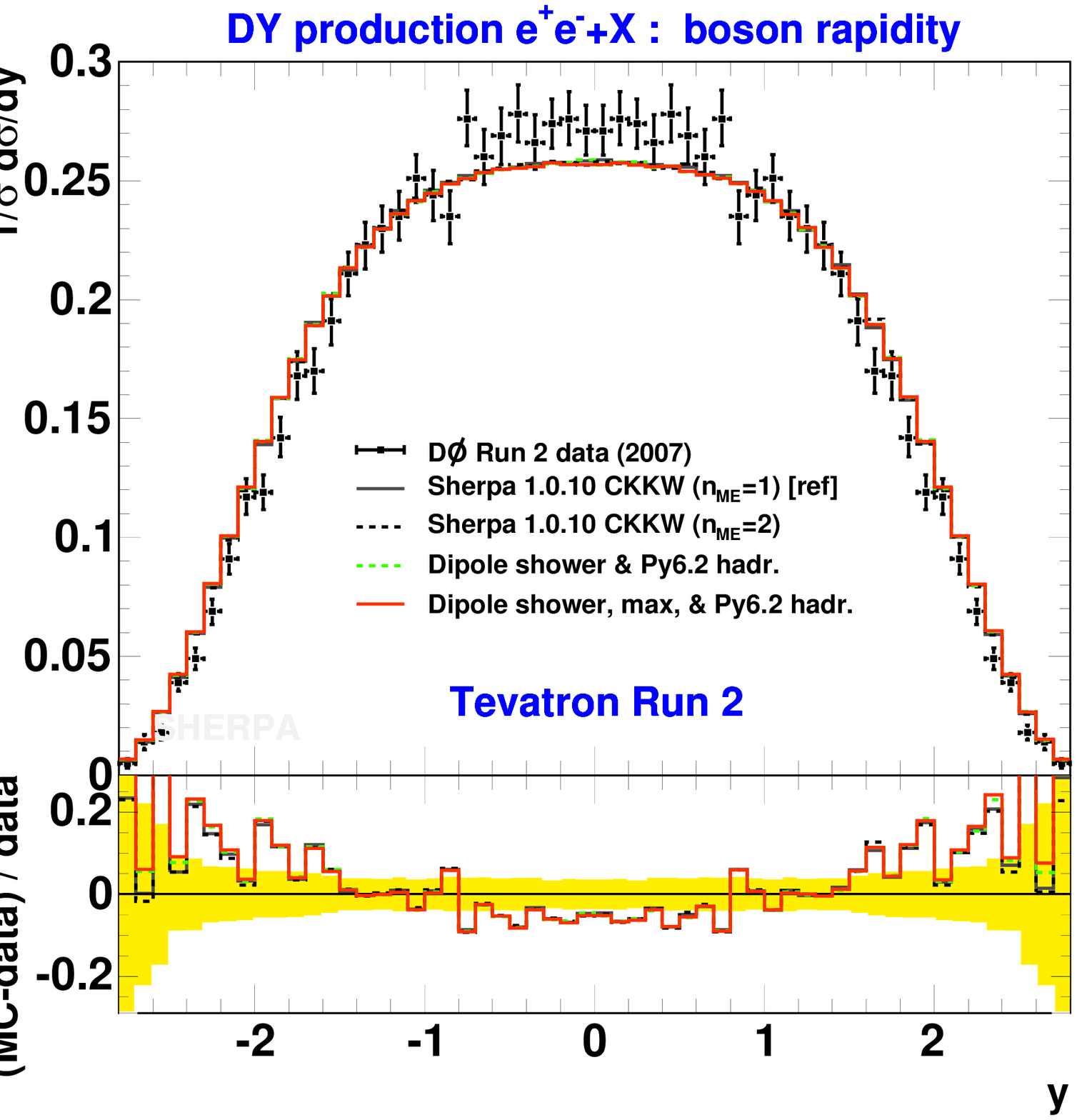}\ec
  \vspace{0mm}
  \myfigcaption{140mm}{Rapidity spectrum of the vector bosons in
    inclusive $Z^0/\gamma^\ast$ production at Tevatron Run II energies
    of $\sqrt S=1.96$ TeV as predicted by \sherpa and the dipole
    shower in comparison to recent D\O\ data \cite{Abazov:2007jy}. The
    grey solid, the black dashed, the green dashed and the red solid
    lines give the \sherpa CKKW $n_\mr{ME}=1$, CKKW $n_\mr{ME}=2$, the
    default and the unconstrained dipole-shower predictions,
    respectively.}
  \label{fig:Tev2YeejHt}
  \vspace{0mm}
\end{figure}
\subsubsection*{Tevatron Run II predictions}
One more validation against data is presented by considering the
rapidity distribution of the decaying vector boson, where the
Drell--Yan lepton-pair mass has been restricted to the interval
$71\mbox{\ GeV}<M_{ee}<111\mbox{\ GeV}$. The QCD NNLO theoretical
prediction for this inclusive observable has been calculated in
\cite{Anastasiou:2003ds} and very good agreement with data from a
recent D\O\ measurement \cite{Abazov:2007jy} has been observed over
the full rapidity range.
\\
Here, hadron-level predictions are presented that have been
obtained from the unconstrained dipole shower -- denoted by ``Dipole
shower, max'' in the plots from now -- as well as from the
$p_{\perp,\mr{ini}}$ restricted dipole shower, which is taken as the
default, since the matrix-element correction of the first emissions
does not apply beyond Drell--Yan processes. The comparison also shows
\sherpa outcomes resulting from the CKKW merging of parton showers and
tree-level matrix elements up to $n_\mr{ME}$ extra partons. This
merging method has been validated in many other comparative studies
\cite{Krauss:2004bs,Krauss:2005nu,Gleisberg:2005qq,Alwall:2007fs} or
even to data \cite{Nilsen:2006}. Here, two such inclusive samples, for
$n_\mr{ME}=1$ and $n_\mr{ME}=2$, were generated with \sherpa using its
version 1.0.10.
All results are displayed in \fig{fig:Tev2YeejHt} and
confronted with the D\O\ data \cite{Abazov:2007jy}. There hardly are
any shape differences visible between the various Monte Carlo
predictions. This nicely confirms that the II dipole kinematics is
eventually well fixed by preserving the rapidity of the final-state
particles, cf.\ \sec{sec:IIKins}. However, compared to data, all Monte
Carlo shapes are somewhat wider showing an excess of up to $20\%$ for
large rapidities.
\begin{figure}[t!]
  \vspace{0mm}
  \hspace*{2mm}
  \includegraphics[width=78mm,angle=0.0]{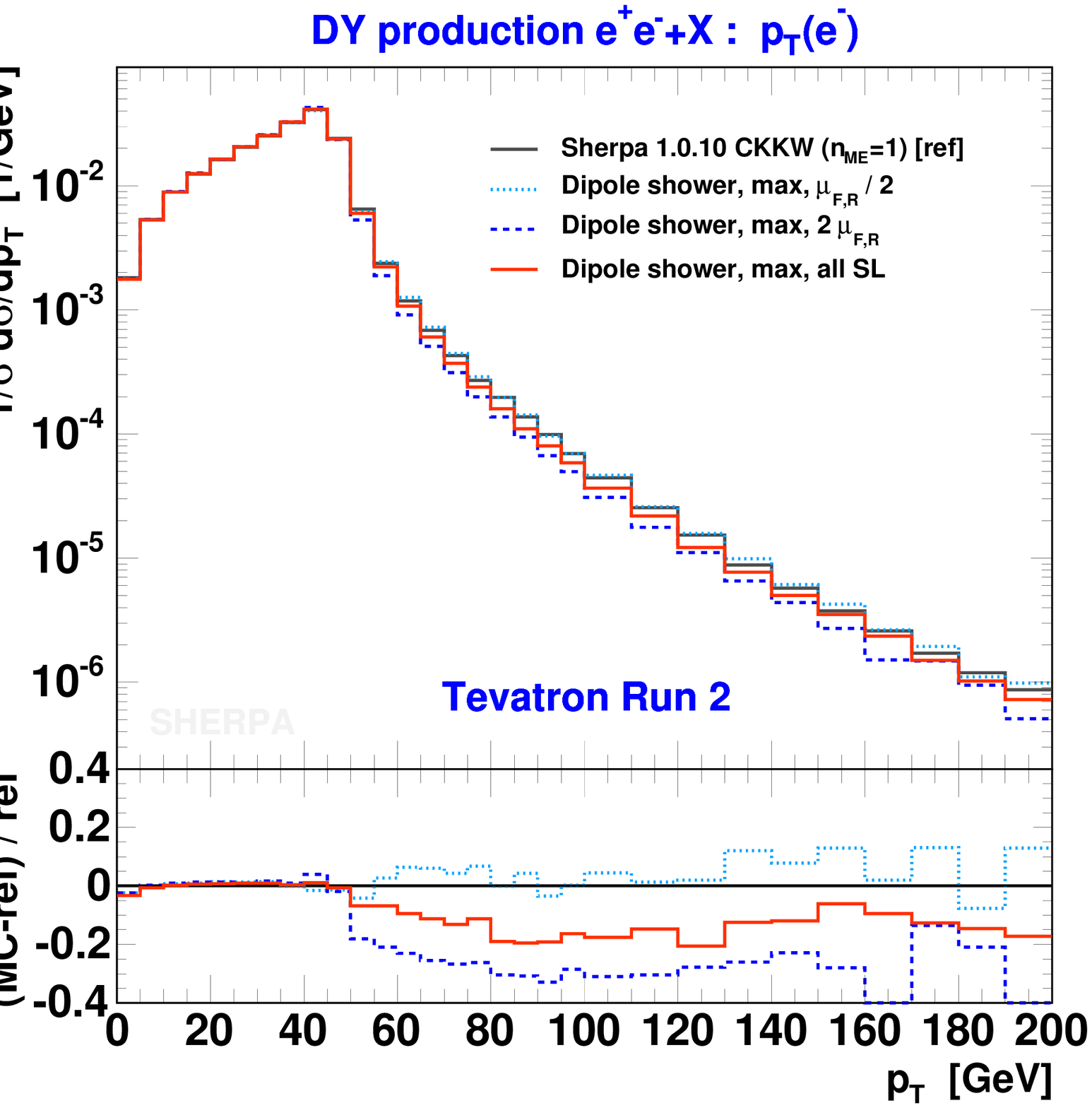}
  \hspace*{4mm}
  \includegraphics[width=78mm,angle=0.0]{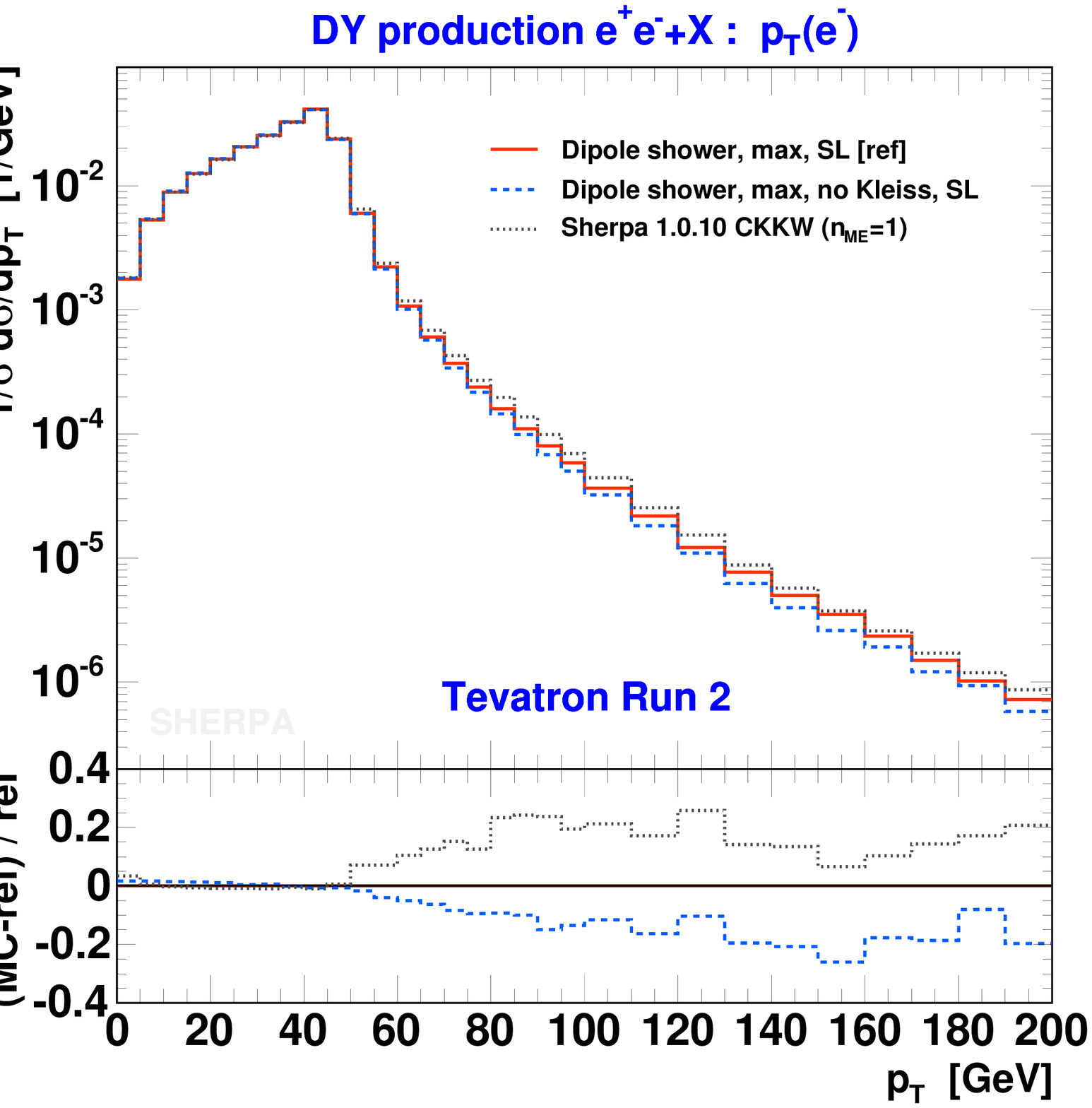}
  \vspace{0mm}
  \myfigcaption{140mm}{Left panel: impact of standard scale variations
    on the shower evolution exemplified by means of the $p_{T,e}$
    distribution. Here, the blue lines reflect the uncertainty of the
    prediction from the unconstrained shower, in both panels given by
    the red solid line. Right panel: impact of the Kleiss correction
    on the transverse-momentum distribution of the $e^-$. Here, the
    blue dashed and the grey dotted lines denote the outcome of the
    unconstrained dipole shower neglecting Kleiss corrections and of
    the \sherpa CKKW $n_\mr{ME}=1$ merging, respectively. Both panels
    depict shower-level (SL) results (lacking hadronization
    corrections).}
  \label{fig:Tev2sv}
  \vspace{0mm}
\end{figure}
\\
A rough estimate for the uncertainty of the shower predictions can be
gained from varying the values taken for the \mufp\ and \mur\ scales
within the shower algorithm. To this end, their defaults were
multiplied/divided by $2$. The \mufp\ scale enters through the PDF
weight, and \mur\ as the scale of the running strong coupling in the
single-emission probabilities, cf.\ \sec{sec:SudAlgo}. This is
exemplified in the left plot of \fig{fig:Tev2sv} for the $p_{T,e}$
distribution, where the uncertainty band for the unconstrained
(intrinsically first-order matrix-element corrected) dipole shower is
shown to cover the same-order-of-accuracy prediction stemming from
CKKW $Z^0+1$ jet merging. Therefore, both descriptions are in good
agreement. The right part of \fig{fig:Tev2sv} is the verification for the
importance of the Kleiss corrections for emissions off $\bar q'_\il
q_\il$ dipoles. Their application yields a hardening and, therefore,
an improvement of the single-lepton $p_{T,e}$ spectrum of about
$20\%$. \sherpa CKKW $n_\mr{ME}=1$ again serves as a good reference,
since it accounts for the full first-order lepton--hadron
correlations.
\subsubsection*{LHC predictions}
The correct energy extrapolation of the dipole shower is now verified
by comparing various approaches at LHC centre-of-mass energies. Therefore, 
the unconstrained dipole shower is studied w.r.t.\ \sherpa's CKKW merging
for $n_\mr{ME}=1,2,3$ and \sherpa's pure showering realized by \apacic
\cite{Kuhn:2000dk,Krauss:2005re}, which is a virtuality-ordered parton
shower in the traditional sense resumming large logarithms in the
collinear rather than the soft limit of QCD radiation.
\\
Most of the observables presented here require the exclusive
definition of jets, which has been attained according to the Run II
$k_T$ algorithm \cite{Catani:1993hr,Blazey:2000qt} using the parameter
$D=1$ and an unconstrained $\eta$\/ range in order to include
forward-jet effects. The jet $p_T$ threshold is given by
$p_{T,\mr{jet}}>20$ GeV. All distributions are simulated at the hadron
level and normalized to unit area, which allows for direct shape
comparisons. Many plots show CKKW predictions for $n_\mr{ME}>1$, which
helps estimate the impact of describing the next-to-first extra
parton emission by matrix elements as well.
\begin{figure}[t!]
  \vspace{0mm}
  \hspace*{2mm}
  \includegraphics[width=78mm,angle=0.0]{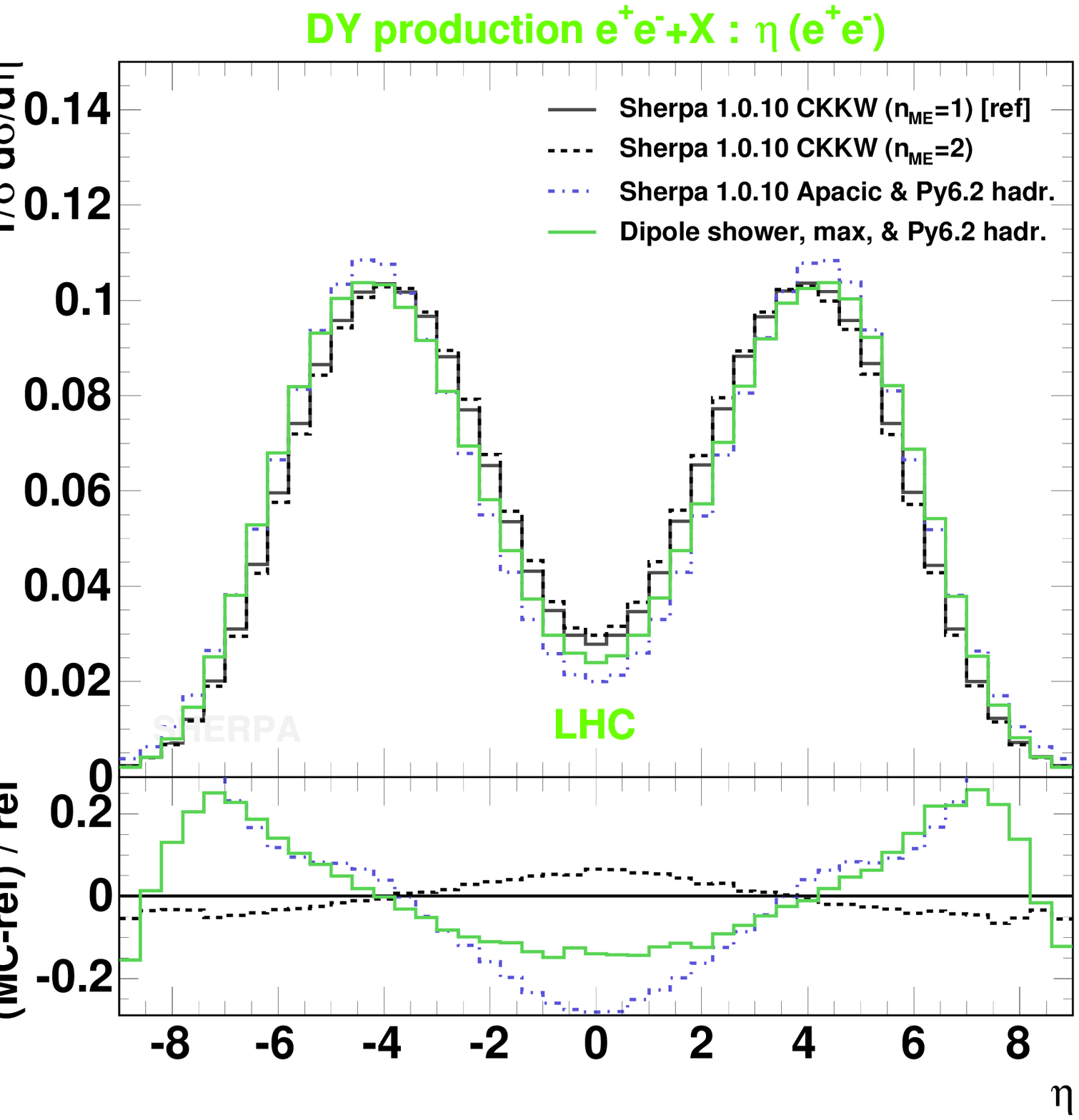}
  \hspace*{4mm}
  \includegraphics[width=78mm,angle=0.0]{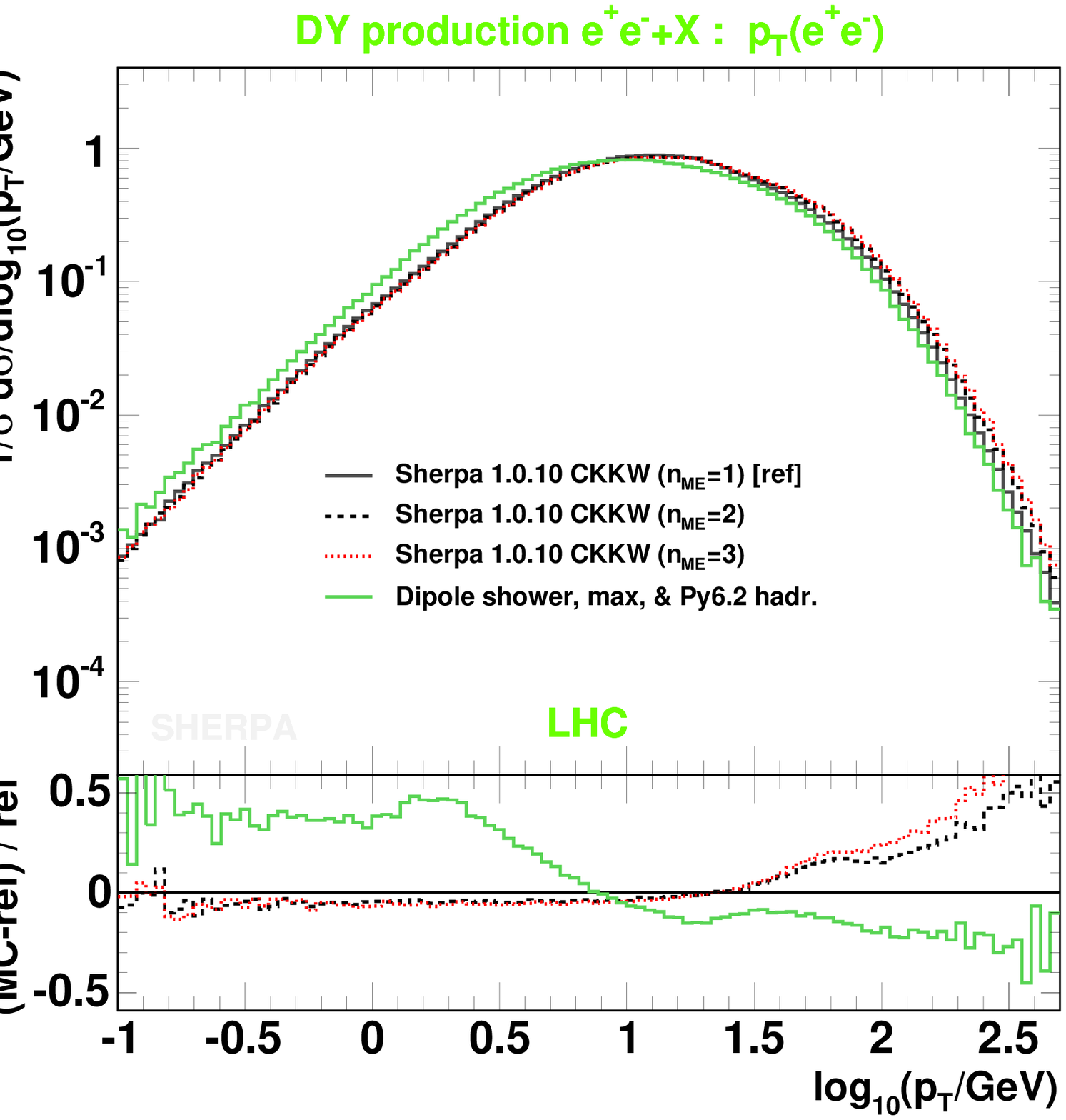}
  \vspace{0mm}
  \myfigcaption{140mm}{Pseudo-rapidity spectrum (left) and $p_T$
    distribution of the lepton pair in inclusive
    $Z^0/\gamma^\ast$+jets production at the LHC. The comparison is
    at the hadron level between the unconstrained dipole shower
    (green solid lines) and various \sherpa results, namely CKKW
    $n_\mr{ME}=1$ (grey solid lines), taken as the reference curve,
    CKKW $n_\mr{ME}=2$ (black dashed lines), CKKW $n_\mr{ME}=3$ (red
    dotted lines) and \apacic (blue dot-dashed lines).}
  \label{fig:LHCEtaPt}
  \vspace{0mm}
\end{figure}
\begin{figure}[p!]
  \vspace{0mm}
  \bc
  \hspace*{6mm}
  \includegraphics[width=67mm,angle=0.0]{dipcas/lhc/%
    AnalysedJets_jet_1_1_eta_1_1.eps}
  \hspace*{14mm}
  \includegraphics[width=67mm,angle=0.0]{dipcas/lhc/%
    AnalysedJets_jet_1_1_pt_1_1.eps}
  \hspace*{6mm}
  \includegraphics[width=67mm,angle=0.0]{dipcas/lhc/%
    AnalysedJets_jet_1_1_eta_2_1.eps}
  \hspace*{14mm}
  \includegraphics[width=67mm,angle=0.0]{dipcas/lhc/%
    AnalysedJets_jet_1_1_pt_2_1.eps}
  \hspace*{6mm}
  \includegraphics[width=67mm,angle=0.0]{dipcas/lhc/%
    AnalysedJets_jet_1_1_eta_3_1.eps}
  \hspace*{14mm}
  \includegraphics[width=67mm,angle=0.0]{dipcas/lhc/%
    AnalysedJets_jet_1_1_pt_3_1.eps}
  \ec
  \vspace{0mm}
  \myfigcaption{140mm}{Pseudo-rapidity (left column) and
    transverse-momentum (right column) distributions of the first
    three jets in inclusive lepton-pair production simulated for LHC
    energies. Dipole-shower results are shown in comparison to those
    obtained by the CKKW method of \sherpa. Labelling is as in
    \fig{fig:LHCEtaPt}.}
  \label{fig:LHCEtajPtj}
  \vspace{0mm}
\end{figure}
\\
The pseudo-rapidity and $p_T$ distributions of the $e^+e^-$ pairs
are shown in \fig{fig:LHCEtaPt}. They are largely determined by 
the pattern of QCD emissions, in particular by the hardest one, and
measure the recoil of the lepton pair against all other final-state
particles. Hence, these inclusive observables are first defined beyond
the leading order Drell--Yan pair production process. For the
$\eta_{ee}$ spectrum shown in the left panel, the maxima and the
central rapidity region are somewhat more and less pronounced by the
dipole shower than by the CKKW predictions, respectively. As it can be
seen, the improved description of second-order emissions results in a
further enhancement of vector bosons that are central in $\eta$\/
space. In contrast, the \apacic prediction features a considerably larger
dip in the central pseudo-rapidity region as a consequence of lacking
sufficiently hard emissions, since this shower's start scale is
constrained by the mass of the lepton pair to ensure evolving in the
collinear and soft phase-space regions only. The right part of
\fig{fig:LHCEtaPt} contains the $p_{T,ee}$ distribution on a
double-logarithmic scale to provide good insight to both soft and hard
$p_T$ domains. In the hard tail the dipole-shower result is $30\%$
below the CKKW reference; the difference in the low $p_T$ part amounts
up to $40\%$, whereas the dipole shower clearly puts emphasis on the
soft region and predicts a slightly lower peak position. The agreement
is still satisfactory and the deviations can be traced back, for the
very soft part, to different parameter settings for the fragmentation
of the partons (including intrinsic $k_T$ smearing), for the range
$1\mbox{\ GeV}<p_T<20\mbox{\ GeV}$, to different radiation patterns
generated by the dipole shower and the vetoed \apacic shower used for
the CKKW merging, and, for the high $p_T$ tail, to differences in
choosing and processing the scales in both approaches. Typically, the
inclusion of next-to-one extra parton emissions at the matrix-element
level leads to a further increase of the high $p_T$ tail. This is
found in \fig{fig:LHCEtaPt} where the effects can become as large as
$40\%$.
\\
\Fig{fig:LHCEtajPtj} presents the jet pseudo-rapidity, $\eta_i$, and
transverse-momentum, $p_{T,i}$, distributions of the first three jets.
These observables directly probe the jet structure of the events. For
$\eta_i$, the dipole-shower predictions are quite similar and in all
cases narrower w.r.t.\ the \sherpa CKKW predictions. For $p_{T,i}$,
the predictions of the unconstrained dipole evolution agree quite well
with the respective ones of \sherpa CKKW $n_\mr{ME}=1$, again on a
$20\%$--$40\%$ level, confirming that the scale setting by the first
(the unconstrained) emission reasonably constrains the subsequent one.
The $p_T$ hardness of the jets predicted by the inclusive two- and
three-jet merging is of course out of reach for the dipole shower.
Such higher-order corrections can only be included by matrix-element
parton-shower merging techniques or a matching with respective NLO
calculations.
\begin{figure}[p!]
  \vspace{0mm}
  \hspace*{2mm}
  \includegraphics[width=78mm,angle=0.0]{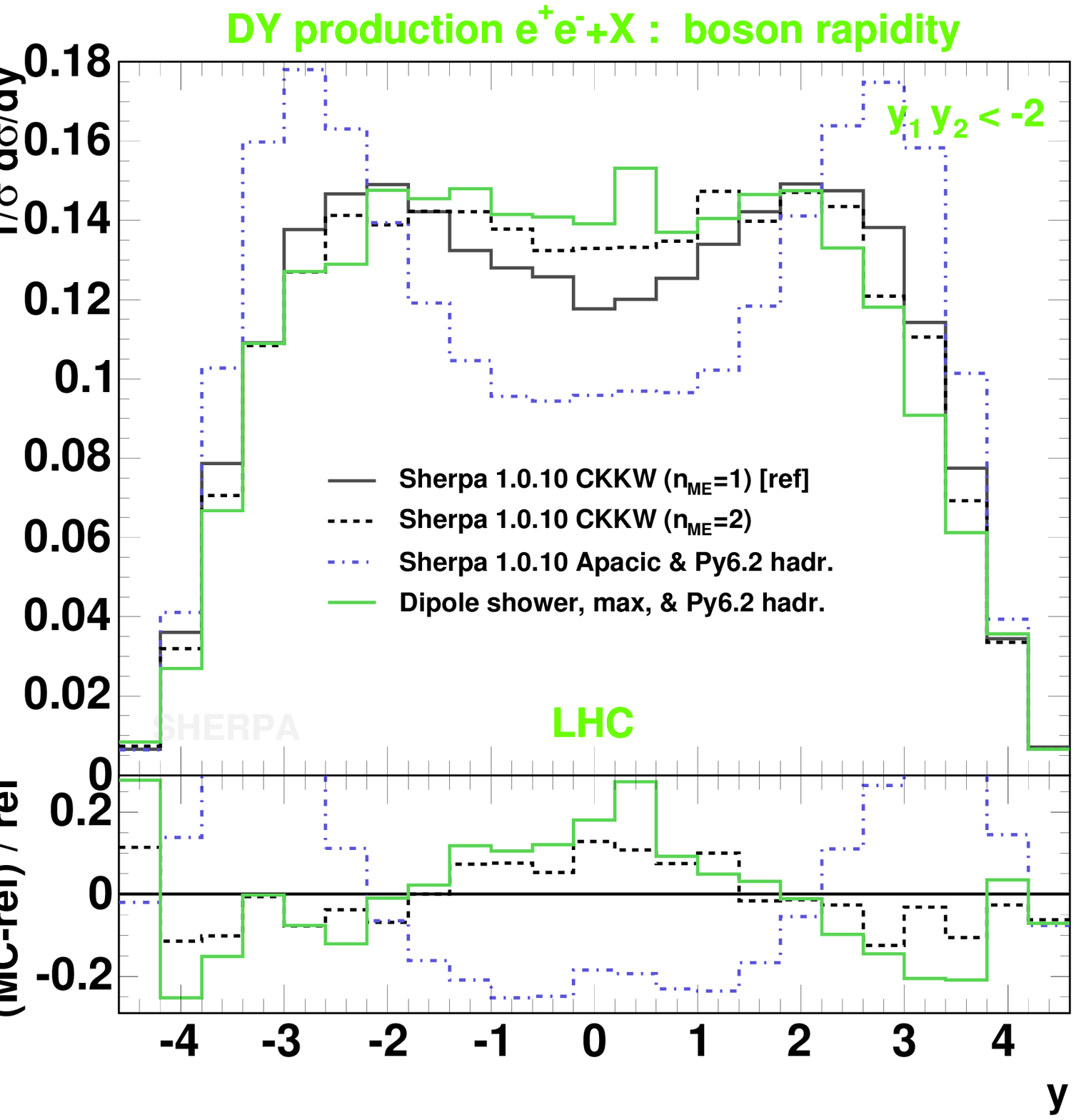}
  \hspace*{4mm}
  \includegraphics[width=78mm,angle=0.0]{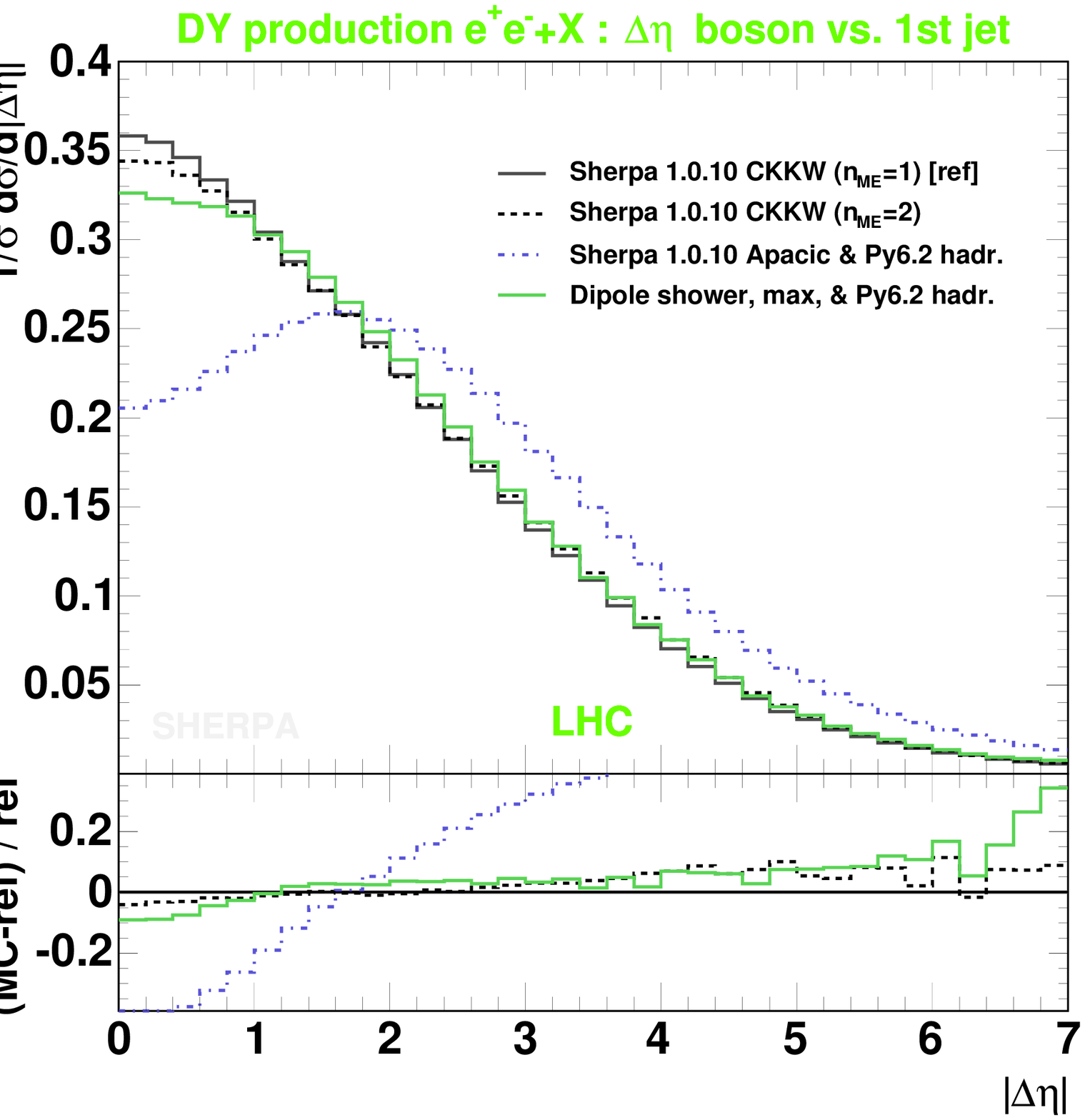}
  \hspace*{2mm}
  \includegraphics[width=78mm,angle=0.0]{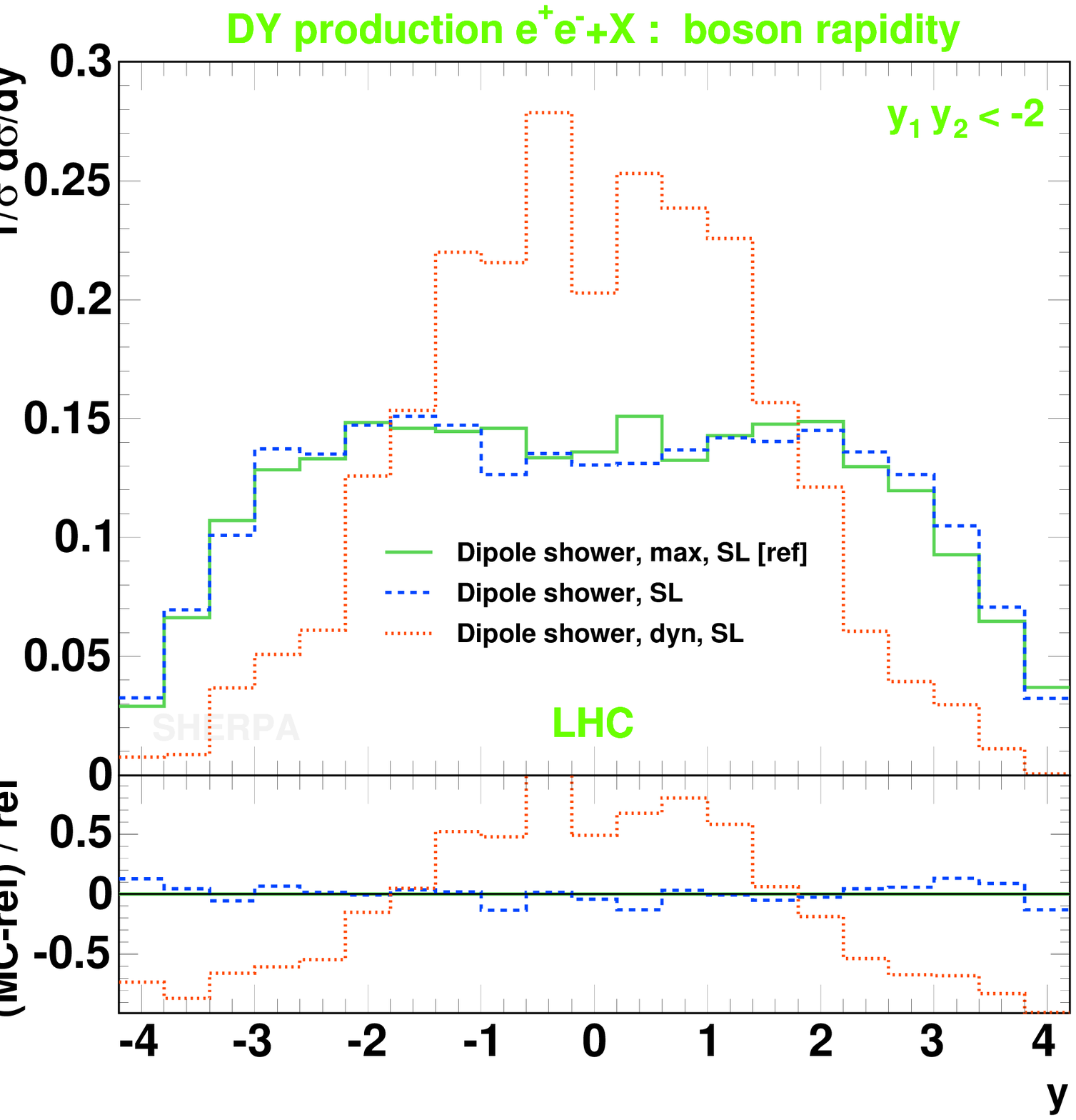}
  \hspace*{4mm}
  \includegraphics[width=78mm,angle=0.0]{dipcas/lhcpsp/%
    AnalysedJets_jet_1_3_cll1_1.eps}
  \vspace{0mm}
  \myfigcaption{140mm}{The rapidity of the lepton pair for the first
    two jets satisfying $y_1y_2<-2$ (top left) and the modulus of the
    pseudo-rapidity difference between the vector boson and the
    leading jet (top right). Both of which is simulated for inclusive
    Drell--Yan $e^+e^-$ production at the LHC. The dipole-shower
    outcome is compared to those received from various \sherpa runs.
    Labelling is as introduced in \fig{fig:LHCEtaPt}.
    \\
    The bottom plots exemplify the impact of the choice of $\hat
    s_\mr{max}$ on the $y_{ee}$ for $y_1y_2<-2$ and $\cos\theta_{13}$
    distributions as predicted by the unconstrained (solid green
    curves), the default (dashed blue curves) and the dipole shower
    where $\hat s_\mr{max}$ is set dynamically (dotted red curves).
    The cosine of the angle between the first and the third jet,
    $\cos\theta_{13}$, is determined in the rest frame of the
    first-plus-second jet system.}
  \label{fig:LHCSpecials}
  \vspace{0mm}
\end{figure}
\\
The top left plot in \fig{fig:LHCSpecials} depicts the vector boson
rapidity spectrum obtained under the additional requirement that the
first and the second jet appear well separated in rapidity according
to $y_1y_2<-2$. Except for \apacic predicting a strong tendency of the
boson to accompany one of the jets, all codes give flat spectra for
central rapidities, and, remarkably, the dipole-shower result agrees
well with that of \sherpa CKKW $n_\mr{ME}=2$. A similar pattern is
found in the $\abs{\Delta\eta_{ee,1}}=\abs{\eta_{ee}-\eta_1}$
distribution shown in the top right plot of \fig{fig:LHCSpecials}. The
dipole-shower curve hardly deviates from the CKKW curves, which
reliably describe this observable owing to their higher-order
contributions. This tellingly highlights the effects of the
improvements of the dipole splitting functions beyond the eikonal
approximation. In contrast, \apacic describes a suppression for low
$\abs{\Delta\eta_{ee,1}}$ values.
\\
In contrast to the shape comparisons above, \tab{tab:DipRatesLHC}
provides some insight concerning inclusive and exclusive jet rates
normalized to the total inclusive cross section. The results given by
the unconstrained dipole shower are close to those of \sherpa CKKW
$n_\mr{ME}=1$. Moreover, it is also found that the default dipole
shower predicts much more soft jets compared to \apacic.
\begin{table}[t!]
  \vspace{0mm}\bc
  \bt{|l||c|c|c||c|c|}\hline    
  & \phantom{aaaaaa} & \phantom{aaaaaa}
  & \phantom{aaaaaa} & \phantom{aaaaaa} & \phantom{aaaaa}
  \\[-2.5mm]
  Monte Carlo &
  \Large$\frac{\sigma_{\ge1\mr{jet}}}{\sigma_\mr{tot}}$ &
  \Large$\frac{\sigma_{\ge2\mr{jet}}}{\sigma_\mr{tot}}$ &
  \Large$\frac{\sigma_{\ge3\mr{jet}}}{\sigma_\mr{tot}}$ &
  \Large$\frac{\sigma_{=1\mr{jet}}}{\sigma_\mr{tot}}$ &
  \Large$\frac{\sigma_{(y_1y_2<-2)}}{\sigma_\mr{tot}}$\\[2.5mm]\hline
  &&&&&\\[-2.5mm]
  CKKW $n_\mr{ME}=1$     &$0.304$&$0.082$&$0.017$&$0.222$&$0.016$\\[1mm]
  CKKW $n_\mr{ME}=2$     &$0.340$&$0.108$&$0.025$&$0.231$&$0.017$\\[1mm]
  CKKW $n_\mr{ME}=3$     &$0.348$&$0.119$&$0.034$&$0.229$&$0.018$\\[2.5mm]\hline
  &&&&&\\[-2.5mm]
  \apacic                &$0.232$&$0.048$&$0.007$&$0.157$&$0.010$\\[2.5mm]\hline
  &&&&&\\[-2.5mm]
  Dipole shower, max     &$0.290$&$0.084$&$0.023$&$0.207$&$0.012$\\[2.5mm]\hline
  &&&&&\\[-2.5mm]
  Dipole shower, max (SL)&$0.296$&$0.087$&$0.024$&$0.210$&$0.013$\\[1mm]
  Dipole shower (SL)     &$0.267$&$0.068$&$0.016$&$0.199$&$0.011$\\[1mm]
  Dipole shower, dyn (SL)&$0.244$&$0.052$&$0.009$&$0.193$&$0.0003$
  \\[2.5mm]\hline
  \et
  \ec\vspace{0mm}
  \mytabcaption{140mm}{Cross section ratios as obtained from the
    various Monte Carlo approaches for inclusive and exclusive
    (hadron- and shower-level ``SL'') jet rates at LHC energies. Jets
    are defined according to the Run II $k_T$ algorithm
    \cite{Catani:1993hr,Blazey:2000qt} and required to have
    $p_{T,\mr{jet}}>20$ GeV.}
  \label{tab:DipRatesLHC}
  \vspace{0mm}
\end{table}
\\
The large phase space available for additional QCD radiation at the
LHC will lead to the copious production of jets. Here, this testbed
provides an excellent means to study the effects of the $\hat
s_\mr{max}$ reduction (see \sec{sec:Scales}). In \tab{tab:DipRatesLHC}
and the bottom row of \fig{fig:LHCSpecials} various predictions are
confronted with each other, namely those of the unconstrained dipole
shower, the default shower where $p^2_{\perp,\mr{ini}}=(1+\sqrt 2)^2
M^2_{ee}$ and the dipole shower denoted by ``dyn'' where additionally
$\hat s_\mr{max}$ is set dynamically according to
$p^2_{\perp,\mr{max}}=p^2_{\perp,\mr{ini}}=(1+\sqrt 2)^2 M^2_{ee}$
(cf.\ \eg \eqss{eq:pperpmax_iifg}{eq:pperpmax_iiig} in
\secss{sec:gf_ii}{sec:gi_ii}, respectively). The default shower loses
hard emissions, such that the ratios of \tab{tab:DipRatesLHC} are
somewhat smaller w.r.t.\ those of the unconstrained shower. For the
``dyn'' variant, the normalized cross sections decrease further,
however, dramatically fall if a rapidity separation for the first two
jets is required and imposed by $y_1y_2<-2$. Furthermore, while the
spectra presented at the bottom of \fig{fig:LHCSpecials} show only
mild differences between the unconstrained and the default shower,
the predictions of the ``dyn'' shower deviate considerably: the
$y_{ee}$ spectrum for $y_1y_2<-2$ is far too peaked in the central
$y$\/ region, which also contradicts the performance of the CKKW
references shown in the top left corner of the same figure. And, in
the $\cos\theta_{13}$ distribution of \fig{fig:LHCSpecials} the third
jet is significantly less collinear w.r.t.\ the first and second jet,
where the angle $\theta_{13}$ between the first and third jet is
taken in the rest frame of the combined first and second jet.
Hence, in all examples, the $\hat s_\mr{max}$ reduction manifests in a
suppression of forward and larger separated emissions (jets), which
can be understood, since, firstly, the $p^2_{\perp,\mr{max}}$ act as
kinematic upper bounds to all start scales $p^2_{\perp,\mr{stt}}$, in
particular initializing scales $p^2_{\perp,\mr{ini}}$. Secondly, for
reduced $\hat s_\mr{max}$, the generation of large $\abs{y}$ is
suppressed for a single emission, eventually avoiding the production
of sets of two-parton invariant masses, $\abs{s_{kg}}=\abs{M}\,p_\perp
e^{-y}$ and $\abs{s_{g\ell}}=\abs{M}\,p_\perp e^{+y}$, where one of
them is very small and the other very large. Taken these findings
together -- also recalling the good agreement with the CKKW results --
it is evident that using $\hat s_\mr{max}\coo{II}=S$ and $\hat
s_\mr{max}\coo{FI}=\mathscr{S}$ is a good choice.
\subsection{Inclusive jet production at hadron colliders}
\label{sec:Jets}
The copious QCD production of jets is a typical and large phenomenon at hadron
colliders, however, from a theoretical point of view, the task of
calculating and/or simulating these processes at higher orders in the
strong coupling is more complicated and rather involved. Clearly, QCD
jet production severely tests the entire shower algorithm and goes
beyond the tasks handled by the dipole shower so far. There are
several reasons for this: the primary state is now given as a
multi-dipole configuration formed by the $2\to2$ hard QCD processes
according to their (large $N_\mr{C}$) colour connections, including
those that link initial- and final-state partons. Possibly, all
dipoles form only one colour singlet or even a ``gluonic ring''.
Matrix-element corrections for the first extra emission in jet
production are absent in the dipole splitting functions; this in turn
requires to carefully constrain the initializing scale, such that the
shower evolution is guaranteed to proceed in the soft and collinear
limits of QCD emission only.
\\
To validate the dipole shower the observables listed below have been
considered in more detail.
\subsubsection*{Dijet azimuthal decorrelations at Tevatron Run II energies}
The dijet-decorrelation observable measured in the transverse plane
between the two hardest jets, $\Delta\phi_\mr{dijet}=\abs{\phi_1-\phi_2}$,
provides good insight to the occurrence of additional soft and hard
radiation. There is no necessity to reconstruct further jets. The
clear full-correlation signature given by $\Delta\phi_\mr{dijet}=\pi$
washes out in the presence of extra emissions. Since, the strength of
the decorrelation rises in dependence on their hardness, the dijet
decorrelation can be used to verify any candidate choice of setting
the initializing scale.\spc\footnote{Showers preferably should
  predict the distribution for small decreases of
  $\Delta\phi_\mr{dijet}=\pi$, the tail may be corrected by the
  inclusion of matrix elements beyond order $\alpha^2_s$.}
\\
The observable was subject of a recent measurement by D\O\ at Tevatron
Run II with the data taken in different $p_{T,1}=p_{T,\mr{max}}$
windows of the leading jet \cite{Abazov:2004hm}. The details of the
analysis are:
\bi
\item Reconstruct cone jets for $R=0.7$,
\item require $p_{T,2}>40$ GeV, and,
\item require central jet rapidities, $\abs{y_{1,2}}<0.5$.
\ei
\begin{figure}[t!]
  \vspace{0mm}
  \hspace*{2mm}
  \includegraphics[width=78mm,angle=0.0]{dipcas/jetdata/%
    ptr1_jet_1_1_dphi2_1.dat_1.eps}
  \hspace*{4mm}
  \includegraphics[width=78mm,angle=0.0]{dipcas/jetdata/%
    ptr2_jet_1_1_dphi2_1.dat_1.eps}
  \hspace*{2mm}
  \includegraphics[width=78mm,angle=0.0]{dipcas/jetdata/%
    ptr3_jet_1_1_dphi2_1.dat_1.eps}
  \hspace*{4mm}
  \includegraphics[width=78mm,angle=0.0]{dipcas/jetdata/%
    ptr4_jet_1_1_dphi2_1.dat_1.eps}
  \vspace{0mm}
  \myfigcaption{140mm}{The dijet azimuthal decorrelations in different
    $p_{T,\mr{max}}$ ranges. Dipole-shower results for different
    choices of the initializing scale are overlaid by data taken by
    D\O\ during Tevatron Run II \cite{Abazov:2004hm}.}
  \label{fig:jetdata2}
  \vspace{0mm}
\end{figure}
\Fig{fig:jetdata2} shows the data overlaid with predictions for
various choices of the initializing scale: besides the default given
in \eq{eq:jinivfr}, two alternatives have been implemented, namely a
geometric mean reading
\be
  p^2_{\perp,\mr{ini}}\left|{\hphantom{A}\atop\sd{\rm{Jets}}}\right.\;=\;
  3\;\mu^2_\mr{QCD}\;=\;
  3\;\frac{2\:\hat s\,\hat t\,\hat u}{\hat s^2+\hat t^2+\hat u^2}\,,
\label{eq:jiniqcd}
\ee
and a more enhanced scale defined as
\be
  p^2_{\perp,\mr{ini}}\left|{\hphantom{A}\atop\sd{\rm{Jets}}}\right.\;=\;
  (1+\sqrt{2}\x)^2\;\frac{\hat u\,\hat t}{\hat s}\,,
  \label{eq:jinihver}
\ee
using the Mandelstam variables of the core process.
The scale denoted by ``II sc.'' is taken according to the latter
equation, \eq{eq:jinihver}; this one denoted by ``QCD sc.''
corresponds to \eq{eq:jiniqcd}, and that marked as ``low default sc.''
follows from dividing \eq{eq:jinivfr} by a factor of $3$. Obviously,
the dipole shower initiated through the low default scale does not
account for enough hard emissions and overshoots for soft ones. The
other predictions are quite similar, with the ``II sc.'' and ``QCD
sc.'' variants giving slightly harder and softer results w.r.t.\ the
default case, respectively. The default performs best and will
therefore be employed in all what follows. Its predictions still tend
to undershoot the data around $\Delta\phi_\mr{dijet}=2.8$ in all
$p_{T,\mr{max}}$ windows of the leading jet, however, keeping in mind
that some gluon splitting processes have not been fully taken into
account yet, the agreement is satisfactory giving evidence that also
other model-intrinsic scales, such as \muf, \mufp\ and \mur, have been
chosen reasonably.
\begin{figure}[t!]
  \vspace{0mm}\bc\includegraphics[width=127mm,angle=0.0]{dipcas/jetdata/%
    jet0_jet_1_1_DJM1.dat_1.eps}
  \ec\vspace{0mm}
  \myfigcaption{140mm}{The dijet mass spectrum as measured by D\O\
    during Tevatron Run I \cite{Abbott:1998wh} compared with the
    prediction of the dipole shower initialized at scales according to
    \eq{eq:jinivfr} and for no phase-space restriction, \ie using the
    default $\hat s_\mr{max}$ settings.}
  \label{fig:jetdimass}
  \vspace{0mm}
\end{figure}
\subsubsection*{Dijet mass spectrum at Tevatron Run II energies}
With the $p_{\perp,\mr{ini}}$ finding in hand, the dipole-shower
prediction for the dijet mass spectrum is confronted with data
measured during Run I by the D\O\ collaboration \cite{Abbott:1998wh}.
The analysis requires:
\bi
\item the reconstruction of jets using a cone algorithm with $R=0.7$,
\item jet transverse energies above $30$ GeV, and,
\item the dijet candidates to satisfy $\abs{\eta_{1,2}}<1.0$.
\ei
As it can be read off \fig{fig:jetdimass}, the comparison versus data
with the dipole-shower results being normalized to the cross section
observed in the experiment shows encouraging agreement.
\begin{figure}[t!]
  \vspace{0mm}
  \hspace*{2mm}
  \includegraphics[width=78mm,angle=0.0]{dipcas/jetdata/%
    full_cuts_jet_1_1_eta_3.dat_1.eps}
  \hspace*{4mm}
  \includegraphics[width=78mm,angle=0.0]{dipcas/jetdata/%
    full_cuts_jet_1_1_Alpha_3.dat_1.eps}
  \vspace{0mm}
  \myfigcaption{140mm}{%
    Colour-coherence tests in inclusive three-jet production at
    Tevatron Run I energies according to a CDF study presented in
    \cite{Abe:1994nj}:
    (left panel) pseudo-rapidity distribution of the third jet and
    (right panel) the angle $\alpha$\/ (defined in the text).
    Experimental errors are statistical only and the histograms are
    normalized to their respective binwidth. For the latter three
    observables, dipole-shower (shower-level) predictions under full
    (blue solid lines) and restricted (black dashed lines) emission
    phase space are shown in comparison with the (detector-level) data
    of the CDF measurement \cite{Abe:1994nj}.}
  \label{fig:jetdata1}
  \vspace{0mm}
\end{figure}
\subsubsection*{Test of colour coherence at Tevatron Run I energies}
An interesting measurement and analysis was carried out by the CDF
collaboration during Tevatron Run I, searching for evidence for colour
coherence in $p\bar p$\/ collisions at $\sqrt S=1.8$ TeV
\cite{Abe:1994nj}.
Discriminatory observables were found for three-jet events
featuring a hard leading jet and a rather soft third jet. They were
shown to be sensitive to the correct treatment of QCD colour
coherence in parton shower simulations. Here, similarly to the
treatment in \cite{Schumann:2007mg}, this CDF analysis is used to test
whether the proposed dipole shower is capable of describing the
colour-coherence effects seen in the data.\spc\footnote{Evolving in
  terms of colour dipoles is said to automatically account for soft
  colour coherence owing to the eikonal structure of the dipole
  splitting cross sections, however, the colour factor ambiguities for
  quark--gluon dipoles (discussed in \sec{sec:FFdips} ff) require more
  serious investigation in this direction. The comparison with the CDF
  data is just a first step.}
The requirements of the CDF study read:
\bi
\item Jets are defined through a cone algorithm, using $R=0.7$,
\item the two leading jets are constrained to $\abs{\eta_{1,2}}<0.7$,
\item they have to be oriented back-to-back within $20$ degrees, \ie
  $\abs{\phi_1-\phi_2}>2.79$,
\item jet $E_T$ thresholds have to be respected for the first jet and
  all next-to-first jets of $110$ GeV and $10$ GeV, respectively, and,
\item for the $\alpha$\/ angle only, a cut on $\Delta
  R_{23}=\sqrt{(\eta_2-\eta_3)^2+(\phi_2-\phi_3)^2}$ has to be
  imposed, namely $1.1<\Delta R_{23}<\pi$.
\item The angle $\alpha$\/ is defined through
  \be
    \tan\alpha\;=\;\frac{\mr{sign}(\eta_2)(\eta_3-\eta_2)}
			{\abs{\phi_3-\phi_2}}\,.
  \ee
\ei
In \fig{fig:jetdata1} the comparison between detector-level data and
dipole-shower predictions obtained at the shower level is shown for
the $\eta_3$ and angle $\alpha$\/ distributions. As pointed out in
\cite{Abe:1994nj}, these two observables receive small detector
corrections only, which is not the case for the $\Delta R_{23}$
separation of the second and third hardest jet in $(\eta,\phi)$ space.
The latter is known to be strongly affected by detector effects,
therefore, not considered here.\spc\footnote{For the same reason, in
  the study of \cite{Schumann:2007mg} $\Delta R_{23}$ has not been
  taken into account either.}
If colour-coherence effects are modelled correctly, $\eta_3$ should
arise broader and feature a significant dip for central values. The
$\alpha$\/ spectrum should be minimal for small $\abs{\alpha}$
followed by a clear rise towards larger positive angles. As can be
seen in \fig{fig:jetdata1}, the dipole shower predicts these
characteristics, providing fairly good evidence that colour-coherence
effects are reasonably modelled. The agreement with data deteriorates
once the prediction is taken from a dipole shower where the $\hat
s_\mr{max}$ setting has been (considerably) reduced. This again
emphasizes that the natural choice is to assign the full phase space
to single emissions by using the default $\hat s_\mr{max}$ settings.
\subsubsection*{Exclusive three-jet final-state challenge}
Recent CDF measurements have found an excess in data of exclusive
three-jet events with small $\Delta R_{23}$, which is not described by
available tools, such as \pythia (Tune A)
\cite{Field:2007,Choudalakis:2007my}. In a first qualitative study the
potential of the new dipole shower to predict $\Delta R_{23}$
differently w.r.t.\ traditional leading-log showers is estimated.
Therefore, the following analysis has been applied:
\begin{figure}[t!]
  \vspace{0mm}
  \hspace*{2mm}
  \includegraphics[width=78mm,angle=0.0]{dipcas/jtev2cj/%
    Cjets1_jet_2_3_eta_3.dat_1.eps}
  \hspace*{4mm}
  \includegraphics[width=78mm,angle=0.0]{dipcas/jtev2cj/%
    Cjets1_jet_2_3_dR2_3_1.eps}
  \vspace{0mm}
  \myfigcaption{140mm}{Pseudo-rapidity distribution of the third jet
    and spatial separation between the second and third jet in
    exclusive three-jet final states simulated for Tevatron Run
    II energies. The plots are obtained requiring a larger jet $p_T$
    threshold for the hardest jet to pass. The \sherpa shower
    prediction generated by \apacic in CKKW scale-scheme mode (black
    dashed curves) is compared to the dipole-shower predictions,
    namely for default (blue solid curves) and lowered initializing
    scales (red dashed curves).}
  \label{fig:jettev2cj}
  \vspace{0mm}
\end{figure}
\bi
\item Require jet reconstruction according to the cone jet algorithm,
  use $R=0.4$,
\item use general cuts on jets of $p_{T,i}>20$ GeV and $\abs{\eta_i}<2.5$,
\item additionally, use $\abs{\eta_1}<1.0$ for the hardest jet, and,
\item consider the trigger-jet effect, \ie demand $p_{T,1}>40$ GeV.
\ei
The results are presented in \fig{fig:jettev2cj} and show the
pseudo-rapidity spectrum of the third jet together with the
$\Delta R_{23}$ distribution mentioned before. Concerning the former,
the dipole-shower variants are found to generate steeper spectra,
while inspecting the latter, the shower starting from the lower
initializing scale prefers populating the region of small jet
separations the most. The \apacic prediction rises towards smaller
separations as well but stays below the dipole shower curves and
features a broader tail. Additionally, by relaxing the $p_{T,1}>40$
requirement, treating all jets likewise, the higher $p_{T,1}$
threshold was identified as a major source in projecting out
the peak for low $\Delta R_{23}$. The different behaviour of the
dipole and parton shower eventually can be seen as a consequence of
generating the full radiation pattern differently. For example,
recalling that the $p^2_\perp$ definition includes a product of
two-parton squared masses $s_{12}s_{23}$, a small $s_{12}$ with a
potentially small angle between parton $1$ and $2$ can still be
compensated by a large $s_{23}$ giving the same $p^2_\perp$. In a
$1\to2$ splitting usually there is no such freedom of compensating a
small $s_{12}$, it might be rather cut away by the parton-shower
cut-off. In conclusion, the tendency of the dipole shower to enhance
the production of spatially less separated jets should be studied in
more detail and more realistically including underlying event
simulation etc.\ and, possibly, a direct comparison to data.
%

%% file: tex/dscon.tex
\section{Conclusions}
%
\label{sec:End}
In this publication, the colour-dipole shower approach based on the
Lund Colour Dipole Model \cite{Gustafson:1987rq,Lonnblad:1992tz} has
been extended. In this model, the description of sequences of QCD
emissions is formulated in an expansion around their soft limit. In
the context of hadronic collisions, a novel, perturbative description
of initial-state showering based on the emission properties of colour
dipoles has been developed, which is in clear contrast to the
corresponding Lund ansatz. In summary, initial-state radiation is
treated directly and not redefined by final-state radiation arising
from final-state dipoles that contain extended colour sources, which,
therefore, are subject of a semi-classically motivated suppression of
high $p_\perp$ emissions. In contrast, in the new model the hadron
remnants are completely kept outside the evolution. The fully
perturbative treatment led to the introduction of new dipole types,
which contain incoming partons.
\\
The description of gluon emissions off colour dipoles has been
generalized to account for all kinematic regions appearing in hadronic
collisions. It centers around a Lorentz invariant generalization of
the definition of the dipole evolution variables.
Splitting functions have been derived for the new dipole types.
Together with the well-known radiation pattern of pure final-state
colour dipoles, their utilization in a complete shower algorithm has
been presented to describe soft and collinear multiple parton
emission. The feasibility of the approach has been shown through its
successful application to electron--positron annihilation into
hadrons, inclusive Drell-Yan pair production and inclusive QCD jet
production at hadron colliders. All comparisons deliver encouraging
results in good agreement with other models and with experimental
data. It is worth to mention that the feature of generating broader
pseudo-rapidity spectra -- often mentioned in connection with
colour-dipole evolution according to \ariadne -- has not been
confirmed by the new model. Moreover, for the first time, results have
been presented for the inclusive production of jets in hadronic
collisions that have been obtained from a shower based on the
colour-dipole approach. First evidences could be given that the model
correctly accounts for colour-coherence effects.
\\
Taken together, an appealing picture of dipole cascading has been
achieved. Future work will concern the full incorporation of gluon
splittings in the initial and final state, and, the generalization to
finite quark masses. In addition, a merging with multi-leg tree-level
matrix elements for additional QCD radiation will be addressed, and, a
matching with full NLO QCD calculations shall be studied.
%

%% file: tex/dsackno.tex
\section*{Acknowledgements}
\addcontentsline{toc}{section}{Acknowledgements}
The authors would like to thank Mike Seymour and Steffen Schumann for
fruitful discussions, and are also thankful to the other members of
the \sherpa team for supporting this work.
\\
J.~W.\ thanks the CERN theory division for great hospitality during
his Marie Curie fellowship period, where parts of this work have been
accomplished. Furthermore, J.~W.\ acknowledges financial support by
the Marie Curie Fellowship program for Early Stage Research Training.
\\
The authors also acknowledge financial support by MCnet (contract
number MRTN-CT-2006-035606) and by BMBF.
\subsubsection*{}